\documentclass{aastex61}

\usepackage{graphicx}
\usepackage{verbatim,xspace,booktabs,xcolor}
\usepackage{natbib}
\usepackage{amsmath,amssymb,upgreek}
\usepackage{wasysym}   

\definecolor{amethyst}{rgb}{0.6, 0.4, 0.8}
\definecolor{green}{rgb}{0.55, 0.71, 0.0}
\definecolor{apricot}{rgb}{0.98, 0.81, 0.69}
\definecolor{auburn}{rgb}{0.43, 0.21,0.1}
\definecolor{babyblueeyes}{rgb}{0.63, 0.79, 0.95}
\definecolor{bittersweet}{rgb}{1.0, 0.44, 0.37}

\newcommand{\ccol}[1]{\textbf{\color{red}{#1}}}
\newcommand{\comm}[1]{\textcolor{black}{#1}}

\newcommand{\comn}[1]{\textcolor{black}{#1}}
\newcommand{\comnn}[1]{\textcolor{black}{#1}}

\newcommand{\mga}[1]{\textcolor{green}{Markus Gaug: #1}} 
\newcommand{\alm}[1]{\textcolor{amethyst}{Alison Mitchell: #1}} 

\newcommand{\ud}{\mathrm{d}}      
\newcommand\ddfrac[2]{\ensuremath{\frac{\displaystyle #1}{\displaystyle #2}}}  
\newcommand{\Lagr}{\mathcal{L}}   

\newcommand{\psic}{\upsilon}
\newcommand{\thetac}{\theta_\mathrm{c}}
\newcommand{\Qpix}{Q_\mathrm{pe}}
\newcommand{\Qpixi}{Q_{\mathrm{pe},i}}
\newcommand{\Qobs}{Q_{\mathrm{obs}}}
\newcommand{\Qobsi}{Q_{\mathrm{obs},i}}

\newcommand{\spi}{\sigma_{\mathrm{ped},i}}
\newcommand{\sit}{\sigma_\theta}

\DeclareMathOperator\acos{acos}
\DeclareMathOperator\sgn{sgn}

\submitjournal{ApJS}
\shorttitle{Muon Rings for Calibration of CTA}
\shortauthors{Gaug M., Fegan S., Mitchell A.M.W., Maccarone M.C., Mineo T., Okumura A.}

\begin{document}

\title{Using Muon Rings for the Calibration of the Cherenkov Telescope Array: A Systematic Review of the Method and its Potential Accuracy}

\correspondingauthor{Markus Gaug}
\email{markus.gaug@uab.cat}

\author{M.~Gaug}
\affiliation{Unitat de F\'isica de les Radiacions, Departament de F\'isica, and CERES-IEEC, Universitat Aut\`onoma de Barcelona, E-08193 Bellaterra, Spain}
\author{S.~Fegan}
\affiliation{Laboratoire Leprince-Ringuet, Ecole Polytechnique, CNRS/IN2P3, F-91128 Palaiseau, France 
}
\author{A.M.W.~Mitchell} 
\affiliation{Physik Institut, Universit\"{a}t Z\"{u}rich, Winterthurerstrasse 190, CH-8057 Z\"{u}rich, Switzerland}
\author{M.C.~Maccarone} 
\affiliation{Istituto di Astrofisica Spaziale e Fisica Cosmica di Palermo, INAF, Via Ugo La Malfa 153, I-90146 Palermo, Italy}
\author{T.~Mineo} 
\affiliation{Istituto di Astrofisica Spaziale e Fisica Cosmica di Palermo, INAF, Via Ugo La Malfa 153, I-90146 Palermo, Italy}
\author{A.~Okumura} 
\affiliation{Institute for Space–Earth Environmental Research and Kobayashi–Maskawa Institute for the Origin of Particles and the Universe, Nagoya University, Furo-cho, Chikusa-ku, Nagoya 464-8602, Japan}





\begin{abstract}
The analysis of ring images produced by muons in an Imaging Atmospheric Cherenkov Telescope (IACT) provides a powerful and precise method to calibrate the IACT 
optical throughput and monitor its optical 
point-spread function (PSF). First proposed by the Whipple collaboration in the early 90's, this method has been refined by
the so-called second generation of IACT experiments: H.E.S.S., MAGIC and VERITAS. We review here the progress made with these instruments and 
investigate the applicability of the method as the primary 
throughput calibration method for the different telescope types forming  the future Cherenkov Telescope Array (CTA). 
We find several additional systematic effects not yet taken into account by previous authors and propose several new analytical methods to include these in the analysis.
Slight modifications in hardware and analysis need to be made to ensure that such a calibration works as accurately as required for the CTA. We derive analytic estimates for the expected muon data rates for optical throughput calibration, camera pixel flat-fielding and monitoring of the optical PSF. 
The achievable statistical and systematic uncertainties of the method are also assessed.
\end{abstract}
%
%
\keywords{gamma rays: general, methods: data analysis, astroparticle physics, atmospheric effects, telescopes  
}

\section{Introduction}

Muon ring calibration for Imaging Atmospheric Cherenkov Telescopes (IACTs) was first suggested by \citet{HILLAS:JPGPP1990a} and \citet{rowell} and further elaborated in detail by \citet{vacanti}.
In addition to the Whipple telescope~\citep{fleury,jiang:1993,rose,rovero}, the method is used to calibrate the optical throughput of practically all currently operating IACTs, including HEGRA~\citep{puhlhofer2003}, H.E.S.S.~\citep{guy,leroy,chalmecalvet2014} and MAGIC~\citep{shayduk,goebel,meyermuons}, and \comm{used as a cross-calibration method for} VERITAS~\citep{humensky,hanna2008}. 

Muon rings 
were also used to calibrate 
the optical point-spread function (PSF) of the MAGIC telescopes~\citep{goebel,meyermuons}.
In-depth studies of the muon method have been carried out in several PhD theses, 
dedicated to the optical throughput calibration of the Whipple telescope~\citep{vacantiphd}, the HEGRA telescopes~\citep{Bolz:Dipl}, the H.E.S.S. telescopes~\citep{bolzphd,leroyphd,mitchellphd}, the CAT~telescope~\citep{iacoucci} and the monitoring of the optical PSF of MAGIC~\citep{GarczPhD}. Detailed muon calibration procedures have been described in internal reports of the MAGIC~\citep{Meyer:2005} and VERITAS~\citep{fegan2007} collaborations.


The Cherenkov Telescope Array (CTA)~\citep{cta,ctaconcept} is the next generation gamma-ray observatory comprising more than one hundred IACTs distributed over two sites (CTA-South at Cerro Armazones near Paranal, Chile, and CTA-North at ``Roque de los Muchachos'' Observatory (ORM) on the Canary Island of La Palma, Spain, both at around 2200~m a.s.l.) that will provide significant performance enhancement with respect to current facilities~\citep{bernloehr2013}\footnote{%
see also \url{https://www.cta-observatory.org/science/cta-performance/}}. 


CTA aims to increase the number of known very high energy (VHE; $E >10~\mathrm{GeV}$) \comn{gamma-ray sources} by a factor of ten and to make measurements of their spectral and morphological properties with unprecedented precision~\citep{ctascience}, for which a well-defined calibration approach, included as an integral part of the CTA design, is paramount~\citep{gaugSPIE2014,maccarone2017}. In order to meet the required accuracy \comn{of 10\% for the determination of the global energy scale}, the Cherenkov photon transmission through all detector components and subsequent conversion to photoelectrons must be known to better than 5\% accuracy, \comn{given that the systematic uncertainties due to the imperfect knowledge of the atmospheric conditions cannot be reduced to smaller than 8\% within reasonable efforts.} Cherenkov light emitted by local muons can serve as a  continuously available and well-understood calibration light source, the intensity of which is known to \comn{the accuracy with which the Cherenkov angle can be determined~\citep{pdg2017}}.
Because muon calibration can be carried out at the same time as the telescopes perform science observations, the muon calibration method has been chosen as the first option for monitoring of the optical bandwidth of all CTA telescopes~\citep{gaugSPIE2014}.



The various telescope types and sizes within CTA~\citep{cta,ctaconcept} are listed in Table~\ref{tab:tellist}~\citep[see][]{CTAspecs}\footnote{These include the Large-Sized-Telescope (LST), the Medium-Sized-Telescope (MST) and the Small-Sized-Telescope (SST).}. 
This article aims to address whether the muon calibration method will fulfill the precision and accuracy requirements for each telescope throughout the $>$30 years of  lifetime of CTA.  
In this paper, we critically review the method, while in a second follow-up paper, we will compare the performance of the muon calibration between telescopes and across different algorithms with dedicated simulations. Moreover, we investigate the accuracy that muons can achieve for the monitoring of the optical PSF and the camera pixel calibration of each telescope.

We will introduce the main concepts for muon calibration in Section~\ref{sec:method}. The selection and reconstruction of muon rings are explained in Section~\ref{sec:reconstruction}. 
The precise reconstruction of their impact distance, ring width, and the telescope's optical throughput is described in Section~\ref{sec:muon_efficiency}. 
Systematic effects are investigated in Sections~\ref{sec:secmuoneff}--\ref{sec:broadening}, and expected muon image rates are calculated in Section~\ref{sec:rates}. We discuss the findings of the previous sections in Section~\ref{sec:discussion} and draw conclusions in Section~\ref{sec:conclusions}.


\begin{table}[h!]
\centering
\begin{tabular}{lcccr@{.}lr@{.}lcr@{.}l} 
\toprule
    &         & \multicolumn{1}{c}{Geometrical}  & \multicolumn{1}{c}{Number} & \multicolumn{2}{c}{Reflector} & \multicolumn{2}{c}{Camera} & Number & \multicolumn{2}{c}{Pixel}  \\
    &  Optics Type & \multicolumn{1}{c}{Mirror Area}  & \multicolumn{1}{c}{ (Primary)}  & \multicolumn{2}{c}{Aperture}  & \multicolumn{2}{c}{FOV} & Camera   & \multicolumn{2}{c}{FOV} \\ 
   &   & \multicolumn{1}{c}{$A_\mathrm{geom}$}  & \multicolumn{1}{c}{Mirror Facets}  & \multicolumn{2}{c}{$2R$}  & \multicolumn{2}{c}{$\Omega$} & Pixels   & \multicolumn{2}{c}{$\omega$}  \\ 
&         & \multicolumn{1}{c}{(m$^2$)}  & \multicolumn{1}{c}{$N_\mathrm{mirrors}$}   & \multicolumn{2}{c}{(m)}  & \multicolumn{2}{c}{(deg) } &  $N_\mathrm{pixels}$ & \multicolumn{2}{c}{(deg) }    \\\addlinespace[0.1cm]
\midrule \addlinespace[0.15cm]
LST  & Tessellated Parabolic 
&  378   
& 198  
&  ~~23&6$^*$ 
&   4&5  
& 1855 
&   0&10$^+$  
\\ \addlinespace[0.1cm]  
MST  & Modified Davies-Cotton$^\ddagger$ 
& 88.5   
& 86 
& 12&5\rlap{$^*$} 
& 7&8\rlap{$^\dagger$}  
&  \llap{$\sim$}1800\rlap{$^\dagger$} 
&  0&175\rlap{$^+$} 
\\  \addlinespace[0.1cm] 
SST (\textit{ASTRI}) & Schwarzschild-Couder 
& 9.2  
& 18  
& 4&1\rlap{$^*$} 
& 10&5     
& 2368     
& 0&19\rlap{$^{\ast\ast}$} 
\\  \addlinespace[0.1cm] 
SST (\textit{GCT})  & Schwarzschild-Couder 
& 9.4  
& 6    
& 4&0  
& 8&3  
& 2048 
& 0&15\rlap{$^{\ast\ast}$} 
\\\addlinespace[0.1cm] 
SST-1M & Davies-Cotton 
& 9.4   
& 18   
& 3&8\rlap{$^*$}   
& 9&1    
& 1296   
&  0&24\rlap{$^+$} 
\\  \bottomrule 
\end{tabular}
\caption{\label{tab:tellist}Current CTA telescope designs \citep{canestrari2013,volpe2014,teshima2014,Inome:2014,schlenstedt2014_2,catalano2014,glicenstein2014,Puehlhofer:2015,Rulten:2016,Glicenstein:2017,Mazin:2017,Sol:2017vcb,White:2017kpl,Samarai:2017}.  The geometrical mirror area has been corrected for the effect of shadowing by the camera or the secondary mirror for the case of the Schwarzschild-Couder design for on-axis incidence of light, but not for camera support structures and other material elements, like e.g. baffles. 
$^*$ Averaged over all azimuth angles, individual values vary around this number because of the hexagonal geometry of the tessellated mirror facets. $^+$ flat-to-flat distance of a hexagonal light funnel in front of the photon detector \comn{without taking into account the thickness of the plastic and the intended gap between funnels, otherwise 0.002$^\circ$--0.003$^\circ$ less}. $^\ast$$^\ast$ Flat-to-flat distance of a square shaped photon detector \comn{without taking into account the $\sim 0.2$~mm gap between pixels}. $^\dagger$ Two MST camera designs are currently developed which differ by 0.2$^\circ$ in their respective camera fields-of-view. $^\ddagger$ The MST optics have been modified with respect to the ``classical'' Davies-Cotton design such that the radius of curvature of the primary reflector does not correspond exactly to the focal length of the individual mirror facets, but instead to 1.2 times their focal length. This setup improves the timing performance of the reflector with respect to the ``classical design''.
}
\end{table}

\clearpage

\section{Review of the Muon-Calibration Method for IACT's \label{sec:method}}

Single muons form naturally as part of hadronic air showers and 
generate Cherenkov light. 
Most muons detected by IACTs are minimum ionizing particles with a long lifetime. 

Muons hence have a large penetration depth, traveling in straight lines toward the ground with Cherenkov radiation emitted under constant opening angle in a cone around the muon direction of travel. As IACTs image in angular space, the Cherenkov light from \textit{local}\footnote{%
Local muons are considered those that either hit the telescope mirror directly or come very close to it.
} muons traveling parallel (or almost parallel) to the telescope's optical axis forms a ring-shaped image in the camera. The ring appears complete provided that the muon passes through the primary reflector of the telescope, such that Cherenkov light is reflected from all azimuthal angles around the muon path.


\begin{figure}
\centering
\includegraphics[width=0.55\linewidth]{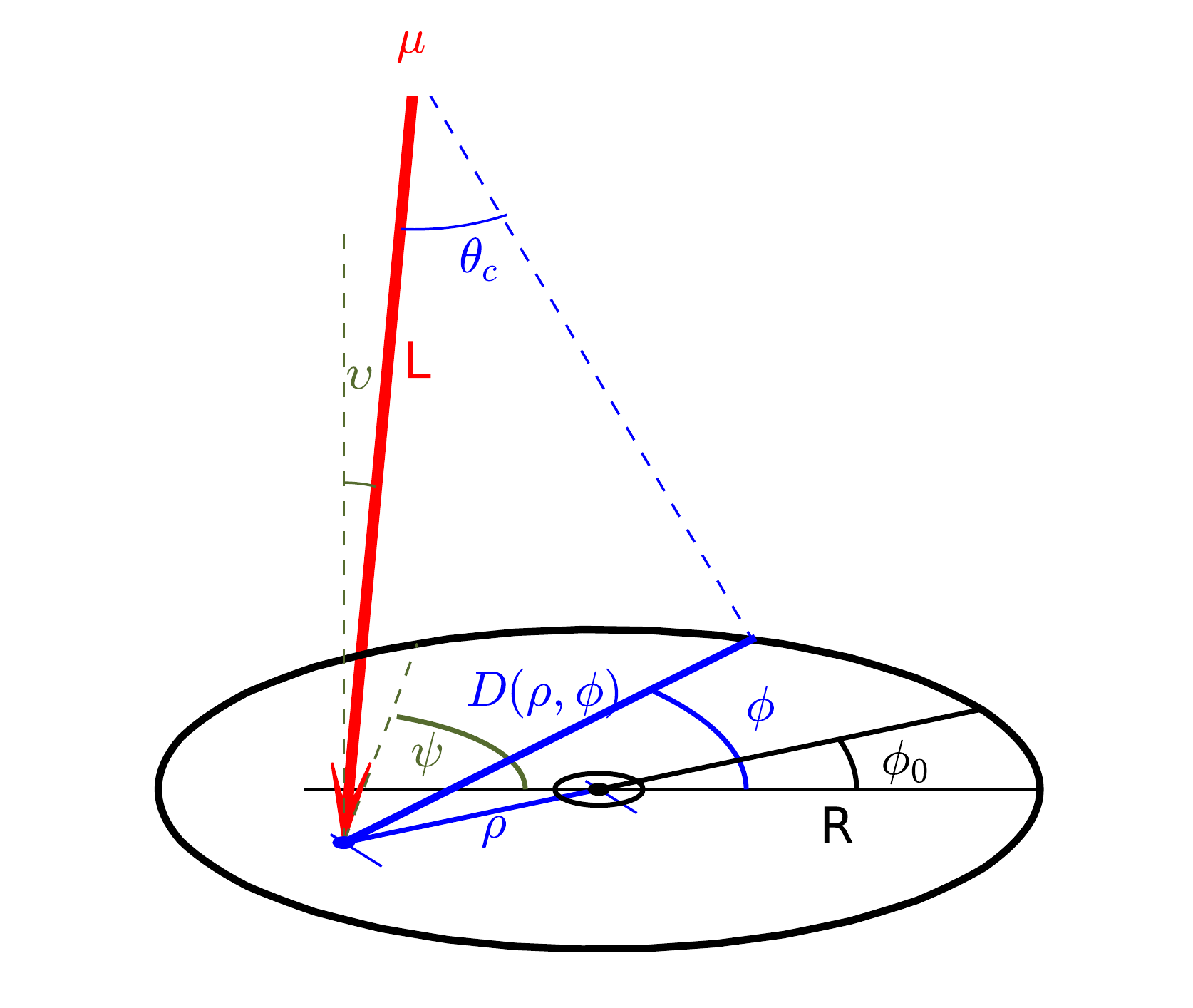}
\caption{Sketch of the parameters introduced to describe the geometry of a local muon $\mu$ and its image in an IACT camera. The muon generates
Cherenkov light at the Cherenkov angle $\thetac$ along its trajectory, which is inclined by the angle $\upsilon$ with respect to the optical axis of the telescope.
It finally hits the telescope mirror, of radius $R$, at an impact distance $\rho$ from its center. \comn{The Cherenkov light illuminates the mirror along the chord $D(\rho,\phi)$, where the azimuth angle $\phi$ runs from zero to $2\pi$. The angle $\phi_0$ denotes the azimuth angle of the longest chord. The angle $\psi$ denotes the azimuthal projection of the muon inclination angle, and the small hole in the center the often-found gap at the center of the mirror \comn{support}.} For reasons of visibility, the angles and lengths are not to scale.
\label{fig:geometry} }
\end{figure}

Figure~\ref{fig:geometry} recalls the geometry of the system and introduces the parameters: the impact distance $\rho$, the inclination angle $\upsilon$, and the
Cherenkov angle $\thetac$. The telescope mirror has a radius $R$. As long as $(\upsilon+\thetac)$ is smaller than the \comn{half} the camera field-of-view (FOV), 
\comn{the telescope} will integrate light along the cord $D$. The cord itself is a function of
$\rho$ and the azimuth angle $\phi$, defined with respect to a reference angle $\phi_0$, normally chosen as the one at which the cord $D$ is largest.

The number of Cherenkov photons $N_c$
emitted in the photon energy range ($\epsilon_1,\epsilon_2$) per unit path length $x$ by a muon in air can then be calculated from the Frank-Tamm formula, provided that 
the muon energy is above the Cherenkov emission threshold~\citep[see, e.g., chapter 34.7.1. of ][]{pdg2017}: 
\begin{equation}
\frac{\mathrm{d}^2N_c}{\mathrm{d}x\,\mathrm{d}\phi} = \ddfrac{\alpha}{hc} \cdot \int_{\epsilon_1}^{\epsilon_2} \left( 1 - \ddfrac{1}{\beta_\mu^2(x) n^2(\epsilon,x)} \right) ~ \ud\epsilon
~,
\label{eq:dNdx}
\end{equation}
\noindent
where $\alpha \approx 1/137$ is the fine-structure constant, \comn{$h$ the Planck constant}, $\beta_\mu$ the velocity of the muon, in units of the speed of light in vacuum $c$, and $n$ the refractive index of the surrounding air.
The velocity of the muon, in turn, is related to the Cherenkov angle via
\begin{equation}
\cos\thetac(x,\epsilon) = \ddfrac{1}{\beta_\mu(x)\cdot n(\epsilon,x)} ~.
\label{eq:costheta}
\end{equation}
Requiring $(\cos\thetac < 1)$ leads to a threshold energy $E_t$ for the muon to emit Cherenkov light,
namely,
\begin{equation}
E_t(x,\epsilon) = \ddfrac{m_\mu \cdot c^2 }{\sqrt{1 - 1/n(\epsilon,x)^2}} ~,
\label{eq:Et}
\end{equation}
\noindent
where $m_\mu \approx 105.7$~MeV is the muon rest mass and 
$E_t \approx 5$~GeV for a reference altitude of 2200~m~a.s.l.

Using the notation $n = 1+\varepsilon$ and assuming $\varepsilon \ll 1$, always valid for air, and $E_\mu \gg m_\mu c^2$, the Cherenkov angle can be approximated to good accuracy as
\begin{align}
\thetac &\simeq \theta_\infty \cdot \sqrt{1 - (E_t / E_\mu )^2  } \label{eq:thetac} \\[0.2cm] 
        & \mathrm{with}  \nonumber\\[0.2cm]
\theta_\infty &\approx  \sqrt{2 \varepsilon} \label{eq:thetainf} ~ . 
\end{align}
\comn{The approximation is better than one in thousand for all cases considered in this paper}.

Because muon Cherenkov light observable by CTA telescopes is always emitted from heights of less than 1.5~km above ground, the refractive index can be assumed constant along $x$ to first order, depending only on the observatory altitude.
We can then simplify Eq.~\ref{eq:dNdx} to the more convenient form~\citep[][chap. 35]{pdg2017}:
\begin{equation}
\ddfrac{\ud ^3N_c}{\ud x\,\ud \epsilon \,\ud \phi} \simeq \frac{\alpha}{hc} \cdot \sin^2\thetac  ~,
\label{eq.dNde}
\end{equation}
\noindent
where we have introduced the photon 
optical bandwidth $\ud \epsilon$.  Note that the \comm{photon} spectrum Eq.~\ref{eq.dNde} shows no explicit dependency on the energy of the Cherenkov photon, in contrast to the conventional parameterization in terms of the wavelength~\citep{vacanti}. The spectrum depends only indirectly on the photon energy through its slight dependency of the refractive index, and hence $\sin\thetac$.

Integrating over the path length $L = D(\vec{\rho},\phi)/\tan\theta_c$, visible for the camera, we obtain a prediction 
for the number of \textit{observed} photoelectrons $N_\mathrm{pe}$, detected by the light sensors of the camera along the ring:
\begin{align}
\ddfrac{\mathrm{d}N_\mathrm{pe}}{\ud\phi}(\rho_R,\phi_0) &= \frac{\alpha}{hc} \cdot \sin^2 (\thetac) \cdot L \cdot \Theta(\vec{\rho},\vec{\upsilon},\phi) \cdot B_\mu \quad, \nonumber\\
                       &= 
\frac{\alpha}{2hc} \cdot  \sin(2\thetac) \cdot D(\vec{\rho},\phi) \cdot \Theta(\vec{\rho},\vec{\upsilon},\phi)\cdot B_\mu\quad.  \label{eq:dNdphi}
\end{align}
\comn{In the case that the telescope mirror may be approximated by a circle of radius $R$, the following analytical form for $D(\rho,\phi)$ may be derived:}
\begin{align} 
D(\vec{\rho},\phi)  &=  R \cdot \left\{
\begin{array}{lll}
                 2 \sqrt{1-\rho_R^2\sin^2(\phi-\phi_0)} \quad & \mathrm{for:}& \rho_R > 1 \\[0.3cm]
                  \sqrt{1-\rho_R^2\sin^2(\phi-\phi_0)} + \rho_R \cos(\phi-\phi_0) \quad & \mathrm{for:} & \rho_R \leq 1 \\
\end{array}
\right.\label{eq:D}\\[0.25cm]
 & \mathrm{and} \quad \rho_R := \rho/R  \quad,\nonumber\\
 & \qquad \,\quad\vec{\rho} ~=~ \begin{pmatrix}\rho_x\\\rho_y\end{pmatrix} ~ =~ \rho_R \cdot \begin{pmatrix}\cos(\phi_0\comn{+\pi})\\\sin(\phi_0\comn{+\pi})\end{pmatrix} \label{eq:vecrho} \\
 & \qquad \,\quad\vec{\upsilon} ~=~ \begin{pmatrix}\upsilon_x\\\upsilon_y\end{pmatrix} ~ =~ \upsilon \cdot \begin{pmatrix}\cos\psi\\\sin\psi\end{pmatrix} \quad.
\label{eq:nphotons}
\end{align}
\noindent
Eq.~\ref{eq:dNdphi} and~\ref{eq:D} neglect the influence of the inclination angle on the observable muon track length but are accurate to $O(\tan\thetac \cdot \tan\upsilon)$ for small \comn{inclination} angles (see Appendix~\ref{sec:inclinedmuons}).
Here  we have introduced the possibility of obscuration of Cherenkov light by shadowing elements (such as the camera) in the parameter $\Theta(\vec{\rho},\vec{\upsilon},\phi)$. In the case of roundish cameras with radius $r$ and small inclination angles, or central holes in the mirror dish, $\Theta(\vec{\rho},\vec{\upsilon},\phi)$ can be approximated as
\begin{align}
\Theta(\vec{\rho},\vec{\upsilon},\phi) &\approx \Theta(\rho_R,\phi) = 1 - \ddfrac{r}{R} \cdot \ddfrac{D(\rho_R\cdot\ddfrac{R}{r},\phi)}{D(\rho_R,\phi)}
\label{eq:Thetaapprox}
\end{align}

Finally, the total optical \textit{bandwidth}\footnote{Note that $B_\mu$ is essentially the same as $I$ in \citet{vacanti}, although their definition is given in 
units of nm$^{-1}$ \comn{and the integration limits of 300~nm and 600~nm have been approximated to a normal range of sensitivity of IACT photomultipliers.}.} 
of the traversed atmosphere and the detector is
\begin{align}
B_\mu &= \int_0^\infty t_\mathrm{\mu}(\epsilon) \cdot \xi_\mathrm{det}(\epsilon) ~\ud \epsilon
\quad, 
\label{eq:Bmu}
\end{align}
\noindent
where $t_\mathrm{\mu}(\epsilon)$ denotes the atmospheric transmission, namely, the probability of a Cherenkov photon of energy $\epsilon$, emitted by the muon, \textit{not} to get absorbed or scattered out of the reach of the detector. The transmission term $t_\mathrm{\mu}(\epsilon)$ has a slight dependency on the impact distance and the telescope pointing zenith angle $\vartheta_\mathrm{tel}$. We will show, however, that the dependency is negligible. Finally,  $\xi_\mathrm{det}(\epsilon)$ is the probability of detecting (i.e. converting to a photoelectron) a Cherenkov photon that has hit the telescope mirror.
Modern IACTs, with a sensitive range from $\sim$2~eV ($\sim$600~nm) to $\sim$4~eV ($\sim$300~nm), show a bandwidth of about $B_\mu \sim 0.65$~eV. 

\begin{figure}
\begin{minipage}{\textwidth}
\centering
\includegraphics[width=0.75\linewidth]{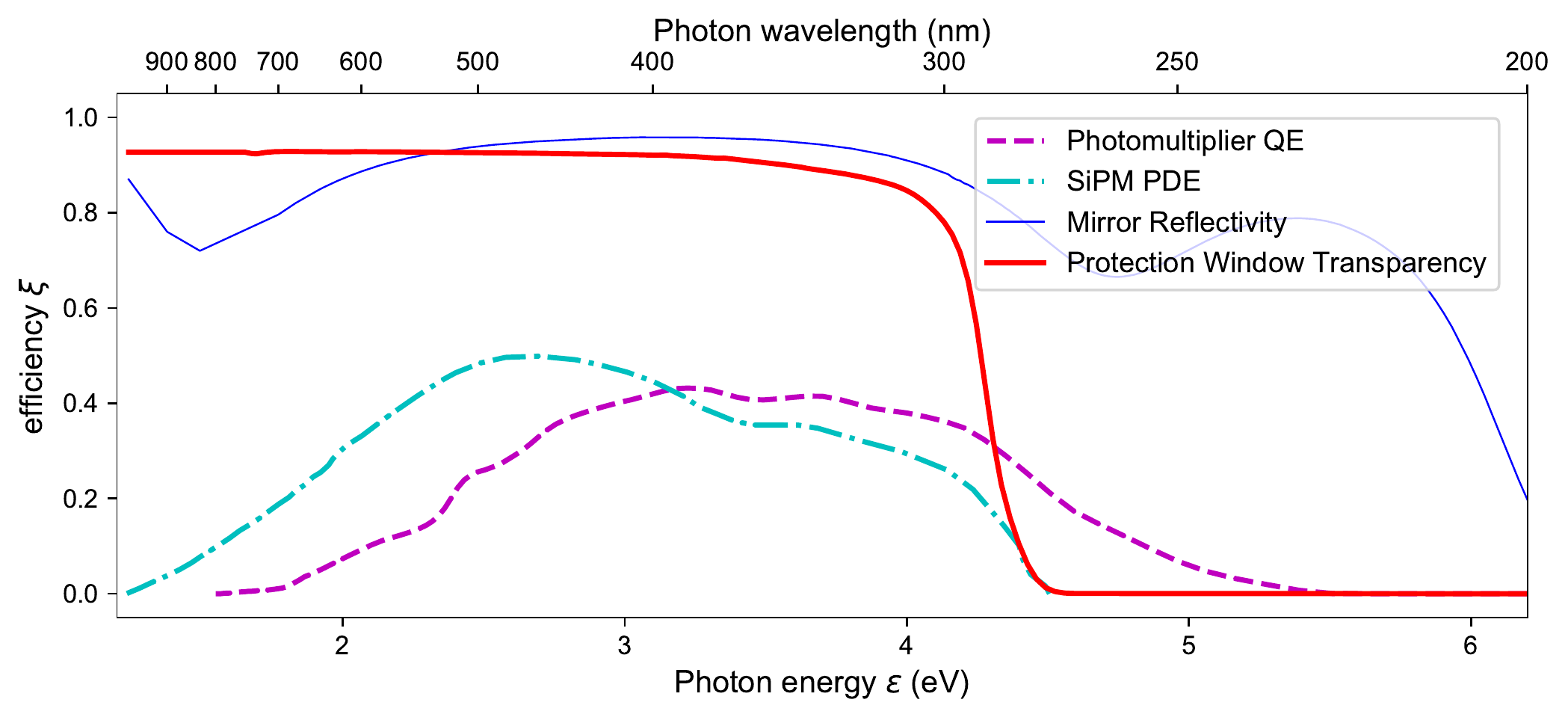}
\caption{\label{fig.xis} Several typical examples of the contributions to the effective optical bandwidth of the telescope: \comn{photomultiplier quantum efficiency (QE),  silicon PM (SiPM), photon detection efficiency (PDE), mirror reflectance, and the camera protection window transparency.
    For the SiPM PDE, the new type with silicone resin coating,\textsuperscript{a},
    has been chosen.}}
\small\textsuperscript{a}{\scriptsize see \url{https://www.hamamatsu.com/resources/pdf/ssd/s13360_series_kapd1052j.pdf}}
\end{minipage}
\end{figure}

The detector efficiency $\xi_\mathrm{det}(\epsilon)$ can be written itself as a combination of mirror reflectance $\xi_\mathrm{mirr}$, camera window transparency $\xi_\mathrm{window}$, light-\comm{concentrator} efficiency $\xi_\mathrm{concentrator}$ and detector photon detection efficiency (PDE) $\xi_\mathrm{pde}$: 

\begin{equation}
\xi_\mathrm{det}(\epsilon) =  \xi_\mathrm{mirr}(\epsilon) \cdot \xi_\mathrm{window}(\epsilon) \cdot \xi_\mathrm{concentrator}(\epsilon) \cdot \xi_\mathrm{pde}(\epsilon) \quad,  
\label{eq:xdet}
\end{equation}
\noindent
\comm{where we have omitted, for simplicity, secondary dependencies of each contribution to $\xi_\mathrm{det}(\epsilon)$ on $\rho, \upsilon$ and $\phi$, which will be treated in later sections of this paper.}  
Figure~\ref{fig.xis} shows a typical example for several of the mentioned contributions.

Muon calibration then means measuring $B_\mu$, with the best possible resolution and accuracy, 
from estimators of $N_\mathrm{pe}$, $\theta_c$, $\rho$ and $\phi_0$. We cannot use muon calibration to 
retrieve any information about $\xi_\mathrm{det}(\epsilon)$, beyond its (weighted) integral. It is, however, possible to determine dependencies of $\xi_\mathrm{det}$ on the impact point of a Cherenkov photon of the mirror (through its dependency on $\rho$), and the camera pixel (through its reconstruction of the inclination angle),
if sufficient statistics have been accumulated.

Many of the gamma-ray parameters are, however, retrieved in an IACT analysis by employing the full shape of $\xi_\mathrm{det}(\epsilon)$ as, for instance, in Monte Carlo (MC) simulations of the detector response to Cherenkov light, $B_\gamma$, emitted from gamma-ray-induced air shower particles:

\begin{equation}
B_\gamma(h,d,\vartheta_\mathrm{tel}) = \int t_\mathrm{\gamma}(h,d,\vartheta_\mathrm{tel},\epsilon) \cdot \xi_\mathrm{det}(\epsilon) ~\ud \epsilon \quad.  
\label{eq:Bgamma}
\end{equation}
Here $t_\gamma(h,d,\vartheta_\mathrm{tel},\epsilon)$ is the atmospheric transmission for Cherenkov light from gamma-ray showers, 
the properties of which depend on the emission height of the shower particles $h$, impact distance $d$ and 
observation zenith angle $\vartheta_\mathrm{tel}$. The detector efficiency $\xi_\mathrm{det}(\epsilon)$ is normally 
measured at or before the construction of an IACT, and particularly its degradation with time requires calibration. 

We can formally relate $B_\mu$ and $B_\gamma$:
\begin{align}
B_\gamma(h,d,\vartheta_\mathrm{tel})  &= B_\mu \cdot C_{\mu-\gamma} (h,d,\vartheta_\mathrm{tel}) \nonumber\\[0.25cm]
                  & \quad \mathrm{with} \nonumber\\[0.25cm]
        C_{\mu-\gamma}(h,d,\vartheta_\mathrm{tel}) &= \ddfrac{\int \xi_\mathrm{det}(\epsilon) \cdot t_\gamma(h,d,\vartheta_\mathrm{tel},\epsilon)~\ud\epsilon}
                              {\int\xi_\mathrm{det}(\epsilon) \cdot t_\mu(\epsilon)  ~\ud\epsilon}  
                              \quad
\label{eq:epsgamma}
\end{align}
\noindent
and need to ensure that $C_{\mu-\gamma}(h,d,\vartheta_\mathrm{tel})$ cannot degrade with time beyond a predefined limit, 
\textit{by design} of the optical elements of the telescope and for all relevant combinations of $h,d$ and $\vartheta_\mathrm{tel}$\footnote{This part of the muon analysis has often been greatly overlooked in the past, sometimes yielding strongly biased results.}. 

Atmospheric transmission of muon light $t_\mu(\epsilon)$ is typically higher than $t_\gamma(h,d,\vartheta_\mathrm{tel},\epsilon)$  and 
extends to larger energies (smaller wavelengths), 
because both the strong extinction of UV light by Rayleigh scattering 
and absorption by ozone are greatly reduced during the short path traversed by the muon Cherenkov light. 


We will hence be obliged to interfere already in the design of the CTA telescopes and cameras to ensure that their possible (chromatic) degradation cannot exceed a predefined limit, as will be shown in Section~\ref{sec:systematic_C}. \\

Whereas the (effective) mirror radius is determined by the chosen hardware, the impact distance has to be retrieved by
the \textit{modulation of light intensity around the ring}. For full rings, the light intensity reaches a maximum at $\phi=\phi_0$, and is proportional to $(R+\rho)$,
while the minimum is reached at $\phi=\phi_0+180^\circ$, where the amount of light scales with $(R-\rho)$. If the impact distance is zero, the modulation along the ring becomes flat.


It follows immediately from Eq.~\ref{eq:D} that for $\rho_R < 1$, a full ring is visible in the camera (i.e. all values of $\phi$ are
allowed), while otherwise only a part of the ring is imaged \comn{within an \textit{azimuth angle range}, fulfilling}:
\begin{equation}
\comn{|\sin(\phi-\phi_0)| < 1/\rho_R \quad \mathrm{and} \quad  \cos(\phi-\phi_0) > 0 }~~.
\end{equation}
Furthermore, a \textit{maximum emission height} $h_\mathrm{max}$ can be derived from which a full muon ring is visible in the camera:
\begin{equation}
h_\mathrm{max} = \left[ R \cdot \cot\thetac, 2 R \cdot \cot\thetac\right] \simeq \left[ R /  \sqrt{2 \epsilon}, R \cdot \sqrt{2 / \epsilon}\right] ~,
\end{equation}
\noindent
where the range applies from muons with zero impact distance to those hitting the edge of the mirror. 

One can also derive the amount of charge (in photoelectrons) received by one single camera pixel, characterized by its \textit{pixel FOV} ($\omega$), which covers an azimuth angle range of $\Delta\phi = \omega/\thetac$:
\begin{align}
\Qpix(\phi|\rho,\phi_0,\omega) &\simeq 
 \frac{\alpha}{2hc} \cdot \frac{\omega}{\thetac} \cdot \sin(2\thetac) \cdot D(\rho,\phi-\phi_0) \cdot B_\mu  ~, \nonumber\\
%
%
 & \simeq  1.0\times 10^2 \cdot \left(\ddfrac{B_\mu}{\mathrm{eV}}\right) \cdot \left(\ddfrac{\omega}{\mathrm{deg}}\right) \cdot \left(\ddfrac{D(\rho,\phi-\phi_0)}{\mathrm{m}}\right) \qquad,
\label{eq:Ipix}
\end{align}
where the last line has been obtained by using a Cherenkov angle of 1.23$^\circ$.


Using the pixel FOV, one can \comn{roughly} estimate the number of pixels hit by the muon light (assuming that the imaged ring width is slimmer than the size of one pixel \comn{and that the hit pixels are aligned on the ring with a constant separation angle}) and in the following a maximum impact distance to trigger the readout:
\begin{align}
N_\mathrm{pix} &= \left\{  \begin{array}{lll}
2 \phi_\mathrm{max} \cdot \thetac / \omega  & \quad\textrm{if}\quad & ~\rho_R > 1 \\[0.17cm]
2 \pi \cdot \thetac / \omega  & \quad\textrm{if}\quad & ~\rho_R \leq 1 \\
\end{array} \right. ~. \label{eq:npix}
\end{align}

The total number of observed photoelectrons can be obtained by integrating Eq.~\ref{eq:dNdphi}:
\begin{align}
  Q_\mathrm{tot}(\theta_c,\rho) &= 2 R \cdot \frac{\alpha}{hc} \cdot \sin(2\theta_c) \cdot B_\mu  \cdot \int_0^\Phi \Theta(\vec{\rho},\vec{\upsilon},\phi) \cdot \sqrt{1 -\rho_R^2\sin^2\phi}~ \mathrm{d}\phi \nonumber\\[0.2cm]
            {}  & \qquad \mathrm{with:} ~ \Phi =  \left\{ \begin{array}{lcl}
    \mathrm{arcsin}(1/\rho_R) & ~~\mathrm{for:~} & \rho_R > 1 \\[0.2cm]
    \pi / 2       & ~~\mathrm{for:}~ & \rho_R \leq 1
\end{array} \label{eq:Qtot}
\right. \\[0.2cm]
        &\approx U_0 \cdot \theta_c \cdot E_0(\rho_R) \label{eq:throughputcalibration}\\[0.2cm]
       {} & \qquad \mathrm{with:} \left\{ \begin{array}{lcl}   
        U_0 &=& R \cdot \ddfrac{\alpha}{\hslash c} \cdot B_\mu \\[0.25cm]
        E_0(\rho_R) &=& \frac{2}{\pi} \int\limits_0^\Phi \Theta(\rho,\upsilon) \cdot \sqrt{1 -\rho_R^2\sin^2\phi}~ \ud\phi \quad, 
        \end{array}
\right.
        \label{eq:e0}
\end{align}\noindent
where we have \citep[following the notation of][]{fegan2007}
split the total muon image size into the total muon size detected at zero impact distance ($U_0$), the reconstructed Cherenkov angle $\thetac$, and
a Legendre elliptic integral of the second kind that can be evaluated numerically. 
In the case of full rings, and neglecting $\Theta(\rho,\upsilon)$, 
$E_0(\rho_R)$ becomes $2/\pi$ times the complete elliptic integral of the second kind $E(\rho_R)$\footnote{%
see, e.g., \url{http://mathworld.wolfram.com/EllipticIntegraloftheSecondKind.html}.}.

\begin{figure}
\centering
\includegraphics[width=0.75\linewidth]{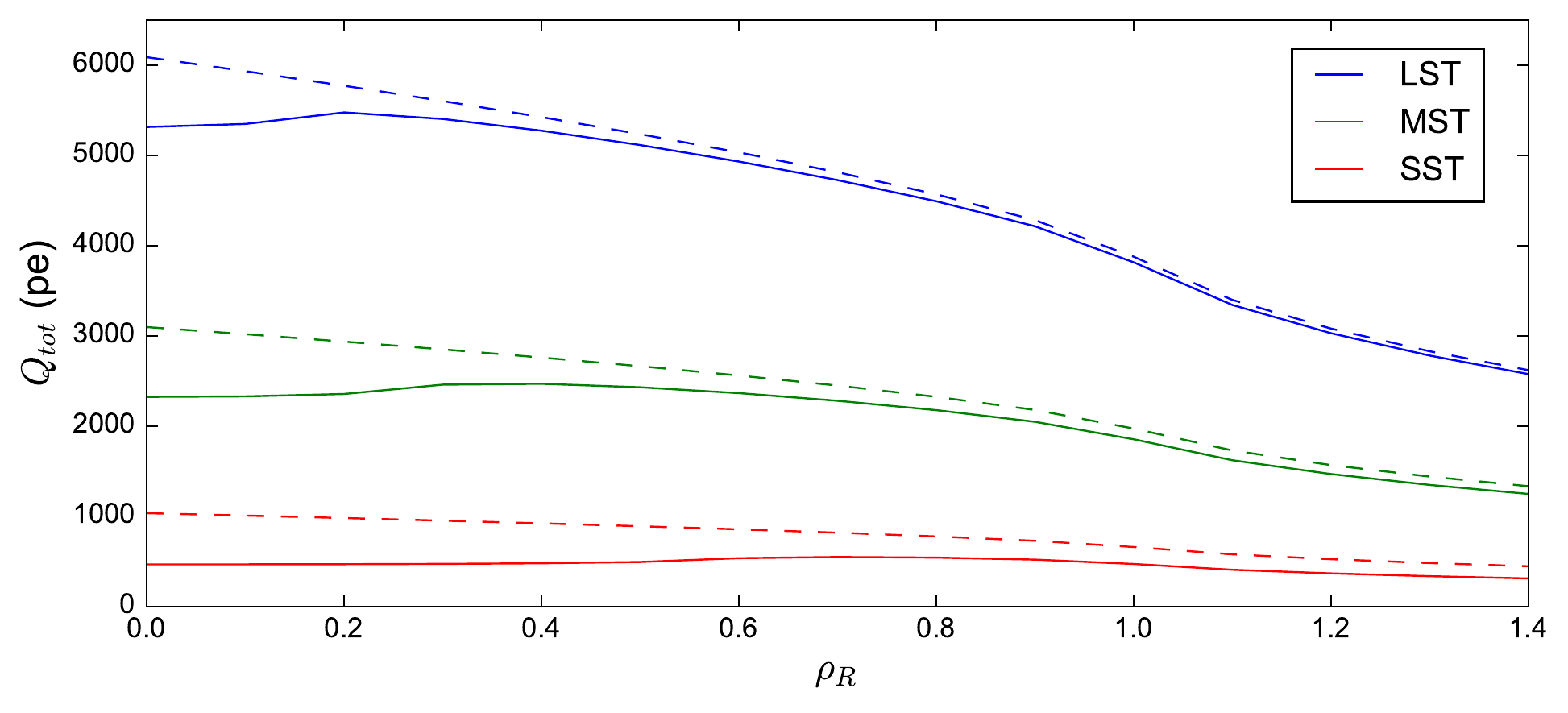}
\caption{Integrated amount of photoelectrons from muon Cherenkov light, \comn{using Eq.~\protect\ref{eq:Qtot}}, 
impinging on the telescope and registered in the camera, 
for the three different telescope designs for the CTA. 
The dashed lines show the case of \comn{no shadows}, while the solid lines show the situation with a corresponding
shadow from either an \comn{approximated roundish} camera or \comm{a possible} secondary mirror for the SST. 
A common value of $B_\mu \simeq 0.65$~eV has been assumed.
\label{fig:Ntot} 
}
\end{figure}

Figure~\ref{fig:Ntot} shows the total number of \comn{photoelectrons} (using $B_\mu = 0.65$~eV) 
impinging on the telescope for each of the \comm{three}  telescope types of the CTA, as a function
of the muon impact distance, according to Eq.~\ref{eq:throughputcalibration}. 
The difference between full and dashed lines shows the effect
of the central hole in the primary mirrors, subtracted from the full mirror estimate using again Eq.~\ref{eq:throughputcalibration}.

For the propagation of uncertainties of the reconstructed impact distance, it is useful to know the derivative of $E(\rho_R)$, namely~\citep[][p. 521]{WhittakerWatson}: 

\begin{align}
 \ddfrac{\delta E(\rho_R)}{E(\rho_R)} &= \ddfrac{1}{E(\rho_R)}\cdot|\frac{\ud E(\rho_R) }{\ud \rho_R}| \cdot \delta \rho_R\nonumber\\[0.15cm]
                         &= \ddfrac{K(\rho_R)/E(\rho_R)-1}{\rho_R} \cdot \delta \rho_R  \nonumber\\[0.15cm]
                         &=      \ddfrac{\delta \rho_R}{F(\rho_R)}  \label{eq:deltaE}   
\end{align}
%
%
where $K$ is the complete elliptic integral of the first kind. The function $F$ ranges from $F=2$ at $\rho_R=0$ to $F \approx 1.34$ at $\rho_R=0.5$ and 
$F=0$ at $\rho_R=1$. 

The total uncertainty of $B_\mu$ relates then to the uncertainties, with which the ingredients of Eq.~\ref{eq:Qtot} can be reconstructed, as

\begin{equation}
\ddfrac{\delta B_\mu}{B_\mu} \approx \sqrt{
  \left(\ddfrac{\delta Q_\mathrm{tot}}{Q_\mathrm{tot}}\right)^2
+ \left(\ddfrac{\delta \thetac}{\thetac}\right)^2
+ \left(\ddfrac{\delta R}{R}\right)^2
+ \left(\ddfrac{\delta \rho_R}{F(\rho_R)}\right)^2 
}\quad,
\end{equation}
\noindent
where we have not yet considered the uncertainty introduced by shadows. 

\clearpage
\section{Muon ring reconstruction\label{sec:reconstruction}}

Geometrical reconstruction of the muon ring in order to recover the Cherenkov opening angle $\thetac$, and  the inclination angle $\vec{\psic}$ of the muon typically involves careful signal extraction, pixel selection, and event cuts to achieve robust and unbiased results. In this section, we present the main ideas, with the associated limitations and accuracy for the CTA telescopes presented later.



The schemes described below are viable for all CTA telescopes, using regular data taken close to or contemporary with normal science observations, but require improving some of the techniques previously applied. At least for the MSTs and SSTs, an efficient single telescope trigger and muon flagging scheme is presumed to identify candidate local muons.

\subsection{FADC pulse integration \label{sec:fadc}}

IACT cameras typically digitize photon detector signals with fast ADCs 
\citep{sitarekdrs,mux,aharonian2004,veritasdaq,puehlhofer2014}, which allow offline pulse integration and arrival time reconstruction (although readout systems without offline time reconstruction have also been built~\citet{barrau,impiombato2013}) 
Algorithms that use sliding search windows or combine the determination of the readout window location with the image reconstruction 
\citep{shayduk2005,shayduk2013} help to 
improve charge resolution in small $\gamma$-ray images 
albeit at the price of a small bias for low pixel signals~\citep{FADCPulsReco2008} 
due to night-sky background. 
As muon ring signals have a low intrinsic time spread of $O(200~\mathrm{ps})$, a fixed pulse integration window used for all pixels around the average arrival time of the muon ring 
allows for the use of a fixed window pulse integrator without bias and eliminates the contribution of spurious night-sky background to the reconstructed number of photoelectrons $N_\mathrm{tot}$. Such a procedure also reduces the muon image contamination by the parent shower.

\subsection{Noise removal\label{sec:tailcut}}

Typically, Cherenkov images are cleaned from spurious signals due to the light of the night sky  using a dual-threshold approach (the so-called ``tail cut''), whereby all pixels with charge 
are retained provided that they are above the lower threshold.
Additionally, time constraints may be applied, if available~\citep{rissiPhD,aleksic2011,Lombardi2011,BirdPhd}. 
Whereas for $\gamma$-ray images it is desirable to remove low, background-dominated signals, for muon calibration the entire signal needs to be integrated to avoid a biased measurement. 
The best results for muon calibration have previously been obtained with two passes: first cleaning the image with the traditional procedure to identify whether a ring is present, 
and then including all pixels with a secure distance from the reconstructed ring~\citep{bolzphd,mitchellphd}, after fixing the ring parameters.

Only clean rings can then be fit in a robust manner (although promising attempts with wavelet filtering~\citep{Lessard2002} and \comm{machine-learning}~\citep{Bird:2018} methods have been made in the past that do not need to rely on a direct removal of night-sky background noise). 

\subsection{Preselection \label{sec:preselection}}

%

For reasons of processing efficiency 
and to eliminate muon ring images obviously unsuitable for calibration purposes, candidate images are usually preselected, before fitting the ring~\citep{bolzphd,humensky}. 
In order to remove frequently triggering small-amplitude images from showers or muons of very low energy, the aforementioned ``tail cuts'' (see section \ref{sec:tailcut}) are also used to clean the image. 
Preselection criteria may cut on the number of remaining pixels after the 
``tail cut'' 
, the mean number of next neighbors $<\!\!\textit{NN}\!\!>$ (a well-focused muon image should contain pixels in a small ring, and hence $<\!\!\textit{NN}\!\!> \approx 2$), the number of pixels located at the camera edges (since those are typically affected by inefficiencies and aberration), the relative intensity variation of the remaining pixels (which is typically smaller for muons than for air shower images), the reconstructed \textit{Width} parameter~\citep{hillas} (smaller for elongated images and wider for ring-like ones), \comm{the \textit{Length} parameter divided by the image \textit{Size} (exploiting the fact that the amount of muon light per arc length is constant in zeroth order)}, or the \comn{standard deviation of the} time spread in the arrival of light in the pixels, which is normally much smaller for muon rings~\citep{MirzoyanSobczynska:2006}, if compared with shower images. 
\newpage
\subsection{Interpolation of non-active pixels \label{sec:interpolation}}

From this moment on, nonactive pixels 
may bias the ring reconstruction, depending on the algorithms used~\citep{bolzphd}.  
The MAGIC Collaboration interpolates the charge and time information of these pixels from their neighboring ones if at least three neighboring pixels are valid~\citep{magicperformance}. For an optimized muon analysis, one might think of interpolations \textit{along the ring}, instead of in a circle around the affected pixel. 

Groups of broken pixels will more seriously impact the muon calibration results, and their effects must be studied with simulations \citep[see, e.g.,][]{bolzphd}.

\subsection{First-level Ring Fit \label{sec:firstring}}

After preselection, typically less than 10\% of all images remain to which a first muon ring reconstruction can be applied.
Several algorithms have been used for this task~\citep{chernov,karimaki1991,Taubin1991,chaudhuri,Tyler:Dipl}, 
all of which calculate the 
distance of each preselected pixel to a tested ring center, using a candidate ring radius, and minimize its chi-square, \comn{possibly being weighted with pixel charges}. It is also common to perform two fits in order to reduce biases due to hadronic shower contamination: a first fit to find a candidate ring radius and center, followed by a second fit that excludes pixels too far from the candidate ring. 
A good initializer for the ring center can be easily derived from the center of gravity of the image exploiting  Eq.~\ref{eq:dc}, and a typical Cherenkov angle (e.g. 1.2$^\circ$) for the radius. 
%
%
The thickness of the ring may be estimated by \citep{fegan2007}:
\begin{equation}
\Delta \thetac = \sqrt{\chi^2/N_\mathrm{tot}}~. \label{eq:deltathetac}
\end{equation}\noindent

Iterated fits can be made more robust after excluding pixels much farther away from the ring center than a minimum distance and closer to the ring center than a certain maximum distance. The total charge excluded may be used as a quality measure.


\subsection{Quality of the Ring Fit}

After the fit(s), it is important to reject rings reconstructed with poor quality.
Common choices for such quality parameters include the $\chi^2/\textrm{NDF}$ of the ring fit, the ring radius (rejecting low-energy muons); and the ring containment.  

Further choices may include the number of inactive pixels on the ring~\citep{bolzphd},
the ratio between image center-of-gravity \comn{distance to the camera center} and the reconstructed ring radius~\citep{fegan2007} and the completeness of the ring (e.g. the number of nonempty azimuthal bins with a certain, loose threshold in pixel size). 
Generally, such cuts should be loose in order not to bias the final distribution of $B_\mu$ and its statistical properties.

Finally, both \citet{bolzphd} and \citet{mitchellphd} suggest 
the application of  a cut on the amount of light found \textit{off the ring},  which may identify muon rings most likely to be contaminated by background air shower light (see also Section~\ref{sec:stereotrigger}), but do not implement it. 

\clearpage
\section{Determination of Impact Distance and Optical Bandwidth\label{sec:muon_efficiency}}



Reconstruction of the impact distance $\rho_R$, the azimuthal phase $\phi_0$ and the number of photoelectrons along the ring $\ud N_\mathrm{pe}/\ud \phi$ (see Eq.~\ref{eq:dNdphi}) allows one
to retrieve the optical bandwidth $B_\mu$ on an event-by-event basis, typically on a selection of high-quality events.

\subsection{One-dimensional Fits to the Light Distribution along the Ring}

Fitting impact distance and optical bandwidth to the distribution of light along the ring, using Eq.~\ref{eq:Ipix}, 
has been explored in~\citet{jiang:1993,rose,shayduk}, and \citet{goebel}. 
All authors claim about 3--5\% overall statistical uncertainties on \comn{their measurements of} $B_\mu$ (which translates to slightly more than 100\% per-event uncertainty, given the large sample sizes used) and around 3\%--12\% attributed to systematic uncertainties. Some obtain additional systematic biases of up to 30\% when comparing MC simulated muons with those from real data. The biases might, however, be related to an incomplete correction of differences between the optical bandwidths of the telescopes to muon and gamma-ray shower light (Eq.~\ref{eq:epsgamma}), or from simulating pure muons instead of muons from hadronic showers. Only \citet{shayduk} provide an estimate of the event-wise precision of the reconstructed impact parameter, namely, $\delta\rho_R \approx 0.07$.  

In general, the difficulty here is how to translate from the individual pixels to the distribution of light as a function of $\phi$, and to make decisions about which pixels belong to the ring and which are considered background dominated instead.

\citet{bolzphd} and \citet{iacoucci} additionally smoothed the pixel-wise charges 
over an azimuthal range $\Delta\phi$ of $N_\mathrm{smooth}=4$ and $N_\mathrm{smooth}=3$ camera pixel widths, respectively (each of 0.17$^\circ$ in the case of H.E.S.S.-I):

\begin{equation}
\Delta\phi = N_\mathrm{smooth}\,\overline{\omega} = N_\mathrm{smooth} \frac{\omega}{\theta_c}
\end{equation}

The obtained light intensity in an azimuthal bin was then renormalized again to the equivalent of one pixel, e.g. for $N_\mathrm{smooth}=4$~\citep{bolzphd}:
\begin{align}
\Qpixi(\phi_i|\thetac,\rho,\phi_0,B_\mu) &= \frac{1}{4} \int_{\phi_i - 2 \overline{\omega}}^{\phi_i + 2 \overline{\omega}} \Qpix(\phi'|\thetac,\rho,\phi_0,B_\mu)\,\ud\phi' \nonumber\\
  &\equiv  \ddfrac{1}{4} \sum_{\phi \geq (\phi_i - 2 \overline{\omega})}^{\phi \leq (\phi_i + 2 \overline{\omega})} \Qpix(\phi|\thetac,\rho,\phi_0,B_\mu)  \label{eq:ib}  
\end{align}

Data points $\overline{Q_\mathrm{pe}}$ were then constructed in distances of 0.5\,$\overline{\omega}$ and fitted to a smoothed version of function Eq.~\ref{eq:Ipix}.

After applying quality cuts on the results of that fit, 
a third iteration was performed including all pixels within $2\omega$ of the ring from the uncleaned image to avoid a bias due to the 
``tail cut.''\footnote{See section \ref{sec:tailcut}} 
For small Cherenkov angles, the allowed ring width was enlarged, according to the expected broadening of the ring. 
This procedure achieved an event-wise precision of $\delta \rho_R \approx 0.02$.  \citet{bolzphd} claims 5\% per-event statistical and only 2\% systematic uncertainty on the reconstructed $B_\mu$\footnote{Generally, the previous analyses used instead $\varepsilon = B_\mu / (hc \cdot \int_{\lambda_1=300~\mathrm{nm}}^{\lambda_2=600~\mathrm{nm}}1/\lambda^2 \ud\lambda)$ with the somewhat misleading name of ``muon efficiency.''}.

\subsection{Center-of-gravity-based Calculations}

An elegant way to estimate $B_\mu$  directly from Eq.~\ref{eq:Ipix}, without the need to pass through (computationally extensive and hence slow) fits, has been outlined in~\citet{fegan2007}.

The impact distance can be related \comn{to the distance of the center of gravity \comn{($\vec{cog}$)} of the image from the camera center \comn{($\vec{\upsilon}$)}, and the reconstructed muon ring radius}, according to~\citet{fegan2007}:
\begin{equation}
\ddfrac{|d|}{\thetac} = \ddfrac{\sqrt{(\textit{cog}_x-\upsilon_x)^2+(\textit{cog}_y-\upsilon_y)^2}}{\thetac} = \ddfrac{1}{2} \ddfrac{\rho_R}{E_0(\rho_R)} \quad.
\label{eq:dc}
\end{equation}

After performing the ring fit, the distance between the center of gravity of the image ($\vec{cog}$) and the ring center ($\vec{\psic}$) is calculated:
\begin{equation}
\vec{d} = \vec{cog}-\vec{\psic} \quad.  \label{eq:d}
\end{equation}
\noindent
With the resulting absolute value $|d|$, Eq.~\ref{eq:dc} is employed, e.g. using tabulated solutions of $\rho_R/E_0(\rho_R)$, to obtain $\rho_R$. The individual components of the impact distance ($\rho_x,\rho_y$) can then be calculated according to
\begin{equation}
\begin{pmatrix}\rho_x\\\rho_y\end{pmatrix} = \ddfrac{2E_0(\rho_R)}{\thetac} \cdot \begin{pmatrix}d_x\\d_y\end{pmatrix}  
\end{equation}

The method achieves a 9\% statistical per-event uncertainty on the retrieved optical bandwidth. An estimate of the precision with which $\rho_R$ could be determined has not been provided.

\subsection{Two-dimensional Fits Including Ring-broadening Effects}



It was first realized by \citet{rovero} that ring-broadening effects, implicitly integrated in Eq.~\ref{eq:Ipix}, 
can be made explicit, and it is possible to fit the measured light in each pixel directly. 
The simplest way to do so leads to a Gaussian model for ring broadening, characterized by the ring width parameter $\sigma_\theta$: 

\begin{align}
\ddfrac{\ud^2 N_\mathrm{pe}}{\ud \phi\, \ud \theta} (\rho,\phi_0,B_\mu,\sigma_\theta) &\simeq 
\frac{\alpha}{2hc} \cdot \sin(2\theta_c) \cdot D(\rho,\phi-\phi_0) \cdot B_\mu  \cdot \ddfrac{1}{\sqrt{2\pi}\sigma_\theta} \cdot \exp\left(-\ddfrac{(\theta-\thetac)^2}{2\sigma_\theta^2}\right)  \quad, \nonumber\\
%
%
 & \simeq  5.0\times 10^1 \cdot \left(\ddfrac{B_\mu}{\mathrm{eV}}\right) \cdot \left(\ddfrac{D(\rho,\phi-\phi_0)}{\mathrm{m}}\right) \cdot \ddfrac{1}{\sigma_\theta} \cdot \exp\left(-\ddfrac{(\theta-\thetac)^2}{2\sigma_\theta^2}\right)\qquad,
\label{eq:Ipixs}
\end{align}
\citet{rovero} reached a statistical per-event uncertainty of $\delta B_\mu\sim$15\% with a function of this type and claimed a systematic uncertainty of $\sim10$\%. \citet{chalmecalvet2014}, using the same function, achieved an overall uncertainty of 5\% on $B_\mu$, but did not specify how this number was obtained.

\citet{rovero} did not assume that $\sigma_\theta$ was a constant, but rather a function of the distance $r$ of a given mirror to the center of the mirror dish and further ring-broadening effects, depending on $\thetac$ and hence $\sigma_\theta(r,\thetac)$. One could improve this approach by making $\sigma_\theta$ a function of $\rho$ and $\phi$ (along the lines started to be developed in \citet{rovero}) and include here analytical models of ring-broadening effects, such as those treated in Section~\ref{sec:broadening}.


%

\subsection{Two-dimensional Fits Including Photoelectron Statistics}
\label{sec:fitspestat}


Two algorithms have been introduced so far in the literature that include the intrinsic Poissonian fluctuations of observed charges around the expectation value $\Qpix$: 

\citet{leroyphd} described a template-fitting procedure, based on~\citet{iacoucci} and \citet{lebohec}. The method constructs an expectation charge value for each pixel on the ring following Eq.~\ref{eq:Ipixs} and constructs a $\chi^2$ function for the summed squared deviations of the observed charges $(Q^\mathrm{obs}_i)$ from the expectation values $\Qpixi(\thetac,\vec{\psic},\rho_R,\phi_0,B_\mu,\sigma_\theta)$, divided by the expected variances, assuming Poissonian statistics\footnote{In \citet{leroyphd} the variance terms were only partly divided between $Q_\mathrm{obs}$ and $\Qpix$, which is rectified here.}: 

\begin{equation}
\chi^2(\thetac,\vec{\psic},\rho_R,\phi_0,B_\mu,\sigma_\theta) = \sum_i \ddfrac{\left(\Qobsi - \Qpixi(\theta_c,\vec{\psic},\rho_R,\phi_0,B_\mu,\sigma_\theta)\right)^2}{\spi^2 + F_i^2/2 \cdot \left(\Qobsi+\Qpixi(\thetac,\vec{\psic},\rho_R,\phi_0,B_\mu,\sigma_\theta)\right)} \quad,
\label{eq:chi2}
\end{equation}\noindent
where $\spi$ represents the baseline fluctuations of pixel $i$ (mainly due to night-sky background) and $F_i$ the excess noise factor of the photon detector (pixel) $i$, respectively (for the case of a photomultiplier tube, see~\citet{Lewis:1987,Bencheikh:1992,mirzoyan1997}). The $\chi^2$ depends here on \textit{seven} parameters and must be minimized numerically, although that not all parameters need to be left free in the fit on an event-by-event basis. For instance, some assumptions for the ring-broadening parameter based on the atmospheric conditions and measured PSF could be made. This line of development still awaits being explored in the future.



\citet{leroyphd} claimed that a fitting procedure minimizing the $\chi^2$, Eq.~\ref{eq:chi2}, \textit{without taking into account ring-broadening effects}, achieves a precision of $\delta \rho_R \approx 0.08$ (with a bias of 0.03), whereas \citet{iacoucci} obtained $\delta \rho_R \approx 0.13$ and $\sim$50\% per-event statistical uncertainty on $B_\mu$. 
The (systematic) uncertainty for $B_\mu$ has not been provided; however, \citet{iacoucci} finds an 80\% discrepancy between measurement and expectations, attributed primarily to a loss of sensitivity of the telescope below 315~nm. 

\citet{mitchell2015m} instead used a log-likelihood fit of a smoothed image to Eq.~\ref{eq:Ipixs}, with $\rho_R,\phi_0, B_\mu$ and $\sigma_\theta$ as free parameters. The ring center $\vec{\psic}$ and radius $\thetac$ are kept as fixed with the values obtained from the previous ring fit. As previously, the observed distribution $\Qobsi$ is first smoothed along $\phi$ (Eq.~\ref{eq:ib}). 
The pixel likelihood is calculated from a Gaussian approximation to Poissonian statistics:

\begin{align}
-2\ln \Lagr  &=  -2 \cdot \ln P(\vec{\Qobs}|\rho_R,\phi_0,B_\mu)   \nonumber\\
           &= \sum_i 
           \Bigg\{ \ln\left(\spi^2+F_i^2\Qpixi(\rho_R,\phi_0,B_\mu)\right) +   \nonumber\\
           & \qquad\qquad\quad +\ddfrac{\left(\Qobsi-\Qpixi(\rho_R,\phi_0,B_\mu)\right)^2}{\left(\spi^2+F_i^2\Qpixi(\rho_R,\phi_0,B_\mu)\right)} ~\Bigg\} ,
           \label{eq:L}
\end{align}
\noindent
where the summation goes over all pixels included in the (cleaned) image. The assumption of asymptotic Gaussian behavior is, however, only reasonable for pixel charges greater than $\sim$10~photoelectrons, which is often not the case, particularly for pixels that do not lie directly on the ring. 

A version of the pixel likelihood that includes Poissonian statistics but assumes a Gaussian pedestal and single photoelectron PMT charge distribution can be written instead as follows:
\begin{align}
-2 \ln \Lagr  &=  -2 \cdot \ln P(\vec{\Qobs}|\rho_R,\phi_0,B_\mu)   \nonumber\\
           &= 2\cdot \sum_i \Bigg\{ \Qpixi(\rho_R,\phi_0,B_\mu)   \nonumber\\
           & \qquad\qquad\quad -\ln 
           \Bigg[ \sum_{n=0}^\infty\ddfrac{\Qpixi(\rho_R,\phi_0,B_\mu)^n}{(\spi^2+F_i^2n)\cdot n!} \cdot \exp\left(-\ddfrac{(\Qobsi-n)^2}{2(\spi^2+F_i^2n)}   \right)
           \Bigg] \Bigg\} .
\end{align}
\noindent


\begin{table}
\centering
\begin{tabular}{lccccp{6cm}}
\toprule
Method  &   \multicolumn{2}{c}{$\hat{\rho}_R$} & \multicolumn{2}{c}{$\hat{B}_\mu$} & Reference  \\
        &   Resolution  &   Accuracy  &  Resolution & Accuracy &  \\
       &    (\%) &    (\%)          &  (\%)  &    (\%)       &  \\
\midrule
One-dimensional fits  &&&&&\\
 \quad Histogramming & 7 & n.a. & 100 & 3--12 & \citet{jiang:1993,rose} (Whipple) \citet{shayduk,goebel} (MAGIC) \\ 
\quad Smoothing and renormalizing & 2 & n.a. & 5 & 2 & \citet{bolzphd}  (H.E.S.S.) \\ 
\quad \textit{cog}-based calculations & n.a. & n.a. & 9 & n.a. & \citet{fegan2007} (VERITAS) \\
\midrule
Two-dimensional fits  &&&&&\\
\quad                & n.a.  & n.a. & 15 & 10 &  \citet{rovero} (Whipple) \\
\quad                & n.a.  & n.a. & \multicolumn{2}{c}{5} &  \citet{chalmecalvet2014} (H.E.S.S.) \\
\midrule
\multicolumn{2}{l}{One-dimensional fits including p.e. statistics} &&&&\\
 \quad Minimize $\chi^2$, Eq.~\ref{eq:chi2} & 8--13 & $>$3 & 50 & n.a. & \citet{iacoucci} (CAT) \citet{leroyphd} (H.E.S.S.) \\
 \multicolumn{2}{l}{Two-dimensional fits including p.e. statistics} &&&&\\
 \quad Minimize $-2\ln \Lagr$, Eq.~\ref{eq:L} & n.a. & n.a. & 16 &  tbd & \citet{mitchellphd} (H.E.S.S.) \\
\bottomrule
\end{tabular}
\caption{Published resolutions and accuracies of the estimated parameters \textit{impact distance} and \textit{optical bandwidth}, using the different methods presented. \label{tab:ipbwresults}}
\end{table}

Table~\ref{tab:ipbwresults} compares the published results in terms of resolution and accuracy of the estimated parameters impact distance and optical bandwidth. One can immediately see that there is a large spread in the resolution obtained, and to a lesser extent in accuracy,
for the estimated optical bandwidth ($\hat{B}_\mu$), even within a similar method. This provides a hint that the fitting routines themselves seem to be less important for achieving precise and robust results than a careful ring reconstruction and event selection. 

\subsection{Quality cuts on the Impact Parameter Fit Results}

Basically all analyses that use Eq.~\ref{eq:Ipix}, or a more sophisticated version of it, cut on the $\chi^2/N_\mathrm{pix}$ of the fit and on an allowed range of the obtained impact distance. Low-impact distances are often excluded because of the shadowing of the camera (which is difficult to model)
or holes in the center of the mirror dish, which are common in this field to host instrumentation for pointing~\citep{gc2010} and camera calibration~\citep{aharonian2004,gaugphd,hanna2008,biland2014}. Typical cut values are hence ($0.2 < \rho_R < 1$), 
where the upper limit excludes the majority of incomplete rings. 

\citet{mitchellphd} also cuts on the reconstructed image width: $0.04^\circ < \sigma_\theta < 0.08^\circ$.

\subsection{\comn{Treatment of Camera Shadows}}

\comn{A correct treatment of shadows becomes important when the size of the shadow is non-negligible with respect to the size of the primary mirror. This is the case for several CTA cameras, particularly the MST, which hosts a 3~m broad camera to provide the required large FOV.
  Moreover, that camera is of quadratic shape, complicating further an analytical expression for its shadow (shown in Appendix~\ref{sec:quadraticcamera}).} 

\begin{figure}[htp]
  \centering
\includegraphics[width=0.55\textwidth,trim={0.2cm 0.2cm 0.2cm 0.2cm},clip]{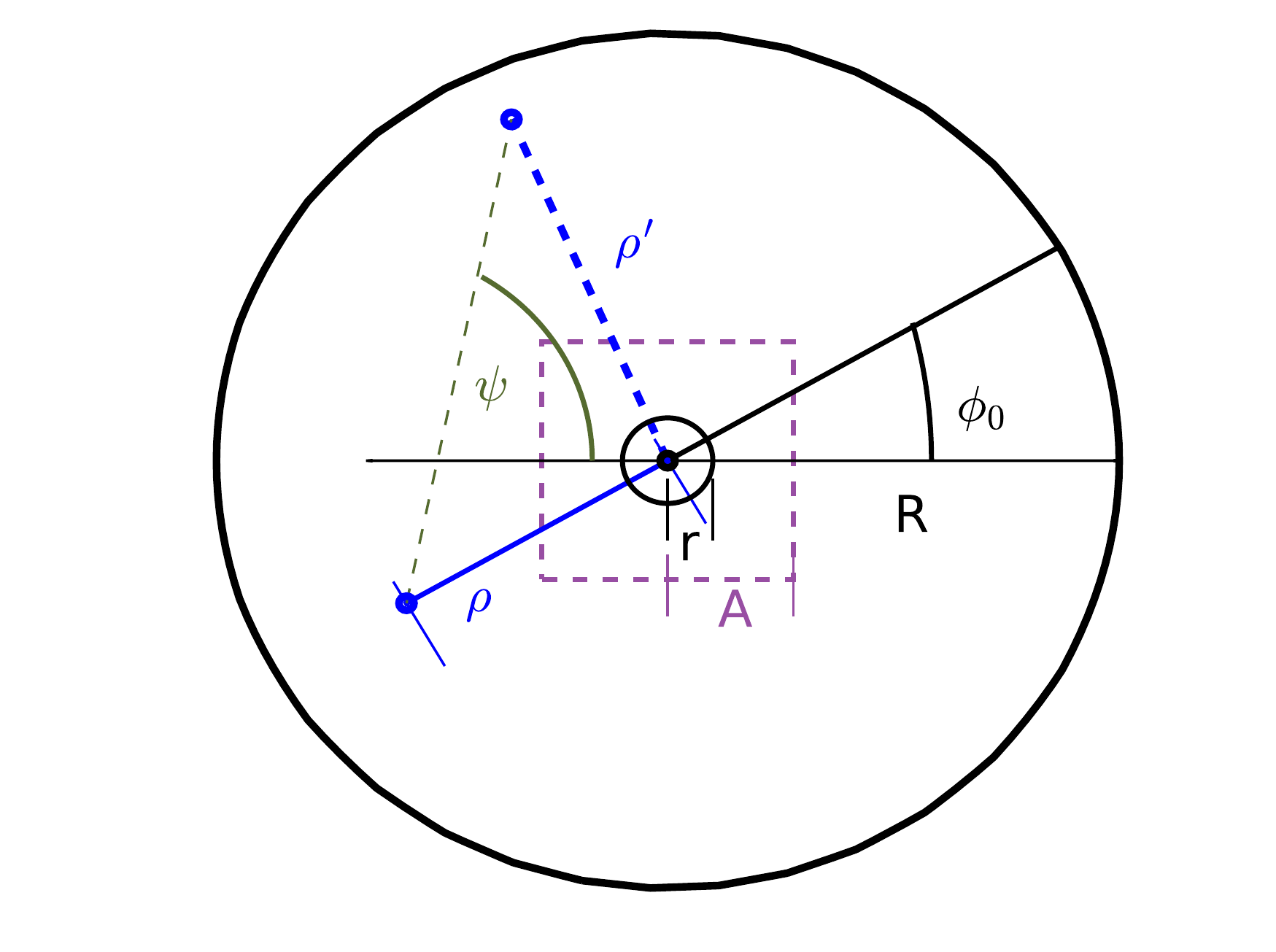}
\caption{\comn{Schematics of the used geometry. The impact point $\rho$ (solid blue line) gets transferred to the camera plane along the green
dashed line, if the muon is inclined (see the definition of $\psi$ in Fig.~\ref{fig:geometry}.) The resulting impact distance with respect to the camera (dashed purple line)
is denoted as $\rho^\prime$. A possible hole in the center of the mirror support structure is drawn as a circle of radius $r$. The half-side length of the camera is denoted by $A$.} \label{fig:camera_shadow}}
 \end{figure}

\comn{In case a muon passes either beside or through the camera, a part of the emitted Cherenkov light from above the camera plane will be shadowed. Instead, light emitted below the camera plane
is \textit{not} shadowed by the camera. \comnn{Its fractional contribution to the total light can ascend to more than 5\% for the case of an MST}.  Both contributions might be affected by additional shadows, like the camera support structure, and the hole in the center of the mirror dish. Figure~\ref{fig:camera_shadow} introduces the relevant contributions and their denomination, used 
in the following formulae to compute the camera shadow. }

\comn{We will first derive the muon impact point at the altitude $F$ of the camera $\vec{\rho}^{\,\prime}$, if the muon is inclined by an angle $\vec{\upsilon}$ (see Eq.~\ref{eq:nphotons}):} 
\begin{align} 
\vec{\rho}^{\,\prime} 
& ~=~ \rho_R \cdot \begin{pmatrix}\cos(\phi_0\comn{+\pi})\\\sin(\phi_0\comn{+\pi})\end{pmatrix}  + 2f \cdot \tan\upsilon \cdot \begin{pmatrix}\cos\psi\\\sin\psi\end{pmatrix} 
\label{eq:rho_moved}  \\
& \qquad \mathrm{with}  \\
& \rho_R^\prime = \sqrt{\rho_R^2 + 4f^2\tan^2\upsilon -4\rho_R f \cdot \tan\upsilon  \cdot \cos(\phi_0-\psi)} \\[0.25cm]
& \phi_0^\prime = \arctan2 \left( \ddfrac{2f\cdot \tan\upsilon\sin\psi -\rho_R\sin\phi_0}{2f\cdot \tan\upsilon\cos\psi -\rho_R\cos\phi_0}  \right) - \pi \quad,
\end{align}
\noindent
\comn{where the $f$-number $f=F/2R$ has been used.}

 \begin{figure}[htp]
  \centering
\includegraphics[width=0.99\textwidth,trim={0.3cm 0.3cm 0.3cm 0.2cm},clip]{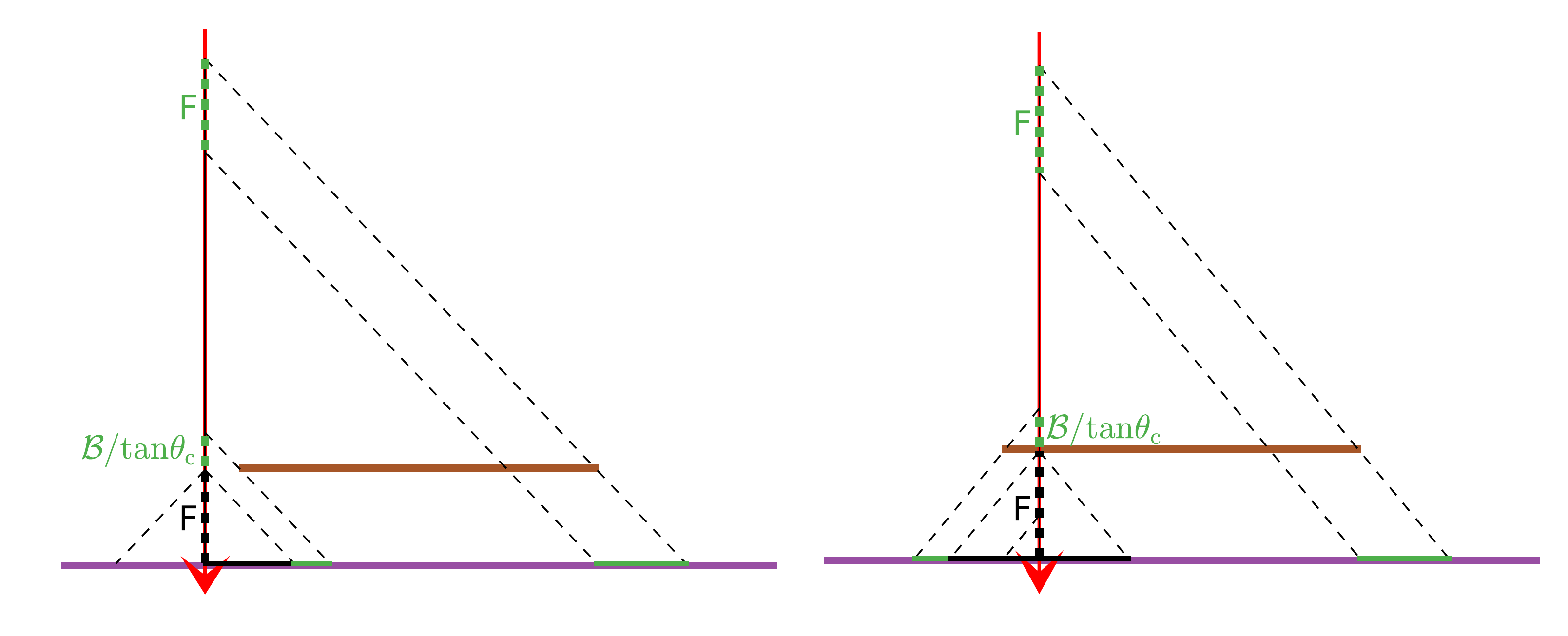}
\caption{\comn{Schematics of the effect of the camera shadow. A muon passes beside (left) or through (right) the camera (in brown), before hitting the mirror (in violet). The Cherenkov light pool emitted below the camera is marked in solid black.
    The projected additional shadows (green) depend on the impact distance and the focal distance $F$ of the camera, nevertheless, the total shadow adds up to exactly the length of the chord, using Eq.~\ref{eq:Dshadow}.  Note that the Cherenkov angle has been exaggerated for better visibility.} \label{fig:camera_shadow2}}
 \end{figure}

 \begin{figure}[htp]
  \centering
\includegraphics[width=0.99\textwidth,trim={0.3cm 0.3cm 0.3cm 0.2cm},clip]{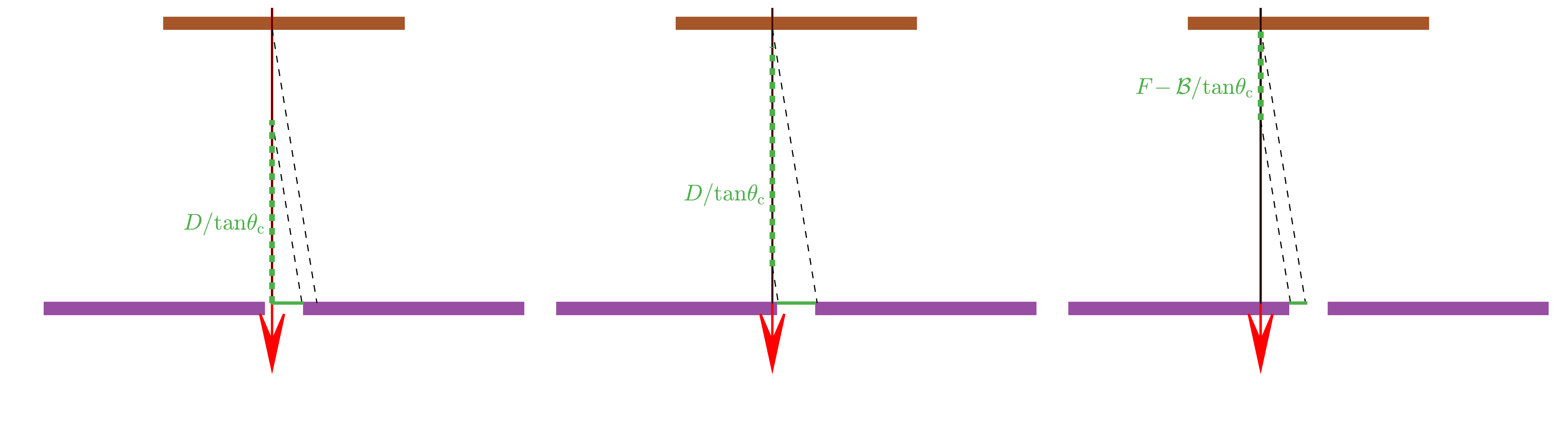}
\caption{\comn{Schematics of the effect of the shadow of the central hole in the mirror. A muon passes 
    through  the camera (in brown), before hitting the mirror (in violet), for three different  impact distances. The Cherenkov half-light pool emitted below the camera is 
    found between the muon (red arrow) and the outer dashed line. 
    The projected additional shadows (green) depend on the impact distance and the focal distance $F$ of the camera, using Eq.~\ref{eq:Dhole}. Note that the Cherenkov angle has been exaggerated for better visibility.} \label{fig:hole_shadow}}
 \end{figure}

\comn{The chord of the shadow on the mirror can be approximated as}
\begin{align}
D_\mathrm{shadow}(\rho_R,\phi_0,\phi;\upsilon,\psi,f,A) &= D\left(\rho_R^{\prime}(\rho_R,\phi_0,\upsilon,\psi,f) \cdot\ddfrac{R}{A} , \phi_0^\prime(\rho_R,\phi_0,\upsilon,\psi,f),\phi)
\right) \quad, \label{eq:Dshadow}
\end{align}
\noindent
\comn{using $A$ either as the radius of a spherical camera shape, used together with Eq.~\ref{eq:D} for $D$, or as the half side length of a quadratic camera, used together with Eq.~\ref{eq:Dquadratic} for $D$.}

\comn{Note that the projected camera shadow on the mirror will be centered on $-2f \cdot \tan\upsilon \cdot (\cos\psi,\sin\psi)$ and might not completely cover the central hole anymore. In that case, the residual contribution of the uncovered part of the hole must be added to the shadow.}

\comn{Eq.~\ref{eq:Dshadow} does not contain any correction for the Cherenkov light $F\cdot \tan\thetac$, emitted below the camera, because its contribution to $D_\mathrm{shadow}$ cancels exactly with the additional shadow projected on the mirror dish from a height $F$ under the Cherenkov angle.} 

\comn{In the case that the central hole is completely covered by the shadow of the camera, its contribution can be estimated as }
\begin{align}
D_\mathrm{hole}(\rho_R,\phi_0,\phi;f) &= 
D(\rho_r,\phi_0,\phi) - \left\{ 
\begin{array}{ll}
\max\left( \mathcal{B}(\rho_r,\phi_0,\phi) -\ddfrac{F}{r}\tan\thetac, 0 \right)  & \rho_r < 1 \\
\min\left( \max\left(\mathcal{B}(\rho_r,\phi_0,\phi)  -\ddfrac{F}{r}\tan\thetac, 0 \right) ,  D(\rho_r,\phi)\right)  & \rho_r\ge 1 \\
\end{array}
\right.  \label{eq:Dhole}  \\
& \mathrm{with} \nonumber\\
\mathcal{B}(\rho_r,\phi_0,\phi) &= \rho_r \cos(\phi-\phi_0) + \sqrt{1-\rho_r^2\sin^2(\phi-\phi_0)}\quad, \\[0.25cm]
\rho_r  &= \rho_R \cdot \ddfrac{R}{r}
\end{align}
\noindent

\subsection{Modifications for Non-radially Symmetric Mirror Geometries}

\begin{figure}[htp]
  \centering
\includegraphics[width=0.9\textwidth,trim={0.2cm 0.3cm 2.5cm 19.5cm},clip]{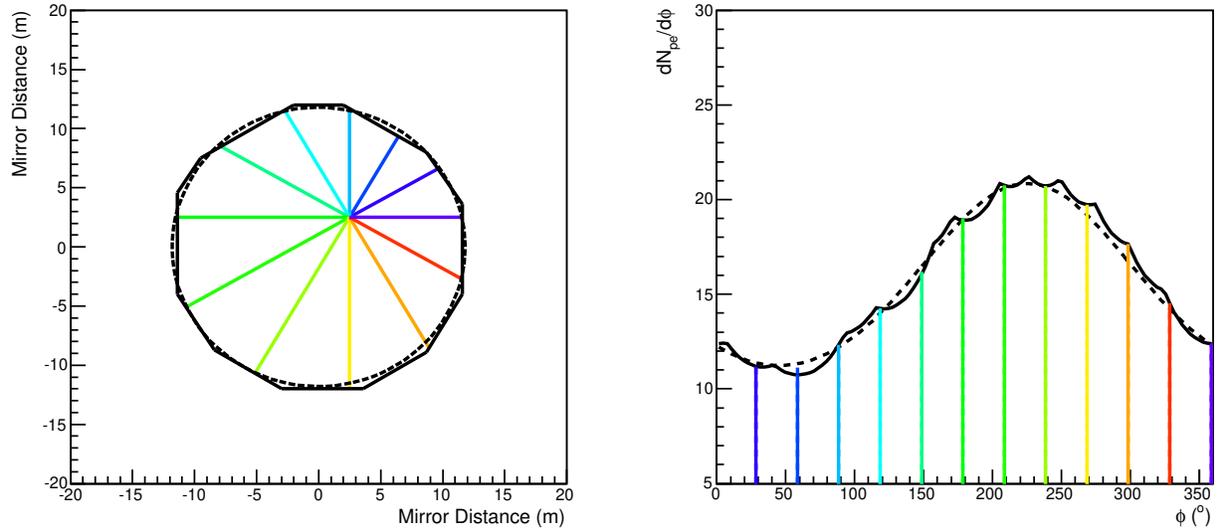}
\caption{Left: line integration along the mirror surface as a function of azimuth angle for a given impact position, shown for the LST with the (still simplified, but more realistic) mirror profile (solid line) and under a circular approximation (dashed line). Right: modulation of the intensity profile along the ring for the given impact distance.  \label{fig:modulation}}
 \end{figure}

The relation Eq.~\ref{eq:D} for the chord $D(\rho,\phi-\phi_0)$ assumes a circle to represent the shape of the mirror dish. Real mirrors, however,  differ from exact circular geometries, some of them considerably, such as e.g. the H.E.S.S.-II telescope~\citep{HESSII:Mirrors}. 
\citet{mitchellphd} has modified the algorithm such that the mirror dish edge is described by a series of points with an interpolation procedure used to find the intersection of the chord $D(\rho,\phi-\phi_0)$ with the edge of the mirror. The effect of this change is shown in Figure~\ref{fig:modulation}. 

\subsection{Modifications for Dual-mirror Telescopes}
\label{sec_dualmod}


\begin{figure}
\centering
\includegraphics[width=0.45\linewidth]{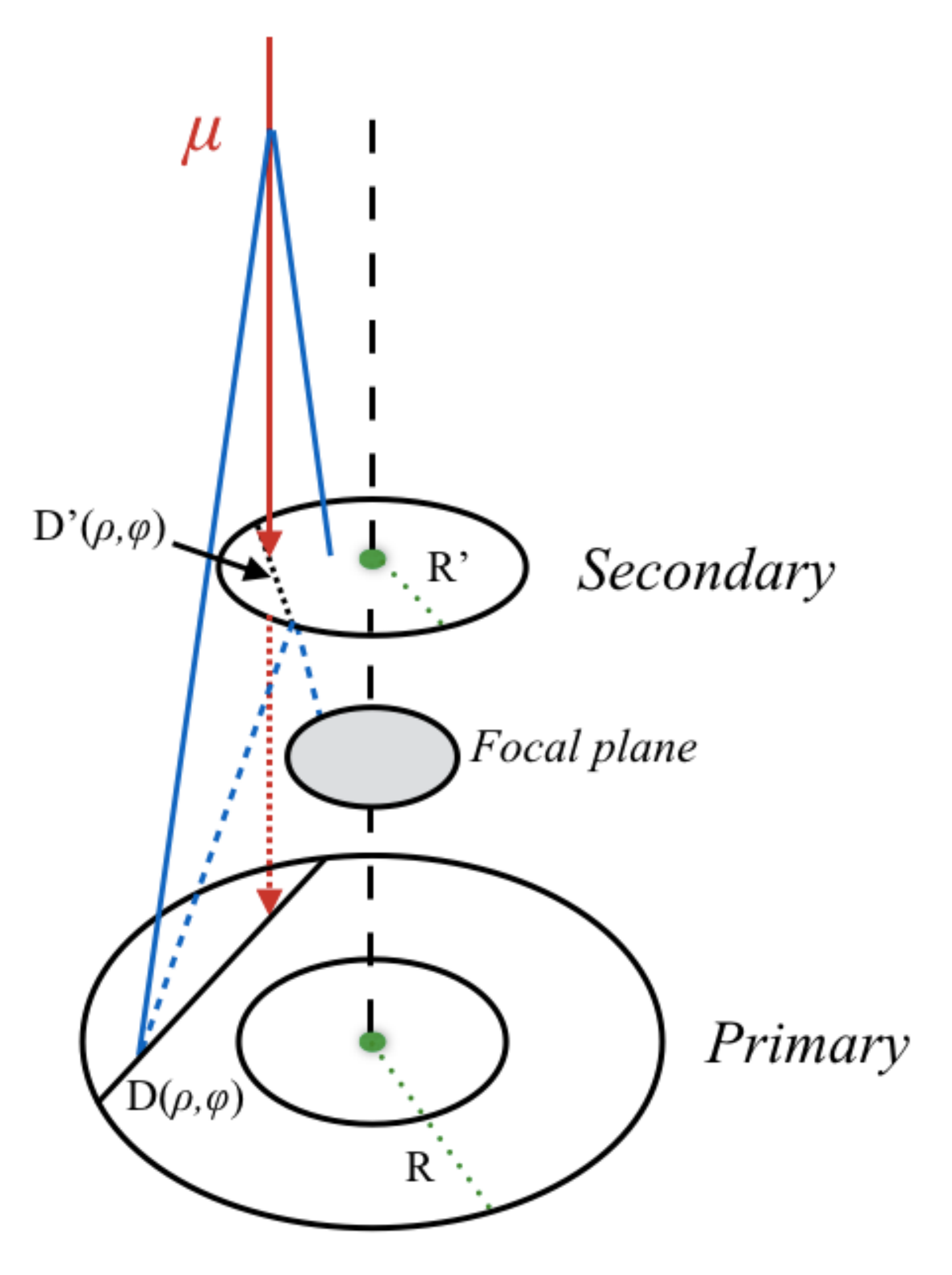}
\caption{\label{fig:muondual} 
Sketch of the parameters introduced to describe the geometry of a local muon $\mu$ and its image in an IACT camera together with a dual-mirror telescope. For reasons of visibility, the angles are not to scale. \comnn{The variation in path length across both mirrors with azimuthal angle must be taken into account.}
}
\end{figure}

The effect of a secondary mirror may be treated as a large shadow, as no light from the muon can pass through the mirror (see Figure~\ref{fig:muondual}).%
This approach, however, neglects the additional condition that the light must reflect off both mirrors in order to reach the camera. 
In predicting the amount of light arriving into a single pixel on the focal plane, the full path length at a given azimuthal angle due to both mirrors needs to be taken into account. 

Preliminary studies on simulations of dual-mirror telescopes found that the sum of the chords ($D_p+D_s$) provided a better description of the Cherenkov light distribution in the camera image from a muon, resulting in a more Gaussian distribution of reconstructed efficiencies and more linear behavior of the correction factor with optical degradation with respect to either neglecting the secondary mirror contribution to the chord or treating the secondary mirror as equivalent to a large hole (causing only loss of light; \citet{mitchellphd}).

In this case, however, to avoid overestimating the total number of photons, a coefficient $\beta$ is introduced to account for the reduction in light due to shadowing by the secondary mirror before the light reaches the primary mirror. 

A precise analytical expression for $D(\rho,\phi)$ in the case of dual-mirror telescopes remains to be fully verified by means of a dedicated study. Investigations of the true light distribution around the muon ring using detailed ray-tracing would be necessary to verify both the dependence of the chord across the secondary mirror on $\phi$ and the correct normalization of the light yield. 

We note also that the shape of the shadow caused by the secondary mirror may change with increasing off-axis angle; the significance of the impact this may have on the muon calibration remains for further study.

\begin{figure}[htp]
\centering
\includegraphics[width=0.57\textwidth]{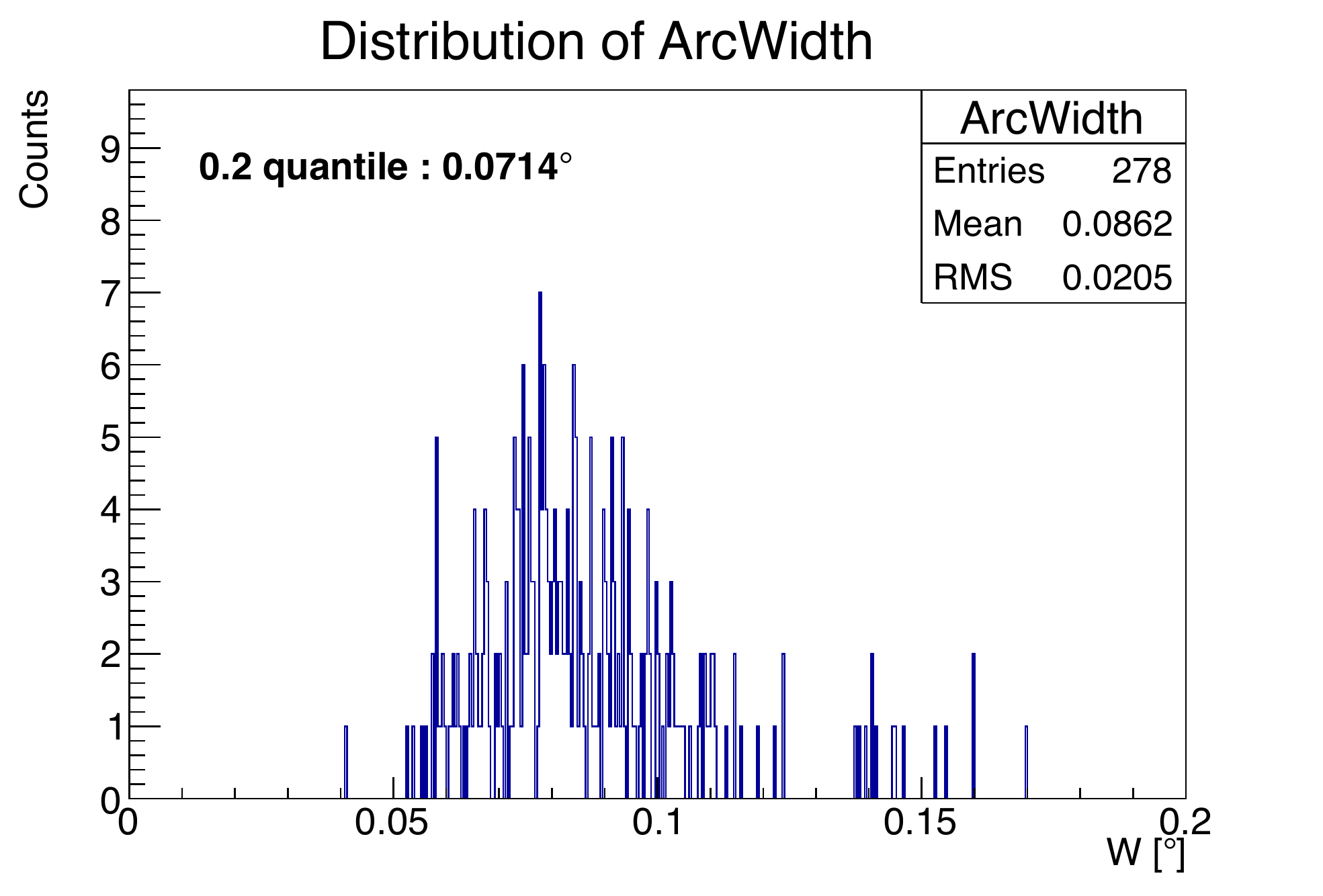}
\caption{
Distribution of the \textit{ArcWidth} parameter (equivalent to $\sigma_\theta$) for a real data sample with bad PSF. The obtained 0.2 quantile of 0.071$^\circ$ is bigger than 
the one expected from the rest of the observation period. 
(Figure courtesy of the MAGIC Collaboration).
\label{fig:quantile}}
\end{figure}

\subsection{Monitoring of the Optical PSF}
\label{sec:optical_psf}
Fits to the equations of type Eq.~\ref{eq:Ipixs} allow
a precise estimator to be retrieved
for the optical PSF size of the reflector, namely, the muon ring width $\sit$. Typically, such a parameter shows a dependency on the muon energy and hence the ring radius; however, this can be easily calibrated~\citep[see, e.g.,][]{bretzPhD}. The distribution of such ``calibrated ring widths''
can then be compared with those from simulated muon images of a detector with different simulated PSFs (see Fig.~\ref{fig:quantile} and \citet{goebel,GarczPhD}. Experience has shown that 
a low quantile (e.g. 0.2 is used by the MAGIC collaboration) of the  distribution should be preferred over its median,  
in order to make the comparison for those rings which are not so strongly affected by other secondary ring broadening effects.

\begin{figure}
\centering
\includegraphics[width=0.43\linewidth]{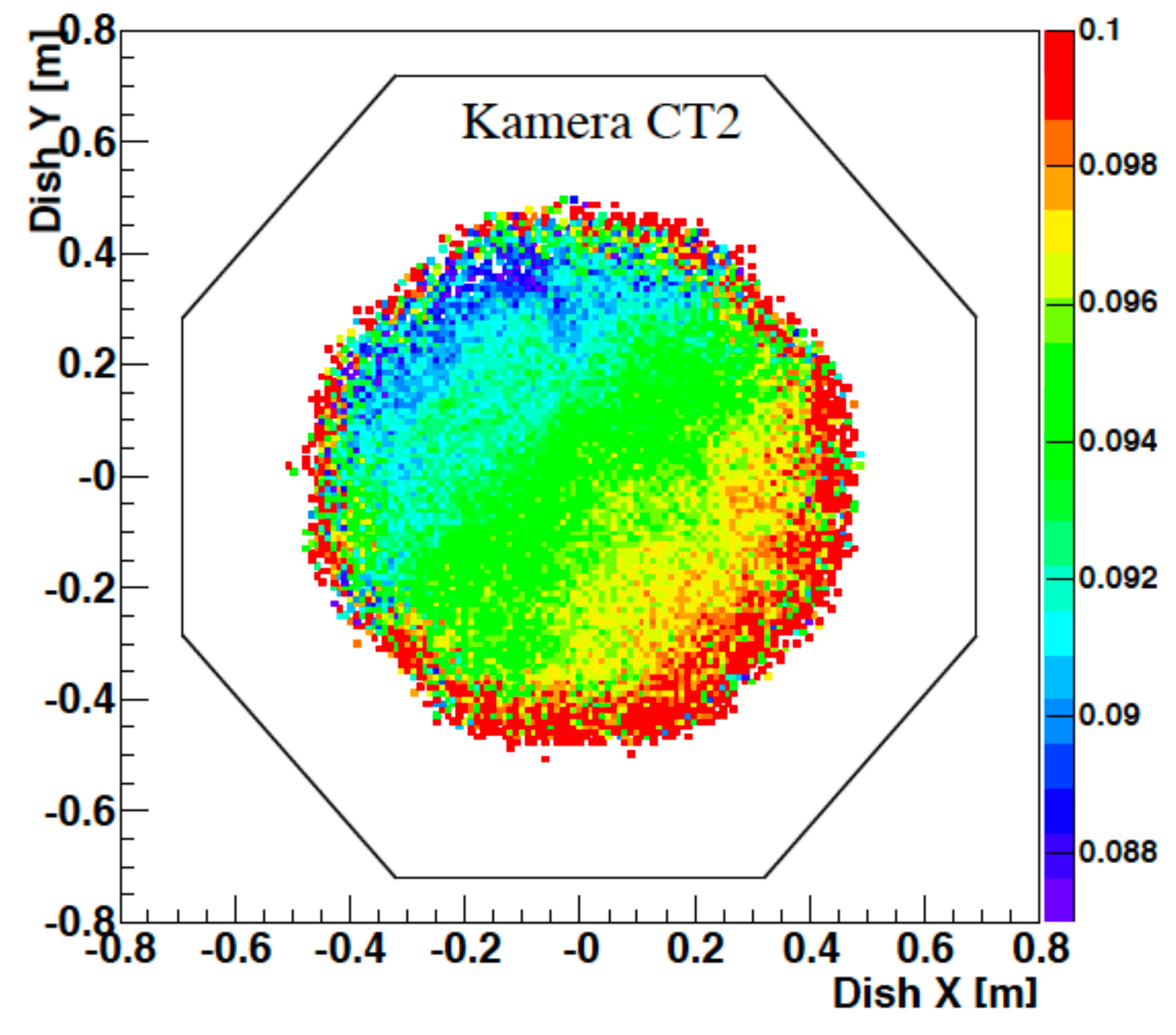}
\includegraphics[width=0.41\linewidth,viewport=0 0 545 540]{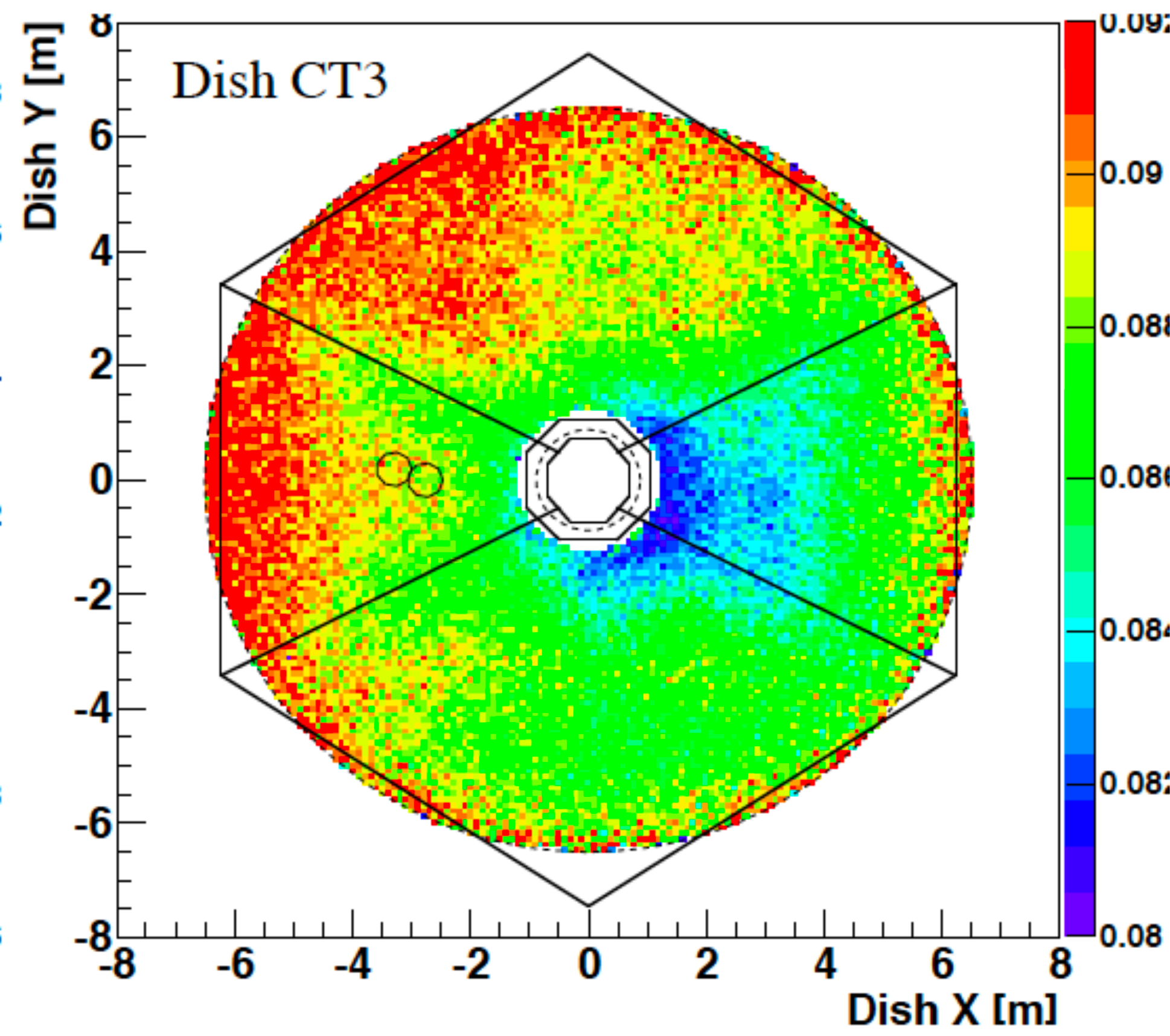}
\caption{\label{fig:effcammirror}Optical efficiency from muons vs. ring center positions in the camera (left) and impact points on the mirror (right) for two selected H.E.S.S.-I telescopes.
The gradients visible may be due to photon detector efficiencies or mirror reflectances due to abrasion by sand particles~\citep{mitchellphd}. The right panel displays also the effects of an asymmetrically opening camera lid (blue area) and the camera support structure (green shadows, visually supported by black lines), which had not been included in the muon model. 
Figure reproduced from~\protect\citet{bolzphd}.}
\end{figure}

\subsection{Monitoring of Individual Camera Pixels}
\label{sec:monitoringpixels}


The muon light intensity seen by one pixel \comn{has been approximated by~\citet{bolzphd}} as\footnote{%
\comn{Note that \protect\citet{bolzphd} introduced a (wrong) factor of two in $k_c$ and overlooked the additional normalization factor $1 + \sit/2\thetac$.}.
}:

\begin{equation}
  Q_\mathrm{pix} = \Qpix (\theta,\rho,\phi) \cdot \ddfrac{1}{\sit\sqrt{2\pi}} \cdot \left( \exp(-\ddfrac{\Delta\theta^2}{2\sit^2}) \cdot \omega \right) k_c^{-1} \quad\textrm{with:} \quad k_c = \left(1 + \ddfrac{\Delta\theta}{\thetac}\right) \cdot \left(1 + \ddfrac{\sit}{2\pi\thetac}\right) \quad, \label{eq:pixel}
\end{equation}\noindent
where $\Delta\theta = \theta-\thetac$ is the radial distance of the pixel from the ring. \comn{Equation~\ref{eq:pixel} approximates the loss of angular coverage $\ud \phi$ of a pixel as its position comes to lie farther away from the ring (and vice versa) by introducing the factor $k_c$. }

Just as for the overall image (Eq.~\ref{eq:dNdphi}), a pixel-wise optical bandwidth parameter $B_\mathrm{pix}$ can be derived that should correlate with the flat-fielding constants from the light flasher calibration~\citep{aharonian2004,gaugphd,hanna2010,gaugSPIE2014,brownICRC2015}. \citet{bolzphd} finds a good correlation for the H.E.S.S.-I telescopes with a spread of 5\% \comn{standard deviation}.

This approach enables the identification of issues 
such as gradients across the camera (see Fig.~\ref{fig:effcammirror}) and differences in photodetector response between vertically incident light (such as used for calibration; \citet{aharonian2004,gaugphd,hanna2008} and \citet{biland2014}) and light arriving at the camera focal plane with a distribution of incident angles.

Although Eq.~\ref{eq:pixel} has been used with some success by \citet{bolzphd}, it is intrinically inconsistent: the infinity at $\Delta\theta = -\thetac$ 
leads to a nonconverging normalization integral.
An improved model should include at least the non-negligible possibility for light scattered across the ring center into the opposite part of the circle and the finite pixel size at the ring center.

\subsection{Monitoring of individual mirror facets}

Plotting the optical bandwidth against the muon impact points $\rho_{x,y}$ enables possible influences of the individual mirrors and shadows on the focused reflectance to be identified (see Fig.~\ref{fig:effcammirror}). Whilst shadowing effects are already taken into account by the telescope model in simulations, effects such as the degradation of individual mirror facets are not usually included. 

\clearpage

\section{Secondary Effects Affecting the Reconstructed Optical Bandwidth \label{sec:secmuoneff}}

\subsection{Instrumental Effects}

\subsubsection{Effects of Non-active Pixels}
\label{sec:nonactivepixels}

Pixels can become broken or be deactivated because of high illumination levels, e.g. due to bright stars~\citep{aharonian2004,magicperformance1}. The latter typically lead to small clusters of 2--3 
deactivated pixels\footnote{%
The number of pixels deactivated due to bright stars mainly depends on the pixel FOV.},
whereas broken pixels tend to be more uniformly spread over the camera. Additionally, groups of 
pixels forming ``clusters'' or ``drawers'' may stop working simultaneously\footnote{%
In the past, clusters of typically 7 or 16 pixels were affected, 
in a camera hosting about a thousand pixels.
}.   
The effect of both has been investigated by \citet{bolzphd} and \citet{mitchellphd} with dedicated MC simulations.  


Both 
find stable results for the reconstructed optical bandwidth up to an average of 10\% of broken pixels on the ring. Interestingly, the large number of deactivated pixels, averaged over neighboring signal amplitudes, leads to an \textit{overestimation} of the optical bandwidth in \citet{bolzphd}, because the reconstructed intensities along the ring become flatter \comn{(if the missing pixel intensities are replaced by the average of their neighbors' intensities, as explained in Section~\ref{sec:interpolation})} and hence the impact parameter is underestimated. \citet{mitchellphd}, who employed the likelihood fit given by Eq.~\ref{eq:L}, found 
robust behavior even up to
30\% of deactivated pixels. Both studies come to the conclusion that clustering of broken pixels 
does not alter the behavior of this systematic effect much
compared to randomly removed pixels. \citet{mitchellphd} showed that the effect of broken pixels on the quality cuts applied later is minor, except for the criterion of a required minimum number of pixels contained in the ring image. 
Nonactive pixels are unlikely to be a problem for muon calibration with CTA, given that all cameras are required to maintain 95\% pixel availability during observations.



\subsubsection{Effects of the Stereo Trigger}
\label{sec:stereotrigger}

If a telescope array is triggered in stereo mode~\citep{hessperformance,VERITASPerformance,magicperformance1}, 
single muons rarely trigger more than one telescope simultaneously. Consequently, the muon events obtained from local muons are often recorded together with the parent air shower, which produces an additional Cherenkov light contribution that may contaminate the muon ring images.


Studies with H.E.S.S. showed that this effect may lead to a positive bias of up to 10\% on the reconstructed optical bandwidth from simulations~\citep{mitchellphd}, whilst only 3\% difference was observed when comparing muon rings obtained from a single telescope trigger to those obtained using a stereo trigger~\citep{bolzphd}. Suggestions for dealing with this effect include simultaneous fits of a Hillas ellipse and a muon ring to the images in order to locate and remove the air shower contribution, although this remains to be demonstrated~\citep{mitchellphd}. 



\subsubsection{Effects of the Pixel Baselines}
\label{sec:baselines}

Drifts in the pixel baselines, which are known to vary with time and temperature~\citep{aharonian2004,sitarekdrs,biland2014}, may lead to biasing the reconstructed number of photoelectrons in Cherenkov images. For CTA, the baseline of each pixel is required to \comn{be measured with an accuracy of better than 0.2 times the equivalent pulse height of one photoelectron} 
and must therefore be regularly monitored. Within the muon analysis, monitoring of the \textit{Off-ring intensity} \footnote{``Off-ring intensity'' means here the summed pixel intensity for all pixels excluded from the muon ring reconstruction~\citep{bolzphd}}, which should peak around zero, may help to identify systematic drifts or issues in the hardware, such as a few slightly defocused mirrors, or to detect possible biases in the muon analysis itself. 



\subsubsection{Effects of the Readout Window}
\label{sec:readoutwindow}

The Cherenkov light pulse received by a camera pixel from a local muon is much shorter $O$(100--200~ps), with a \comn{standard deviation} arrival time spread across the camera of $< 0.7$~ns, than typical signals from air showers, which may spread for several nanoseconds across the camera and show a typical light pulse width of 0.8--3~ns for $\gamma$-ray shower images~\citep{MirzoyanSobczynska:2006}. The choice of telescope optics and differences in the transit time spread of the photodetectors~\citep{FADCPulsReco2008} may further smear the arrival time distribution. The MST telescope structure has been modified to reduce the $O$(5~ns) time spread of a traditional Davies-Cotton reflector~\citep{Oakes2015}, whilst for dual-mirror telescopes the arrival time spread exhibits an angular dependency, ranging from isochronous for on-axis rays to \comm{$O$(1~ns)} at $\sim 5^\circ$ off-axis~\citep{Vassiliev2007,Rulten:2016}. 
Shifts of up to 3~ns in mean intensity maxima between mono and stereo muons have been observed~\citep{bolzphd}. It is important to ensure that the readout window fully contains the muon signal. 
The readout window may, however, be further optimized to reduce contamination by an associated air shower signal.
Additionally, the gamma-ray analysis should integrate the full gamma-ray pulse such that an unbiased conversion from the optical bandwidth obtained from muons to those of gamma-rays is obtained. Otherwise, correction factors need to be applied that take into account the difference in pulse coverage.

\subsubsection{Optical Aberrations}

Aberrations affect the estimated muon parameters, apart from the global broadening of the ring width. In particular, coma and distortion~\citep{Schliesser:2005,Vassiliev2007} can move the image \textit{centroid} away from its ideal position and hence bias the reconstructed ring width, if not properly taken into account. 
The apparent radius and center of the ring image on the focal plane are hence not completely independent of the impact point and inclination angle of the muon. 
For example, rays from an on-axis muon with $\rho_R = 0$ preferentially impact the mirrors at smaller angles than is the case for a muon with large impact distance producing a different pattern of aberrations for each case.



The tangential image centroid coordinate $<\!\xi\!>$ can be approximated for parallel rays, uniformly distributed over the reflector  \comn{of diameter $\mathcal{D}$ and focal length $F$, with an angle $\upphi$ with respect to on-axis incidence}, as~\citep{Vassiliev2007,Fegan:2018}: 
\begin{align}
  <\xi> & \approx \upphi \cdot \left( 1+ \ddfrac{3-2\kappa}{32}\left(\ddfrac{\mathcal{D}}{F}\right)^2 \right) ~,  \label{eq:xi:all}
\end{align}
\noindent
where  $\kappa=1$ corresponds to the ideal classic Davies-Cotton design~\citep{DaviesCotton}, $\kappa=1/2$ to the case of a parabolic reflector, and $\kappa = 0.83$ to the ``modified Davies-Cotton'' reflector proposed for the MST~\citep{MST-TDR}. A similar approximation can be used for the Schwarzschild-Couder telescope~\citep{Vassiliev2007}.

Eq.~\ref{eq:xi:all} translates into a $\sim$4\% shift with respect to the \comn{$\upphi$-scaling} for the parabolic LST\footnote{even $\sim$7\% for the MAGIC telescopes~\citep{GarczPhD}.} and a $\lesssim$2\% deviation for the MST/SST-1M.  
Corrections could be implemented straightforwardly through an adequate redefinition of the plate scale.
Such a ``plate-scale calibration'' fits the observed image centroid (e.g. of astrometric standard stars) in camera pixel coordinates to the tangential and sagittal angular distances $(\xi,\eta)$. The distribution of light becomes, however, strongly asymmetric along the tangential coordinate, at least for coma aberration, as one moves to high incidence angles. In such a case, a Gaussian ring width model, as used in Eq.~\ref{eq:Ipixs}, may result in biased ring parameters, as the muon inclination angle increases. Such biases need to be investigated with carefully simulated muon ring images.  
Cuts on the reconstructed impact parameter and/or muon inclination angle may be necessary to limit this effect. 

Moreover, Eq.~\ref{eq:xi:all} has been obtained by integrating parallel light rays uniformly covering the reflector, an assumption that does not hold for the conical wavefront of Cherenkov light from local muons, which hit the reflector preferentially toward the muon impact point. The plate-scale correction will hence be different in our case
and probably depend on muon impact distance and incidence angle. 

We will conservatively assume that the error made for the determination of $\thetac$ by the (gamma-ray-optimized) plate-scale calibration is less than half of the calibration correction itself, until a dedicated study on this issue using ray-tracing has been carried out. Such a study will be part of our second paper on this topic.

\subsubsection{Finite Camera Focuses}

Large IACTs have rather limited depths of field and are unable to focus all parts of shower images at the same time~\citep{hofmann2001}. This applies particularly to the LST and the MST, which will focus their cameras at the mean altitude at which air shower maxima are observed, normally chosen as 10~km distance above the telescope.  

For an ideal telescope of focal length \comn{$F$}, focused at infinity, i.e. with the camera placed at \comn{$z_f = F$}, the Cherenkov angle of the emitted rays equals the radius of the imaged ring. For a telescope \comn{focused} at a distance \comn{$x_f$}, i.e. with the camera placed at $z_f$ with $1/z_f = 1/F - 1/\comn{x}_f$, the rays from an on-axis muon radiating at distance \comn{$x$} will form a ring of radius
\begin{equation}
\tan\theta_\textit{zf} = \tan\thetac \cdot \left(1-\comn{x/x_f}\right) \quad.
\label{eq:tanvf}
\end{equation}
The values of \comn{$x=0$} and $\comn{x_f} = \infty$ give $\theta_\textit{xf} = \thetac$ as expected. For \comn{$x=x_f$} we get $\theta_\textit{xf} =0$; this is, of course, the definition of what it means for the telescope to be focused at \comn{$x_f$}. At its highest point, the Cherenkov light from a local muon is imaged into a smaller ring with radius smaller than $\thetac$, by the time the muon impacts the mirror \comn{$(x = 0)$} the angle is the nominal value. So this results in a systematic error if the ring radius is directly interpreted as the mean Cherenkov angle. 
The bias introduced on the reconstructed Cherenkov angle amounts to (see Appendix~\ref{sec:varfocus})

\begin{equation}
B \approx  - \ddfrac{R}{2\thetac} \cdot \ddfrac{1}{\comn{x_f}} \cdot E_0( \rho_R) ~.\label{eq:biasfinitefocus}
\end{equation}

For the MST with $\comn{x_f} = 10$~km, the bias ranges from $\sim -2\%$ to $\sim -4$\%, depending on the muon energy, whereas the LST can suffer biases of as high as $-8\%$.  

\subsubsection{Shadowing Effects}

For the MST, the large 3\,m camera and masts cause an average shadowing in excess of 15\% of the muon light. 
The effect of the central shadow (or of a hole in the mirror dish) may be approximated by subtracting from Eq.~\ref{eq:dNdphi} a chord of the same shape, but using the radius of the hole (or the camera), instead of the mirror. 
Fig.~\ref{fig:Ntot} has been obtained in this way; however, the square shape of the MST camera makes this approach inaccurate (see Section~\ref{sec:biasinclination}). 
For dual-mirror SSTs, shadowing in excess of 35\% is anticipated \comm{mainly} owing to the secondary mirror, 
\comm{but also telescope masts~\citep{Rulten:2016}}, although the exact value and its variation (between muons and air showers, and with $\rho$ and $\psi$) remain to be studied with dedicated simulations. 





\subsubsection{Nonuniformities of the Camera Acceptance and Mirror \comn{Reflectance}}
\label{sec:nonuniformities}


Nonuniformities of the camera acceptance can be controlled by a pixel-wise reconstruction of the optical bandwidth (see Section~\ref{sec:monitoringpixels}).
However, acceptance gradients and particularly changes of the dependency of the pixel response with the incidence angle of light may bias the obtained optical throughput correction for gamma-ray shower images. 
Gradients of up to 3\% have been found by \citet{bolzphd} for the H.E.S.S.-I telescopes, after plotting the mean $B_\mu$ against the inclination azimuth angle.

\subsubsection{Nontrivial Dependencies of the Optical Bandwidth on Incidence Angle\label{sec:incidence}}

Whereas $\gamma$-ray-induced showers may be assumed to illuminate the reflector homogeneously, muons illuminate the reflector preferentially close to the impact point. Together with the variation in efficiency of the photodetectors (PMTs or SiPMs) with incidence angle, this can lead to a nontrivial dependence of the reconstructed $B_\mu$ on $\rho_R$.

\begin{figure}[h!]
\centering
\includegraphics[width=0.99\linewidth]{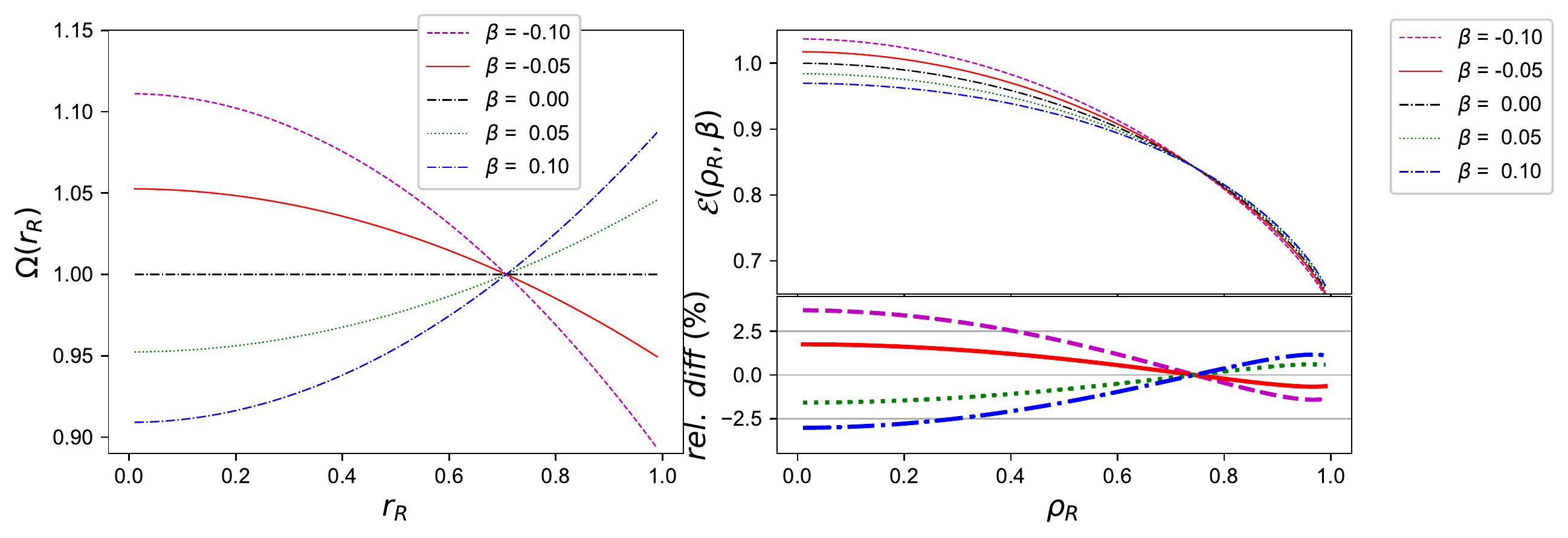}
\caption{Left: model of the relative light-collecting efficiency of the camera for photons arriving from various parts of the reflector. Right: relative total photon yield as a function of muon impact distance obtained for the different models displayed on the left side.  \label{fig:Qpixbeta} 
}
\end{figure}

 Photo-detection devices, such as PMTs~\citep{Okumura:2017} or SiPMs~\citep{Otte:2017}, used for IACT cameras exhibit quite pronounced dependencies of their photon detection efficiencies on incidence angle. IACT camera pixels are, moreover, often preceded by light \comm{concentrators} (or funnels, often approximating as much as possible the form of a \comm{compound parabolic concentrator (CPC);~\citet{Winston:2004}} \comm{or B\'ezier curves~\citet{Okumura:2012}}) to reduce dead spaces and enhance overall optical efficiency~\citep{Henault:2013,Aguilar:2015,HAHN:2017,Okumura:2017}. Such light concentrators achieve \comm{only approximately} uniform photon detection efficiencies up to a carefully chosen maximum incidence angle. Residual dependencies of several percent per degree remain and need to be taken into account.
Moreover, light \comm{concentrators} in front of the photodetector plate may degrade with time \comm{(unless they are kept in the camera housings and are hence not  exposed to the outer environment~\citet{Okumura:2017})}, particularly the dependency of their acceptance with respect to light incidence angle can change. 

The dependence of the photon yield as a function of impact point and nonuniformity
has been modeled by an analytic approximation (see Appendix~\ref{app:beta}), where a parameter $\beta$ describes the enhancement $(\beta > 0)$ or suppression $(\beta < 0)$ of the response for
photons reflected from the edge of the reflector. 
This is shown in Figure~\ref{fig:Qpixbeta}: for a muon with low impact parameter ($\rho_R \approx 0$), the difference in total muon light received (and hence the systematic error) scales roughly as \comn{$-\beta/3$}, 
until $\rho_R \approx 0.3$, whereupon the error reduces to zero at $\rho_R = 1/\sqrt{2} \approx 0.7$ (by construction) and finally reaches \comn{$\approx \beta/9$} at $\rho = 1$.
Hence, the muon light yield varies  by about the same amount as $\beta$ across the possible ranges of $\rho_R$, whereas by construction the telescope's response to gamma-ray shower light remains the same, and independent of $\beta$. 

\paragraph{Incidence-angle-dependent Photon Detection Efficiencies}\hspace{0pt}\\

Typical values of $\beta$ are $\sim$0.07 for the PMT-based cameras, mainly introduced by the light funnels in front of each PMT\footnote{However, a quite pronounced wavelength dependency of this value is expected~\citep[see Fig. 10 of][]{Hayashida:2017}. }, and $\sim$-0.06 for the solutions involving SiPMs. 
 
\paragraph{Incidence-angle-dependent Transmission of the Protecting Camera Entrance Window}\hspace{0pt}\\

The camera entrance window will introduce further incidence-angle-dependent effects: the amount of light lost to \textit{reflections} at the two window surfaces changes with incidence angle such as the \textit{optical path length} through the window. The latter will introduce a \comm{decrease in transmissivity with increasing incidence angle and hence a \textit{shift of the ``cutoff'' wavelength} of the window transmission spectrum}. The conversion of the optical bandwidth to muon Cherenkov light to that of gamma-ray-induced Cherenkov light (see Section~\ref{sec:systematic_C}) becomes then slightly dependent on incidence angle. Simulations show that these effects can cause an additional contribution to $\beta$ of up to $-0.1$ and will be treated in more detail in our second paper.  

A \textit{flat camera entrance window} shifts the light ray in tangential direction, as a function of the the relative distance of the impinging light ray from the camera center $r_R$, the window thickness \comn{$d$} and its refractive index $n$ as
\begin{align}
\Delta \upphi &= \ddfrac{d}{F} \cdot \tan\alpha\cdot \left(  1 - \ddfrac{\cos\alpha}{\sqrt{n^2 - \sin^2\alpha}}     \right)  \qquad \mathrm{with:}\\[0.2cm] 
  \tan\alpha &:= \tan\theta_1  \approx \ddfrac{r_R}{2} \cdot \ddfrac{D}{F} + \xi \quad.
\end{align}

\noindent
Inserting realistic numbers for $d \lesssim 6$~mm and $n \approx 1.5$ for plexiglas, one can see that the angular shift $\Delta \upphi$ is always smaller than 0.005$^\circ$ and hence negligible. 

\begin{figure}[htp]
\centering
\includegraphics[width=0.99\linewidth]{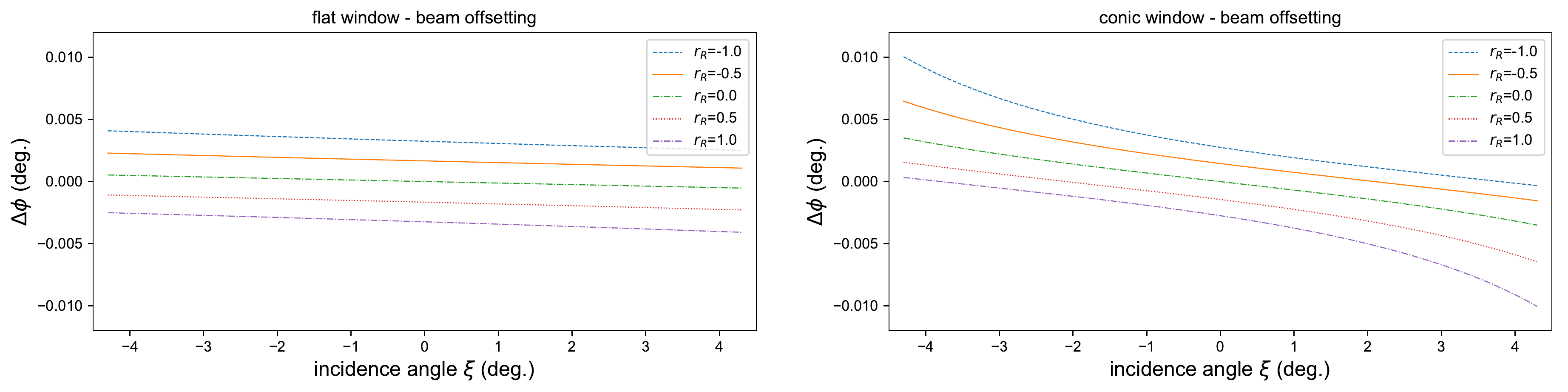}
\caption{Light ray offsetting due to a flat (left) and curved camera protection window for a typical MST camera. \label{fig:offsetting} 
}
\end{figure}

The situation changes slightly with a \textit{conic camera protection window}, as foreseen for the LST~\citep{LST-TDR} and one proposed MST camera solution~\citep{NectarCam-TDR}. In that case, the incidence angle of the light ray on a conic window becomes
\begin{align}
 \sin\alpha &:= \sin\theta_2  \approx \sin\theta_1 \cdot\ddfrac{ R_\mathrm{window}-h_\mathrm{offset}}{R_\mathrm{window}} + \cos\theta_1 \cdot \ddfrac{\xi\cdot F}{R_\mathrm{window}} \quad,
\end{align}
\noindent
where $R_\mathrm{window}$ is the window's curvature radius and $h_\mathrm{offset}$ the distance of the window's center from the focal plane. 
Inserting realistic values of $R_\mathrm{window} \approx 3.4$~m and $h_\mathrm{offset} \approx 0.4$~m yields now a maximum shift of up to 0.01$^\circ$. Figure~\ref{fig:offsetting} shows an example. On average, such an offset will translate, however, into an error of the reconstructed muon Cherenkov angle of considerably less than 1\%. 

Additionally, deflection of the rays occurs owing to the conic window shape, and the incidence and exit direction of the ray to and from the window are now different. The net effect seems to be negligible, though for all proposed solutions so far.

\subsubsection{Cherenkov Light Emitted by the Muon between the Primary and Secondary Mirror \label{sec:secondary} }

Additional considerations for a Schwarzschild-Couder telescope include the amount of light emitted between the secondary and primary mirrors if the muon travels through the secondary mirror. This contribution amounts to \comn{a maximum of} about 3\% for the dual-mirror SSTs, considering a distance of about 3~m between primary and secondary mirror~\citep{canestrari2013}. However,  
the small opening angle of the Cherenkov light cone may result in a large proportion of this light being lost through the hole in the primary mirror for muons passing near the center of the dish. This contribution is hence applicable only to muons with medium impact distance, which avoid the hole in the primary mirror but still hit the secondary. A careful analysis might detect that contribution and correct for it; otherwise, a suitable impact distance cut will remove it.    

\subsubsection{Trigger and Selection Biases \label{sec:triggerbiases}}

If the mirror area is small, the reflected muon light \comn{intensity} may not be strong enough to ensure a stable trigger efficiency above a certain muon energy threshold.
In that case, only Poissonian upward fluctuations will launch a trigger, and the distribution of reconstructed optical bandwidths $B_\mu$ will become asymmetric and its mean value biased.
Mirror degradations of \comn{reflectance} will not scale linearly with $<B_\mu>$ anymore, and a bias correction needs to be applied. The corresponding correction
factor needs to be retrieved from simulations; however, \comn{CTA will probably not be able to continuously monitor reflectance of each single mirror, and} simulations \comn{will} therefore tend to level out real fluctuations 
\comn{among the different mirrors.}
The correction factors may hence suffer from a considerable systematic uncertainty.

\begin{figure}[htp]
\centering
\includegraphics[width=0.95\linewidth]{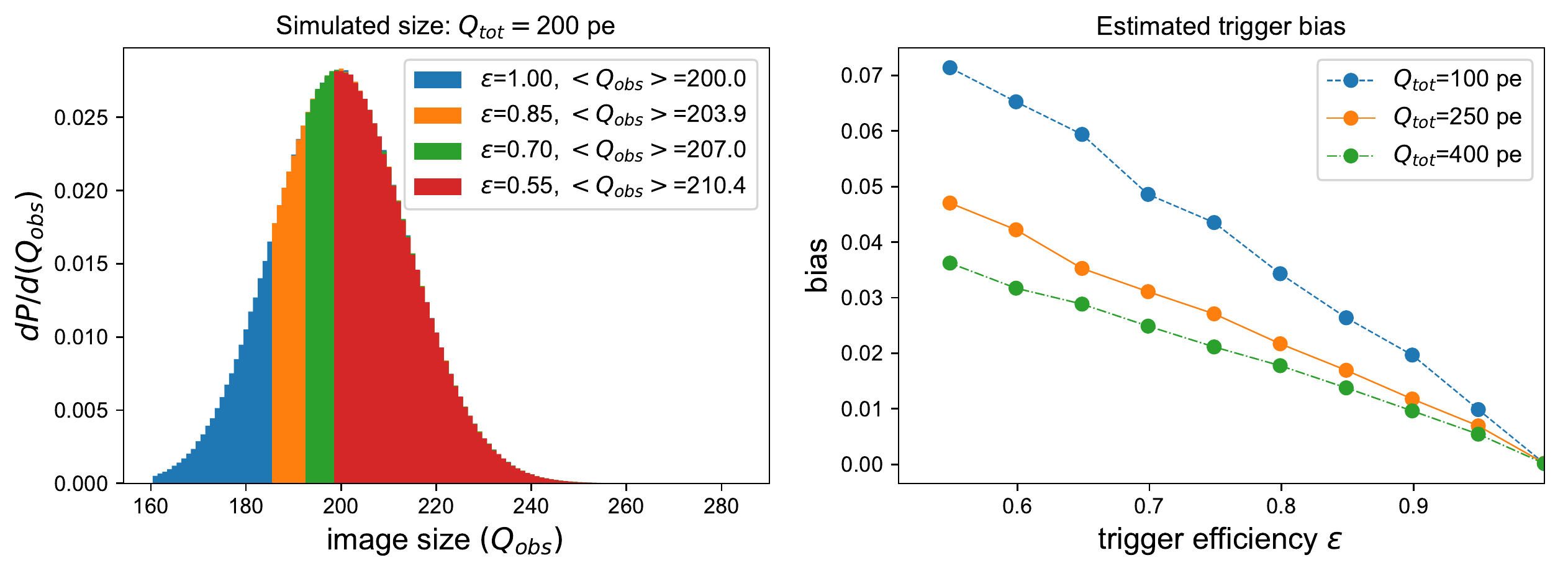}
\caption{Left: simulated distributions of image sizes $Q_\mathrm{tot}$, for a mean size of 200 photoelectrons, a typical value for SST telescopes. The left part of the distribution has been subsequently cut, in order to obtain 
the efficiency numbers provided in the legend. The corresponding difference between statistical mean and the position of the Gaussian peak may give a hint of the expected 
trigger bias. Right: distribution of biases as a function of trigger efficiency, for different mean muon image sizes. \label{fig:triggerbias} }
\end{figure}

We have tried to simulate the effect using a Gaussian distribution of a certain mean muon image size and its square root as Gaussian width. Then, the 
lower parts of the distribution were removed until a given trigger efficiency was reached (see Fig.~\ref{fig:triggerbias} left panel). The difference between the statistical mean of the 
new distribution and the original image size has then been defined as a toy-model trigger bias (Fig.~\ref{fig:triggerbias} right panel). 
This extremely simplified model may give an order of magnitude of the expected real trigger biases.

However, if the distribution of $B_\mu$ shows an identifiable cut toward lower values, one can try to fit only the peak of that distribution, instead of
calculating its mean or median. If such an approach can help, it needs further investigation.
In any case, all CTA telescopes are required to ensure that their triggers are sensitive enough to local muon light, such that they can ensure that the trigger bias becomes negligible, even if the telescopes' optical throughput degrades within the permitted ranges.

\subsubsection{Mirrors}


The mirror \comn{reflectance} degrades with time, most probably in a wavelength-dependent way. The overall \textit{average} degradation is a free parameter of the calibration procedure and is part of the obtained optical throughput. A general worsening of the \textit{PSF} of the mirror should be retrieved by a careful analysis of the reconstructed muon ring width.

\comn{More difficult is the estimation of the effect of a few very misaligned mirrors, possibly on very short time scales, due to malfunctioning of the Active Mirror Control for that particular set of mirrors.  In that case, the muon light gets reflected into the margins of or even outside the ring by those particular mirrors and possibly misinterpreted as off-ring contribution by the muon analysis. }
In this case, the disentanglement 
of the worsened PSF from the effects of the mirror \comn{reflectance} becomes less accurate.
For the case of the LST, one misaligned mirror out of $\sim$200~\citep{Ambrosi:2014a} may create a systematic bias of $\lesssim$0.5\%, while for the MST one mirror out of 86~\citep{schlenstedt2014_2} creates a bias of $\lesssim$1\%.


\subsection{Atmospheric Effects}

\subsubsection{Bending of the Muon Trajectory in the Geomagnetic Field}

Bending of the muon trajectory in the geomagnetic field has not been considered important so far for local muons. It is not even mentioned as a source of ring broadening by~\citet{vacanti}. 
A muon of momentum $p$ propagating a distance $D$ perpendicular to the field of strength $B$ will be deflected by an angle $\Delta\theta$~\citep[see, e.g., chap. 35.12 of][]{pdg2017}:
\begin{align}
\Delta\theta &= \left(\ddfrac{\ud\theta}{\ud x}\right) \cdot x \nonumber\\
             &= 6.9\times 10^{-5} ~\mathrm{(deg/m)}
\cdot \left(\ddfrac{10~\mathrm{GeV}}{p}\right) 
\cdot \left(\ddfrac{B_\perp(\vartheta,\varphi)}{40~\mu\mathrm{T}}\right) 
\cdot \left(\ddfrac{x}{\mathrm{m}}\right) \label{eq:muonbending} 
\end{align}
\noindent
where $\vartheta,\varphi$ are the local muon direction angles and $B_\perp$ the magnetic field component perpendicular to its velocity.  The maximum achieved at La Palma at a zenith angle of $\sim 40^\circ$ is of $B \approx 40~\mu$T~\citep{commichau2008}, whereas at Paranal maximum vertical $B$-fields of $\sim 25~\mu$T are achieved~\citep{Hassan:2017}. The angle under which the muon Cherenkov light is imaged in the camera gets then slightly shifted by (see Appendix~\ref{sec:varmagnetic})
\begin{align}
\Delta \thetac(x,\phi)  
                &\simeq     \cos(\varphi-\phi) \cdot \left( \ddfrac{\ud\theta}{\ud x}\right) \cdot x ~. \label{eq:deltatheta_magn}        
\end{align}
The shift, averaged over the ring, has an expectation value $E[\Delta \thetac]$ of
\begin{align}
E[\Delta\thetac] &= \ddfrac{1}{2\pi} \int_0^{2\pi} \!\!\ddfrac{1}{L(\phi)} \int_0^{L(\phi)}\!\!\!\! \Delta \thetac(x,\phi)\, \ud x\,\ud\phi  \nonumber\\
           &\approx \rho_R \cdot \ddfrac{1}{\comn{4}\thetac} \cdot \left( \ddfrac{\ud\theta}{\ud x}\right) \cdot R \cdot \cos(\phi_0-\varphi) \label{eq:muonbending2}
\end{align}
The bias for the reconstructed Cherenkov angle $B(\thetac) = (E[\Delta\thetac])/\thetac$ can reach up to \comn{2}\% for the worst case of a full muon ring imaged into an LST camera, under the maximum impact distance and the azimuthal component of the $B_\perp$  parallel azimuth component of the impact parameter.   

Geomagnetic field effects can be studied straightforwardly  by comparing data from pointings toward the south (where the perpendicular magnetic field is strongest) and the north (where it is weakest, almost reaching zero; \citet[see][]{commichau2008}) in the case of La Palma. 

\subsubsection{Atmospheric Transmission \label{sec:transmission}}

A photon of energy $\epsilon$ emitted at a certain distance $r$ above the telescope suffers molecular and aerosol extinction before reaching the mirror. 
One can write the atmospheric transmission for that photon very generically as 
\begin{equation}
T(\epsilon,r;\vartheta)  = \exp \big(  - \int_0^{r\cdot \cos\vartheta} \!\!\!\!\alpha_\mathrm{mol}(\epsilon,h) + \alpha_\mathrm{aer}(\epsilon,h) ~\ud h / \cos\vartheta\big)  \quad,
\end{equation}
\noindent
where $\alpha_\mathrm{mol}$ and $\alpha_\mathrm{aer}$ are the volume extinction coefficients from molecular and aerosol extinction at an altitude $h$, respectively, and the telescope itself points to the sky under a zenith angle $\vartheta$.

Assuming that the muon light emitted along the track $L$ from $r_\mathrm{max}=D(\vec{\rho},\phi)/\tan\thetac$ to the telescope mirror dish, we can write
\begin{align}
t_\mu(\epsilon,\rho_R;\vartheta) &\approx \ddfrac{1}{2\pi} \int_0^{2\pi} \frac{1}{r_\mathrm{max}(\vec{\rho},\phi)} \int_0^{r_\mathrm{max}(\vec{\rho},\phi)}\!\! T(r,\epsilon;\vartheta) ~\ud r \,\ud\phi\nonumber\\[0.3cm]
              &\approx \ddfrac{1}{2\pi} \int_0^{2\pi} \frac{1}{r_\mathrm{max}(\vec{\rho},\phi)} \cdot \nonumber\\[0.3cm]
              & \qquad \cdot \int_0^{r_\mathrm{max}(\vec{\rho},\phi)} \!\!\!\!\exp \left(  - \int_0^{r\cdot \cos\vartheta}\!\!\!\!\!\!\!\! \alpha_\mathrm{mol}(\epsilon,h) + \alpha_\mathrm{aer}(\epsilon,h)~\ud h / \cos\vartheta \right) ~\ud r \,\ud\phi\label{eq:transtot}\\[0.35cm]
              &\approx \quad t_{\mu,\mathrm{mol}}(\epsilon,\rho_R;\vartheta) ~ \cdot ~ t_{\mu,\mathrm{aer}}(\epsilon,\rho_R;\vartheta) \qquad\textrm{with:}\nonumber\\[0.4cm]
t_{\mu(\mathrm{mol,aer})}(\epsilon,\rho_R;\vartheta) &= \ddfrac{1}{2\pi} \int_0^{2\pi}\!\!\!\! \frac{1}{r_\mathrm{max}(\vec{\rho},\phi)} \int_0^{r_\mathrm{max}(\vec{\rho},\phi)} \!\!\!\!  \!\!\!\!\exp \left(  - \int_0^{r\cdot \cos\vartheta} \!\!\!\!\!\!\!\! \alpha_\mathrm{(mol,aer)}(\epsilon,h) ~\ud h/ \cos\vartheta  \right) ~\ud r  \,\ud\phi\label{eq:transmolaer}
\end{align}

The transmission expressed in Eq.~\ref{eq:transtot} is independent of the azimuthal coordinate of the impact distance $\phi_0$, because of the radial symmetry of the system. It must then be folded with the telescope efficiencies to obtain the optical bandwidth of the full system: 
\begin{align}
B_\mu(\rho_R;\vartheta) &= \int t_\mathrm{\mu}(\epsilon,\rho_R;\vartheta) \cdot \xi_\mathrm{det}(\epsilon) ~\ud \epsilon
\quad. 
\label{eq:Bmut}
\end{align}

\paragraph{Molecular Transmission \label{sec:moltransmission}}\hspace{0pt}


The volume scattering cross section for molecular extinction in the atmosphere can be written as~\citep{mccartney,bucholtz,tomasi,gaug2014}:
%
%
\begin{equation}
\alpha_\mathrm{mol} (\epsilon,h) \approx 5.764 \cdot (n_s(\epsilon)-1)^2 \cdot \epsilon(\mathrm{eV})^4
\cdot \frac{P(h)}{P_s}\cdot \frac{T_s}{T(h)}~\mathrm{m^{-1}} \quad ,  \label{eq:Rayleigh1}
\end{equation}
%
%
\noindent
where $P(h)$ and $T(h)$ are the atmospheric pressure and temperature at altitude $h$, and $P_s$ and $T_s$ are the reference values of the U.S. standard atmosphere~\citep{uss}.

\begin{figure}[h!]
\centering
%
%
\includegraphics[width=0.7\linewidth]{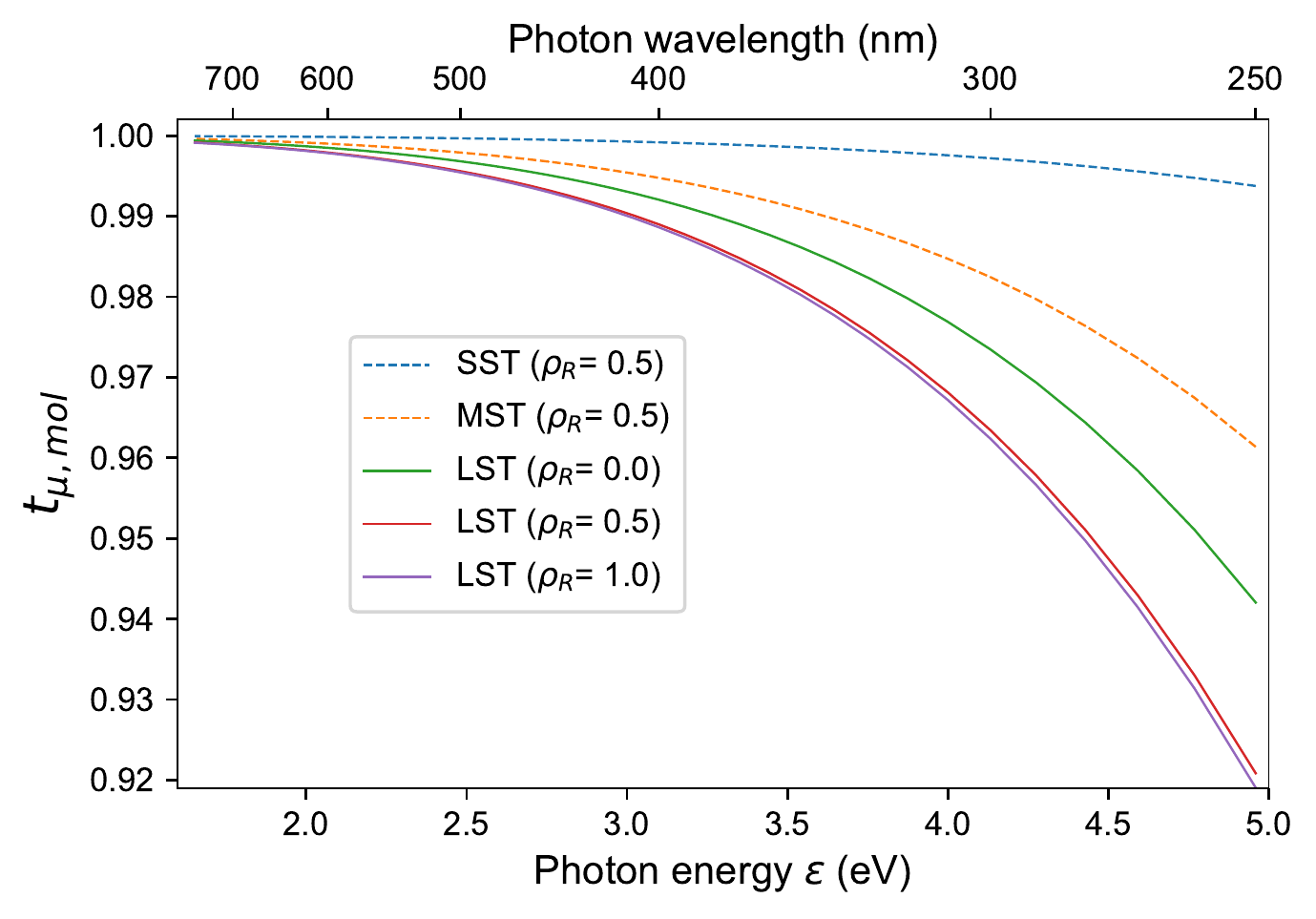}
\caption{Approximate molecular transmission of muon light for different telescope types, derived from Eqs.~\ref{eq:Rayleigh2} and~\ref{eq:transmolaer} for an observatory altitude of 2200~m a.s.l. For the LST, also different impact distances, used in Eq.~\ref{eq:transmolaer}, are shown. A Cherenkov angle of 1.23$^\circ$ was used to produce all data points. Shadows have been included according to Eq.~\ref{eq:Thetaapprox}.
\label{fig:Tmol}}
\end{figure}

Assuming a typical winter atmosphere at the ORM~\citep{gaugatmohead2016}, the term $\left(P(h)/P_s\right) \cdot \left(T_s/T(h)\right)$ can be described by a scale height of $H_0 \approx 9.7$~km; hence,
\begin{equation}
\alpha_\mathrm{mol} (\epsilon,h) \approx 5.764 \cdot (n_s(\epsilon)-1)^2 \cdot \epsilon(\mathrm{eV})^4 
\cdot \exp(-h/H_0)~\mathrm{m^{-1}} \quad .  \label{eq:Rayleigh2}
\end{equation}
%
The result using Eq.~\ref{eq:Rayleigh2} for $\alpha_\mathrm{mol} (\epsilon,h)$ to calculate $t_{\mu,\mathrm{mol}}(\epsilon,\rho_R;\vartheta)$ (Eq.~\ref{eq:transmolaer}) is displayed in Fig.~\ref{fig:Tmol} for three CTA telescope types. Since the LST is most affected by molecular extinction, the effect of different impact distances is also shown. We studied also different telescope observation zenith angles; however, their effect was even smaller, i.e. well below 0.1\%, and is hence not displayed. Averaging $B_\mu(\rho_R)$ over the different impact distances yields $<B_\mu> = 0.6348\pm 0.0006$ for the LST, hence a systematic effect of less than 0.1\%. 

%
%
\begin{table}[h!]
\centering
\begin{tabular}{lccr@{.}l} \toprule
Telescope  &  \multicolumn{1}{c}{$h_\mathrm{max}$}   & \multicolumn{1}{c}{$t_{\mu,\mathrm{mol}}(300~\mathrm{nm})$} & \multicolumn{2}{c}{$B_\mu$}  \\
           &  \multicolumn{1}{c}{(m)}              &  \multicolumn{1}{c}{}  &  \multicolumn{2}{c}{(eV)} \\
\midrule
LST (Hamamatsu PMT)  & [550,1100] &  0.968  & 0&6348$\pm$ 0.0006\\ 
MST (Hamamatsu PMT) & [280,570] &  0.982 & 0&6401$\pm$ 0.0006 \\  
SST      & [95,190]  & 0.997 & 0&6406$\pm$ 0.0002 \\  \addlinespace[0.15cm]
\bottomrule
\end{tabular} 
\caption{\label{tab:Tmol} Approximate molecular transmission and optical bandwidths for observed Cherenkov light from muons for the different telescope types. 
The first column shows the range of possible maximum emission heights (depending on the muon impact distance, for full rings), the second column the molecular transmission for an averaged Cherenkov light ray from muons at an impact distance of $\rho_R = 0.5$, and the last column the obtained optical bandwidths for a generic telescope. The uncertainties stem from varying the impact distance of the muons. 
All values have been derived for an observatory altitude of 2200~m a.s.l. }
\end{table}


One can see that only the LST shows non-negligible molecular extinction of muon light; however, its contribution to $B_\mu$ can be safely estimated to lie always below 3\%, independent of the impact distance of the muon. In order to assess the range of variation of $t_{\mu,\mathrm{mol}}$ due to changes in the atmospheric profile for the LST, we used the data of one full year from the ``Global Data Assimilation System'' (GDAS) ``Final Analysis'' database\footnote{\url{https://rda.ucar.edu/datasets/ds083.3/}} for the closest grid point to La Palma, whose temperature and pressure predictions show excellent correlation with the measurements of the MAGIC weather station~\citep{Gaug:2017site}, apart from a constant 2$^\circ$C bias, which can be corrected. The full height-resolved temperature and pressure profiles were used to derive $B_\mu$ for fixed telescope efficiencies $\xi_{\mathrm{det}}(\epsilon)$ for the LST. The \comn{resulting} distribution of $B_\mu$ shows a peak-to-peak variation of less than 0.2\%.  
We can conclude that the uncertainty of the molecular profile on $t_\mathrm{\mu,mol}$ is absolutely negligible, \comm{at least for CTA-N, once the correct average atmospheric profile has been selected.}


\paragraph{Aerosol Transmission \label{sec:aertransmission}}\hspace{0pt}

Astronomical sites, like Armazones and La Palma, are characterized by extremely clean environments and small aerosol content close to the ground. 
\citet{patat2011} find a median aerosol extinction of 0.045~mag~airmass$^{-1}$, and the semi-interquartile range is 0.009~mag~airmass$^{-1}$ at 400~nm wavelength for 
the VLT site at Paranal (about 500~m higher than the CTA-S site). 
The height profile of the aerosol extinction is not provided, however. We can nevertheless assume that stratospheric aerosol accounts for about 0.005~mag, and the 
rest forms part of the nocturnal boundary layer close to ground. Hence, about $(4 \pm 1)$\% of the Cherenkov light from gamma-ray showers is expected to be scattered 
out of the FOV in the first kilometer on the ground, probably slightly more on the CTA-S site, due to the lower altitude. 

At La Palma, \citet{garciagil2010} measure a median extinction of $0.129\pm 0.002$~mag airmass$^{-1}$ in the $V$ band, throughout the whole year, except for the summer months, at 2369~m a.s.l. (about 100--150~m higher than CTA-N). Subtracting molecular extinction of 0.079~mag airmass$^{-1}$ and stratospheric ozone absorption of 0.023, an aerosol extinction of about $\approx$0.027~mag~airmass$^{-1}$ is obtained. 

The wavelength dependency of the aerosol extinction is typically expressed by the {\AA}ngstr\"om coefficient $A$, where
\begin{equation}
\ddfrac{\alpha_\mathrm{aer}(\lambda_1)}{\alpha_\mathrm{aer}(\lambda_2)} = \left( \ddfrac{\lambda_2}{\lambda_1} \right)^{A} \label{eq:angstrom} \quad,
\end{equation}
with $A \approx 1.4$ for Paranal~\citep{patat2011}, and ranging from $A \approx 0.8$ to~2.0 for La Palma~\citep{whittet}. 
Similar values have been obtained for Tenerife by \citet{maring}, ranging from $A \approx 1.5$ to~2 during nondusty nights, and 
\citet{andrews}, which tend toward $A \approx 1.2$ for small scattering coefficients. 

The MAGIC Collaboration 
has found that the altitude profile of the aerosol extinction coefficient 
for clear nights is exponential 
with a scale height of $H_\mathrm{aer} \approx $(500--700)~m~\citep{Gaug:2017site}, with an aerosol optical depth (AOD) on the ground of about (AOD$\approx 0.02$) on average, however ranging from practically zero to almost 0.1 for normal clear nights. 

Stronger aerosol extinction is possible only during Saharan dust intrusions
(called ``calima'' in the Canary Islands). 
Such nights show severely enhanced optical depths (up to close to or greater than 1.0), \AA ngstr\"om coefficients close to zero, and constant aerosol extinction from the ground to approximately 5~km altitude~\citep{Lombardi2011}.

\begin{figure}[h!]
\centering
%
%
\includegraphics[width=0.7\linewidth]{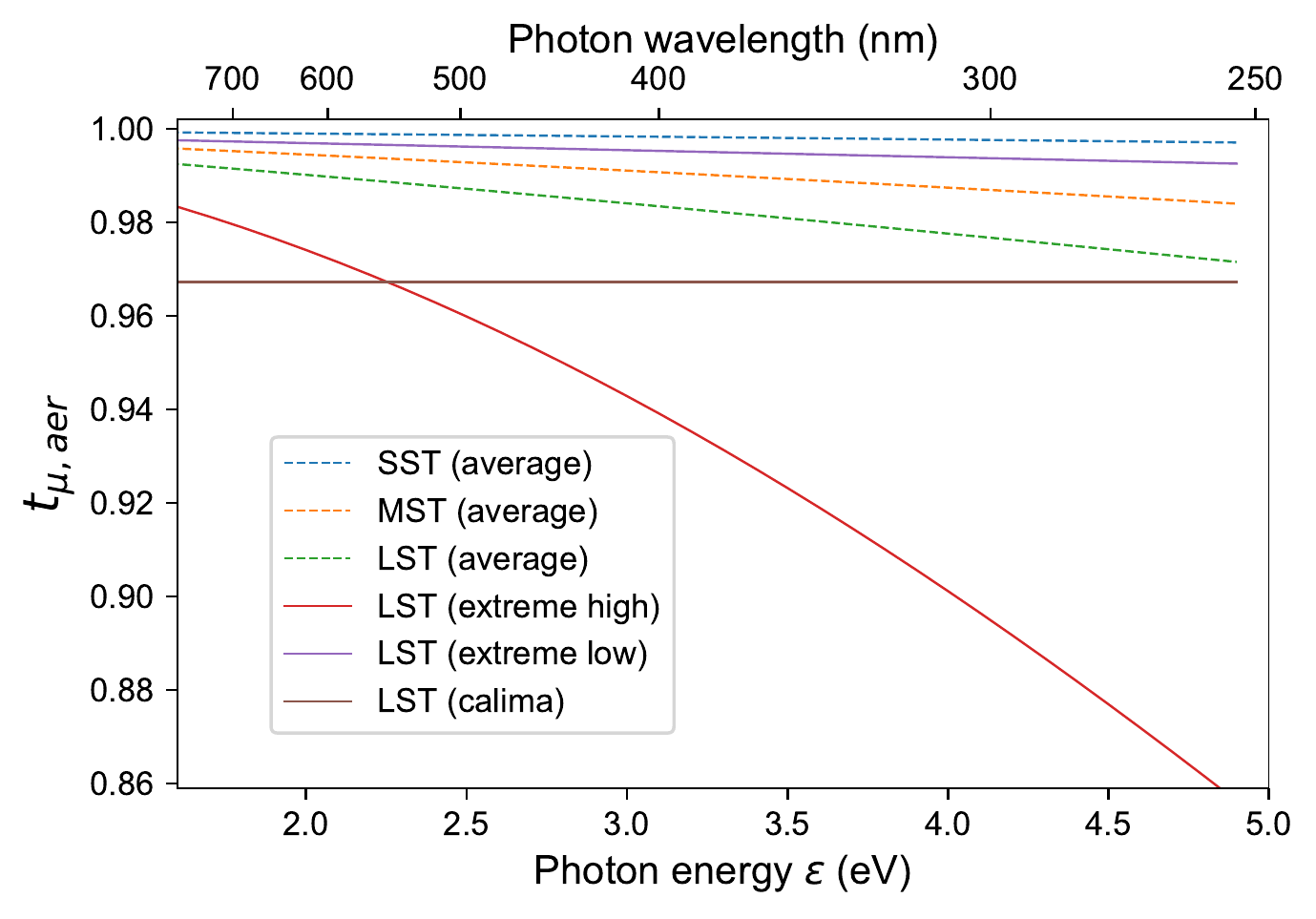}
\caption{Approximate aerosol transmission of muon light for different telescope types, derived from Eqs.~\ref{eq:angstrom} and~\ref{eq:transmolaer} for an observatory altitude of 2200~m a.s.l. For the LST, also different impact distances, used in Eq.~\ref{eq:transmolaer}, are shown. A Cherenkov angle of 1.23$^\circ$ was used to produce all data points. Shadows have been included according to Eq.~\ref{eq:Thetaapprox}.
\label{fig:Taer}}
\end{figure}

We have calculated the aerosol transmission for muons $t_{\mu,\mathrm{aer}}$ for an average case (AOD$_\mathrm{532~nm}=0.03$, $H_\mathrm{aer}=600$~m, $A =1.2$) 
and two extreme cases of clear nights with the exponential profile, created to yield strong extinction of muon Cherenkov light (AOD$_\mathrm{532~nm}=0.08$, $H_\mathrm{aer}=500$~m, $A =2.0$)  and tiny extinction (AOD$_\mathrm{532~nm}=0.01$, $H_\mathrm{aer}=700$~m, $A = 1.0$), just to show the range within which $t_\mathrm{\mu,aer}$ can vary for clean nights. The case of Saharan dust intrusions (``calima'') is treated as another extreme example: AOD$=0.5$ (supposing that this is the absolute limit for observation), $A = 0.$ and constant aerosol extinction up to 3~km above the ground. 
The results are shown in Figure~\ref{fig:Taer} and Table~\ref{tab:Taer}.

%
%
\begin{table}[h!]
\centering
\begin{tabular}{lcccc} \toprule
Telescope  &  \multicolumn{1}{c}{$t_\mathrm{\mu,aer}$(300 nm)} & \multicolumn{1}{c}{$t_\mathrm{\mu,aer}$(300 nm)} & \multicolumn{1}{c}{$t_\mathrm{\mu,aer}$(300 nm)} & \multicolumn{1}{c}{$t_\mathrm{\mu,aer}$(300 nm)} \\
           &    \multicolumn{1}{c}{Average} & \multicolumn{1}{c}{Extremely} & \multicolumn{1}{c}{Extremely}  & \multicolumn{1}{c}{Extremely} \\
           & (1)   & Low & \multicolumn{1}{c}{High (no calima)}  & \multicolumn{1}{c}{High (calima)} \\
\midrule\addlinespace[0.1cm]
LST (
PMT) 
& 0.977  & 0.994 & 
0.897 &  0.968 \\ 
MST (
PMT)  
& 0.987  & 0.997 & 
0.938 & 0.984  \\  
SST (SiPM)                   & 0.998 & 0.999 & 0.988      & 0.998   \\  \addlinespace[0.15cm]
\midrule\addlinespace[0.15cm]
 &  \multicolumn{1}{c}{$B_\mu$}  & \multicolumn{1}{c}{$B_\mu$}  & \multicolumn{1}{c}{$B_\mu$}  & \multicolumn{1}{c}{$B_\mu$}   \\
           &    \multicolumn{1}{c}{average} & \multicolumn{1}{c}{extremely} & \multicolumn{1}{c}{extremely}  & \multicolumn{1}{c}{extremely} \\
           &    & low & \multicolumn{1}{c}{high (no calima)}  & \multicolumn{1}{c}{high (calima)} \\
           &  (eV) &  (eV) &  (eV) &  (eV) \\\addlinespace[0.1cm]
\midrule\addlinespace[0.1cm]
LST (
PMT)  & 0.624  & 0.632 & 
0.592
 & 0.614 \\ 
MST (
PMT)  & 0.634  & 0.638 & 
0.614 & 0.630  \\  
SST (SiPM) & 0.639 & 0.640 & 0.636   & 0.639   \\  \addlinespace[0.1cm]
\bottomrule
\end{tabular}  
\caption{\label{tab:Taer} Aerosol transmission $t_\mathrm{\mu,aer}$ at 300~nm wavelength
for an averaged Cherenkov light ray from muons at an impact distance of $\rho_R = 0.5$, for the different telescope types and average and extreme aerosol conditions (see text for details). 
The obtained optical bandwidth $B_\mu$ has been calculated for a generic telescope and average molecular transmission. All values have an uncertainty of $\Delta B_\mu \lesssim 0.001$ due to the residual impact distance dependency. 
Those entries which deviate by more than 2\% from the average are marked as bold. 
All values have been derived for an observatory altitude of 2200~m a.s.l.}
\end{table}


One can see that only in the case of strong aerosol densities, very close to the ground and with an unusually high contribution of the accumulation mode particles, does a correction of less than 3\% 
need to be applied \textit{for the MSTs and LSTs}. This problem can be either circumvented either by adequate data selection (e.g., using LIDAR data)
or by applying a bias correction, which can be calculated once the aerosol profile is assessed. Such a correction is expected to take place only rarely, when the observatory operates under nonoptimal conditions.
Variations due to different muon impact distances are smaller than $\pm 0.2$\% for the average aerosol case, and smaller than $\pm 0.5$\% for the extreme high case. 

\subsection{Effects Related to Analysis and Reconstruction of the Muon Rings}

\subsubsection{Biased Signal Estimators \label{sec:systbias}}


Many signal estimators show a small bias toward low signal amplitudes~\citep{extractor2008}, which will distort the amount of reconstructed muon light. It is recommended to switch to unbiased estimators, at the latest in the last analysis step prior to determination of the optical bandwidth and impact parameter. The error introduced by keeping biased estimators (typically around 0.5--1 p.e. at signal expectations below 2~p.e.;~\citet{FADCPulsReco2008}) can be estimated as follows: assuming that on average one pixel on either side of the ring is affected by this bias at each azimuthal angle, the total bias is roughly 1.5~p.e. per number of pixels on the ring ($N_\mathrm{pix}$, Eq.~\ref{eq:npix}), hence $1.5\cdot 2\pi \cdot \thetac/\omega$, where $\omega$ is the pixel FOV. For $\thetac \approx 1.23^\circ$ and $\omega = [0.1^\circ,0.24^\circ]$, systematic biases in the range from 8\% to~15\% of the total signal are obtained.



\subsubsection{Non-negligible Inclination Angles \label{sec:biasinclination}}

Muon inclination angles were so far limited to $|\psic| \lesssim \comn{1^\circ}$ for $\lesssim 5^\circ$ FOV cameras, with associated effects scaling as \comn{$\approx \tan\thetac \tan\psic + \tan^2\psic/2 \comn{\lesssim 0.05\%}$ for $|\psic| < \comn{\lesssim 1^\circ}$} (see Appendix~\ref{sec:inclinedmuons}) and hence negligible. However, wide-field cameras for CTA with up to 10$^\circ$ FOV can  regularly obtain muon images with ring centers up to \comn{$3.5^\circ$} from the camera center, resulting in a maximum effect on the modulated signal along the ring of \comn{0.3\%}. 



The situation becomes more complicated when central shadowing objects are considered, like the camera. Traditionally, relatively small cameras (with small focal ratios and hence small  dimensions compared to the reflector) have been employed by IACTs~\citep{barrau,Cogan:2005,KILDEA2007,Anderhub:2009, magicperformance1,Giavitto:2017}. In these cases, the shadowing function $\Theta(\vec{\rho},\vec{\psic})$ can be approximated by Eq.~\ref{eq:Thetaapprox}. For telescopes with a  focal ratio of $f = F/D$, the shadow produced by the camera moves with inclination angle \comn{approximately} as
\begin{align}
\Delta \rho_R &\approx f \cdot |\psic|  
\quad, \label{eq:DeltarhoRpsic}
\end{align}
\noindent
slightly more than the intrinsic resolution with which $\rho_R$ can be obtained (see Section~\ref{sec:fitspestat}). \comn{The apparent shift} of the camera shadow can, however, be included in the reconstruction of the impact parameters and optical bandwidth using Eq.~\ref{eq:rho_moved}. 

\comn{Furthermore, the square shape of some camera designs (like the MST and the LST) introduces dependencies of the reconstructed optical bandwidth on the inclination angle; see Figs.~\ref{fig:totphotonsMST_edge} and~\ref{fig:totphotonsMST_corner}. Using the procedure described in Appendix~\ref{sec:quadraticcamera}, the effect of such a quadratic shape can be included in the analytical model, as well as the additional shadow of the central hole (Eq.~\ref{eq:Dhole}).}

 \comn{We considered the \comn{MST} as the worst-case \comnn{scenario for} 
 the effects of camera shadows.} The MST will host a camera of $\sim 8^\circ$ FOV in an $f/D \sim 1.3$ configuration~\citep{Glicenstein:2013,Oakes2015,Puehlhofer:2015}, \comn{resulting in a 3~m diameter square shadow on a 12~m diameter optical mirror structure. Figures~\ref{fig:totphotonsMST_edge} and~\ref{fig:totphotonsMST_corner} (top) show the number of photons for a muon incident on the mirror at different impact distances with $0^\circ$ and $2^\circ$ inclination angle ($\upsilon$), the latter under various azimuthal projection angles ($\psi$). The predictions have been produced by ray-tracing simulations of the entire telescope, including all shadows and the Cherenkov light of the muon emitted below the camera.  One can clearly see that the amount of expected photoelectrons depends not only on $\upsilon$, but also sensitively on $\psi$, throughout a range of impact distances from $\rho_R \approx 0.1$ to $\rho_R \approx$ 0.7. If only the solution by \citet{vacanti} (Eq.~\ref{eq:Qtot}) was used, both for the chord on the mirror and the chord of the shadow (Eq.~\ref{eq:Thetaapprox}), peak-to-peak differences of up to 25\% are observed for the simulated amount of light, \textit{only varying the $\psi$ angle}. Accordingly, an error of the same magnitude can be obtained in the reconstruction (middle panels). If instead the displaced square camera shadow is used for the model, the maximum error is reduced to below 5\% (bottom panels). The latter includes now all the additional shadows from ropes, the camera support structure, etc., which have not been taken into account in the analysis. One can also see in both figures that an \textit{average} error of less than 2\% can be expected in that case. }

\begin{figure}
\centering
\includegraphics[width=0.63\linewidth]{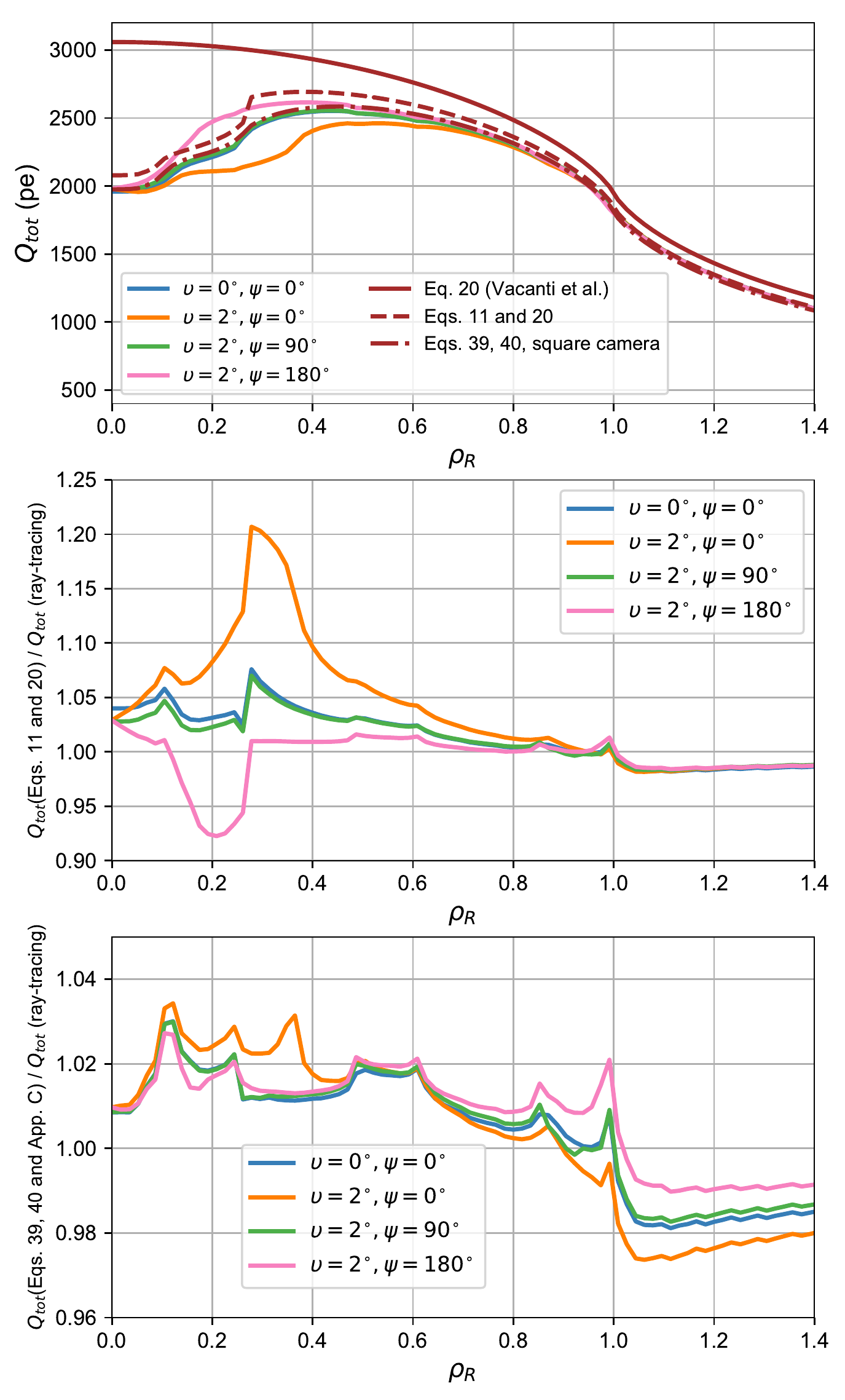}
\caption{Expected effect of the shadow of the square camera of the MST on the total amount of photons collected (top). In \comn{solid} brown, Eq.~\ref{eq:Qtot} is plotted without considering the shadow; in \comn{dashed} brown a roundish analytic camera model (following Eq.~\ref{eq:Thetaapprox}) has been subtracted from the previous one. \comn{The dot-dashed line uses instead an analytical solution for the shadow of a square-shaped camera (Appendix~\ref{sec:quadraticcamera}) and uses Eqs.~\ref{eq:Dshadow} and~\ref{eq:Dhole}}. The solid blue lines show the true value for an on-axis muon, and the orange, green, and pink lines show instead the case for a 2$^\circ$ inclined muon, inciding with azimuth angles of $0^\circ, 90^\circ$ and $180^\circ$. A $\phi_0$ value of 180$^\circ$ has been used for all data points shown. The \comn{middle} panel shows the relative values of the latter cases, compared to the analytical roundish camera model, \comn{and the bottom panel shows those with the analytical square camera model and the improved shadow models. Note the different y-axis scales of the center and bottom panels. With the full shadow model, a peak-to-peak difference in modeled and simulated photon yield between the investigated impact distances and muon inclination angles of only about 4\% is observed, if fully contained rings are considered.} 
\label{fig:totphotonsMST_edge}}
\end{figure}

\begin{figure}
\centering
\includegraphics[width=0.63\linewidth]{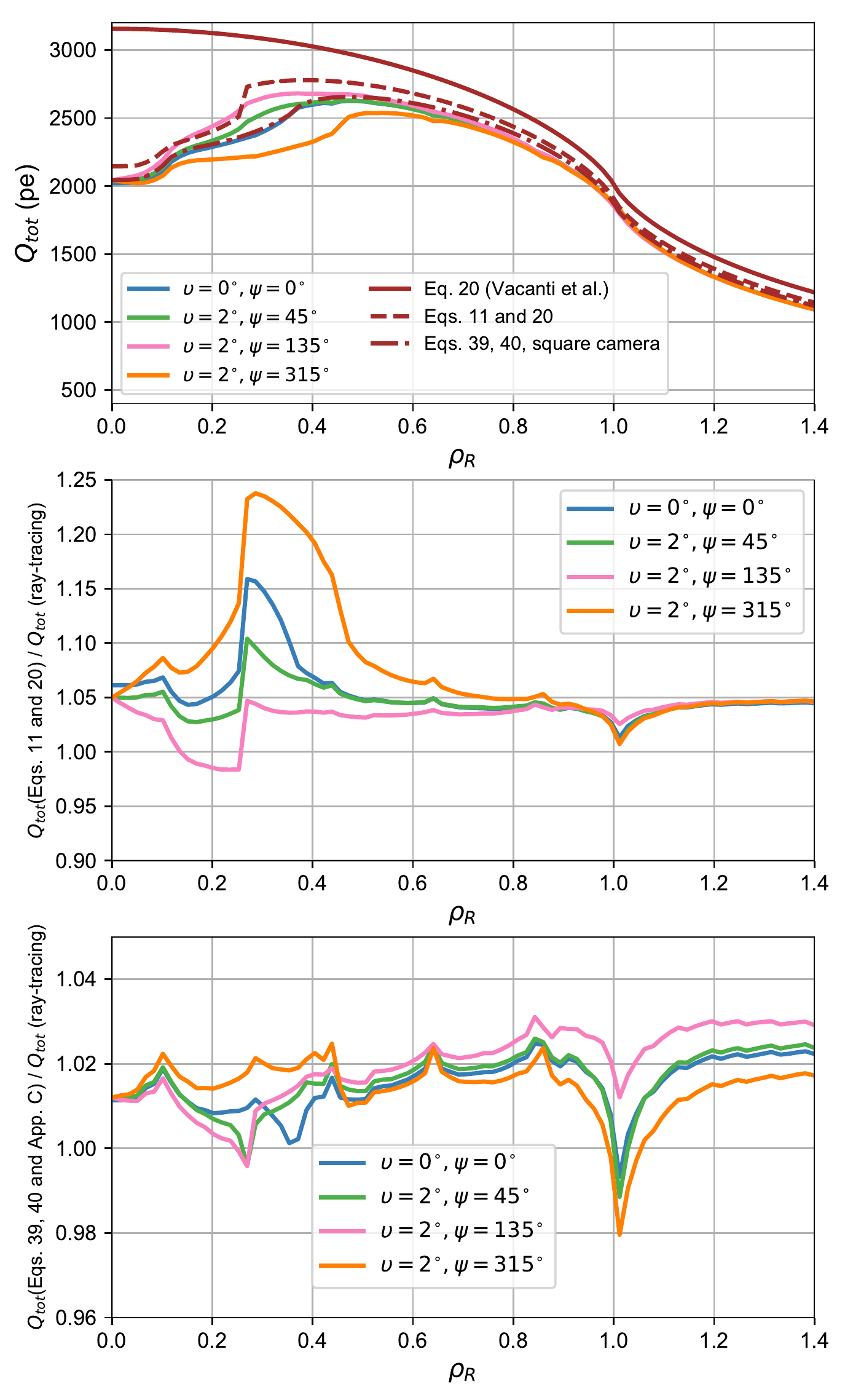}
\caption{\comn{Same as Figure~\ref{fig:totphotonsMST_edge}, now with a  $\phi_0$ value of 135$^\circ$ and azimuthal projections of the inclination angle of 45$^\circ$, 135$^\circ$, and 315$^\circ$. 
 With the full shadow model, a peak-to-peak difference in modeled and simulated photon yield between the investigated impact distances and muon inclination angles of only about 4\% is observed. }
\label{fig:totphotonsMST_corner}}
\end{figure}

\clearpage

\section{Systematic Effects due to the Different Optical Bandwidth of the Cherenkov Light Produced by Muons and  Gamma-ray Showers \label{sec:systematic_C}}

Given the different energy spectra of the Cherenkov light from local muons and (distant) \comn{gamma-ray} showers after traversing the atmosphere, unrecognized chromatic changes in the telescope optics and camera may lead to unrecognized 
changes in the conversion factor $C_{\mu-\gamma}$ (defined in  Eq.~\ref{eq:epsgamma}). This factor, defined as the ratio between $B_{\gamma}$ and $B_{\mu}$, depends on the detector efficiency $\xi_{det}(\epsilon)$ and on the atmospheric transparency to Cherenkov light from muons $t_\mu$ and gamma-ray showers $t_\gamma$, respectively.

\begin{figure}[htp]
\centering
\includegraphics[width=0.85\linewidth]{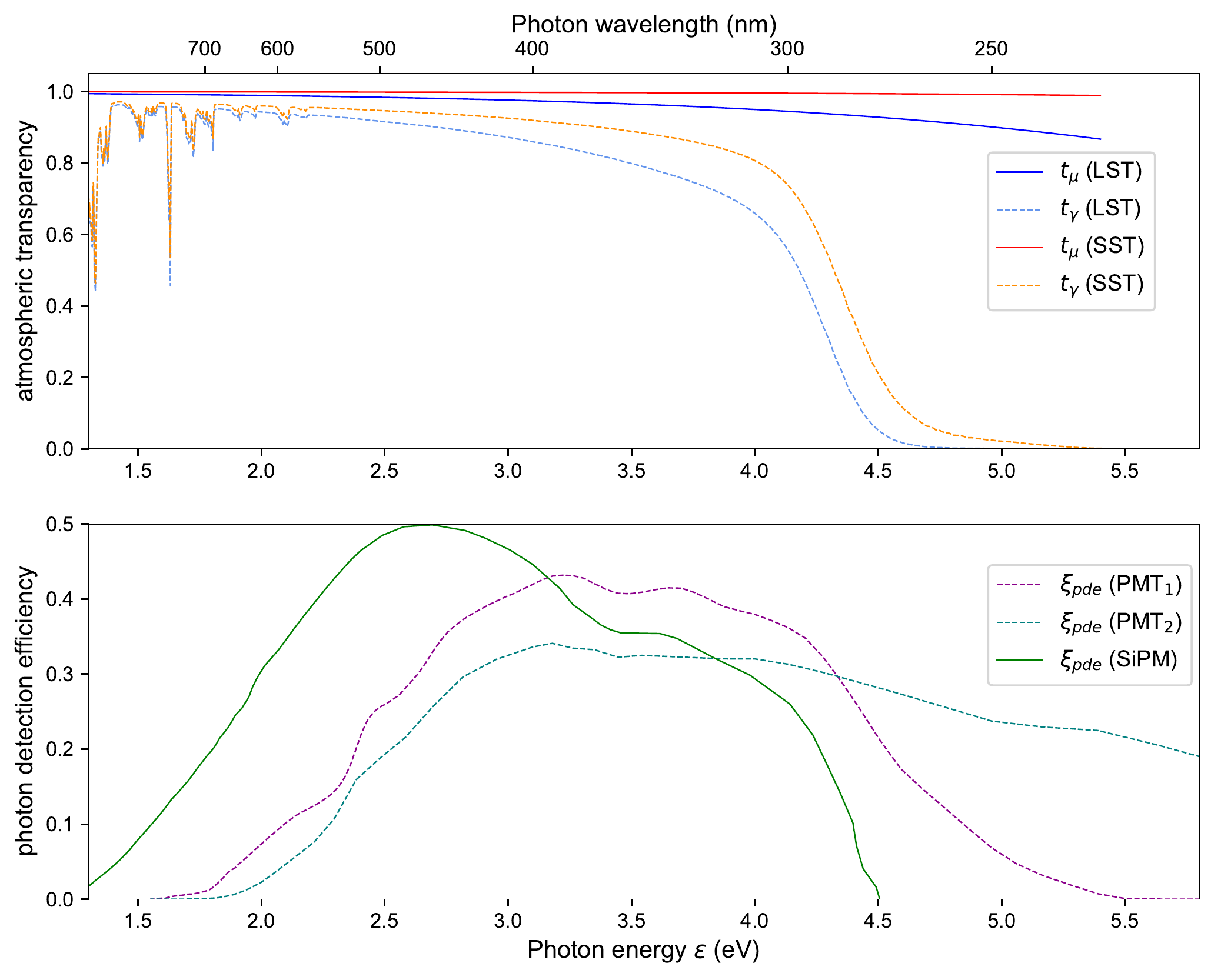}\\
\caption{Top: atmospheric transparency for Cherenkov light from muons $t_\mu$ and gamma-ray showers $t_\gamma$, for typical conditions for the LST and the SST. Bottom: the photon detection efficiency of the Hamamatsu R11920 (PMT$_1$), of the Electron Tube ETE~D569/3SA (PMT$_2$)~\protect\citep{TOYAMA2015280}, and of a silicon PM. 
\comn{Note, however, that the latter scales with operating voltage and recent products using silicone resin for the protection window (coated on the SiPM surface with 100~$\mu$m thickness) show even better UV transmittance than old epoxy resins.}
\label{fig:Cherenkov} }
\end{figure}

Figure~\ref{fig:Cherenkov} shows that
at 
wavelengths lower than $\sim$290~nm (photon energies greater than $\sim 4.3$~eV) the Cherenkov light from gamma-ray showers is almost completely absorbed by the atmosphere, at variance with that from muons. 
On the other hand, PMTs with super-bialkali photocathodes, protected by \comn{borosilicate} glass, can feature considerable quantum efficiency even 
below 250~nm. 
This is the range in which muon calibration was most strongly affected by systematic errors in the past~\citep{leroyphd,fegan2007}.

Because $B_\gamma$ and $B_\mu$ contain only an integrated detector spectral efficiency $\xi_\mathrm{det}(\epsilon)$, a recalibration of the telescope optical throughput would lead to wrong results for gamma-ray analyses, if 
the telescope loses efficiency 
at less than 300~nm wavelength degrading $B_\mu$. Since there is practically no Cherenkov light from gamma-rays in this energy band, 
corrections based only on $B_\mu$ will overestimate $B_\gamma$. 

\begin{figure}[h!]
\centering
%
%
\includegraphics[width=0.9\linewidth]{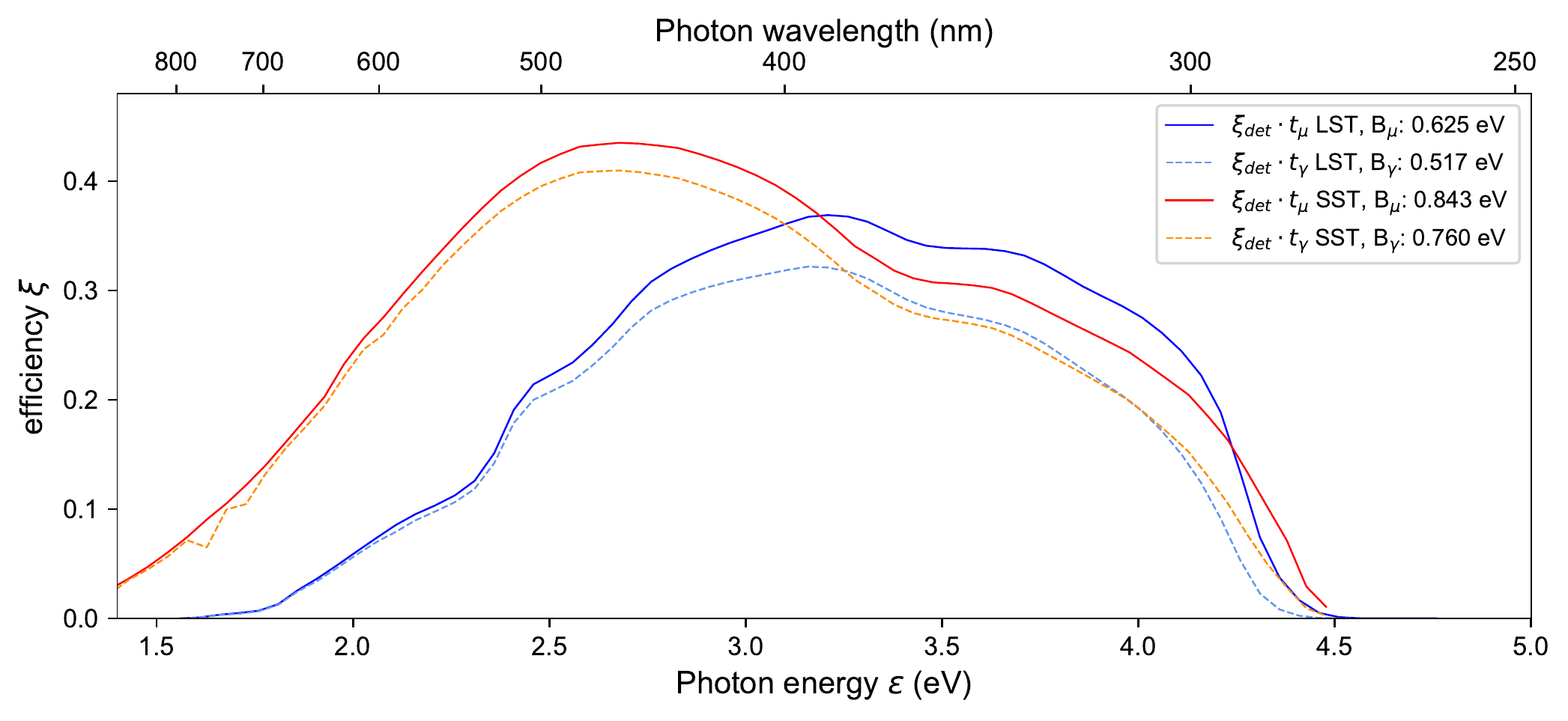}
\caption{Approximated combined detector and atmospheric efficiency for Cherenkov light from muons or gamma-rays, for a generic LST or an SST. 
The different shapes for the two telescopes are mainly due to the different light detectors used: PMTs for the LST and a SiPMs for the SST.   
Gamma-ray shower emission from 10~km a.s.l. on average was assumed for the LST, and 6~km for the SST. 
\label{fig:Bandwidth2}
}
\end{figure}

Figure~\ref{fig:Bandwidth2} compares the combined detector and atmospheric efficiency for Cherenkov photons from local muons and  gamma-ray showers, for two different telescope types. The loss of efficiency for gamma-ray shower light is clearly visible for the LST, which uses PMTs,
whereas the effect is much lower for the SST, where the camera sensors are SiPMs. Also visible is the carefully chosen cutoff around 
  290~nm 
 for the LST, produced by the camera protection window~\citep{LST-TDR}.

The CTA Consortium has  made a considerable effort to eliminate exposure of the PMT cameras to light with wavelengths below 290~nm, through the \comm{choice} of an adequate  protection window. That window,\footnote{Shinkolite \#000 from Mitsubishi Rayon,
see \url{https://www.m-chemical.co.jp/shinkolite/}.} \comn{made of polymethyl methacrylate} (PMMA), filters out such short wavelengths, while maintaining a high transmittance to photon wavelengths \comm{above}  300~nm (see Figure~\ref{fig.xis}). 
We treat in the following the residual systematic effect due to chromatic degradation of individual elements, after taking into account the spectral cut produced by the camera protection window.

Since optical elements tend to degrade in a chromatic manner, losing efficiency rather toward the blue and UV part of the spectrum,  
we consider here the extreme case of a complete sensitivity loss (``blindness'') below a given wavelength while maintaining full efficiency throughout the rest of the spectrum.
Such a complete unrecognized loss of sensitivity is very improbable, compared to partial losses. In that sense, this 
procedure provides an upper limit on the possible error that can be made by neglecting the effect of chromatic degradation of optical elements, neglecting additional  possible,
though unexpected and less harmful, chromatic  effects at  longer wavelengths. Note that an additional degradation at longer wavelengths normally reduces that error because the sensitivity loss will become less chromatic. Also, the difference between muon- and gamma-ray-generated Cherenkov light spectra becomes negligible  at long  wavelengths. 

In that hypothetical extreme scenario, the measured optical bandwidth degrades to
\begin{equation}
B_\mu(\epsilon_\mathrm{blind}) = \int_0^{\epsilon_\mathrm{blind}} \xi_\mathrm{det}(\epsilon') \cdot t_\mu(\epsilon') \ud\epsilon'\quad.
\end{equation}
\noindent
An analyzer would (wrongly) conclude that the telescope's optical throughput will need to be corrected by the factor
\begin{equation}
B_\mathrm{\gamma,cal}(\epsilon_\mathrm{blind}) = B_\mu(\epsilon_\mathrm{blind}) \cdot \ddfrac{\int_{\epsilon_\mathrm{min,orig}}^{\epsilon_\mathrm{max,orig}}  t_\gamma(\epsilon') \ud\epsilon'}{\int_{\epsilon_\mathrm{min,orig}}^{\epsilon_\mathrm{max,orig}}  t_\mu(\epsilon') \ud\epsilon'} \quad, 
\end{equation}
\noindent
where $\epsilon_{\mathrm{min,orig}}$ and $\epsilon_{\mathrm{max,orig}}$ are the originally assumed detector sensitivity limits (or atmospheric cutoff, if applicable). 
The error made by the analyzer is then
\begin{equation}
 \ddfrac{\Delta B_\gamma}{B_\gamma} = \ddfrac{B_\gamma(\epsilon_\mathrm{blind})-B_\mathrm{\gamma,cal}(\epsilon_\mathrm{blind})}{B_\gamma(\epsilon_\mathrm{blind})} \quad. 
\label{eq:blind}
\end{equation}

\begin{figure}[htp]
\centering
\includegraphics[width=0.9\linewidth]{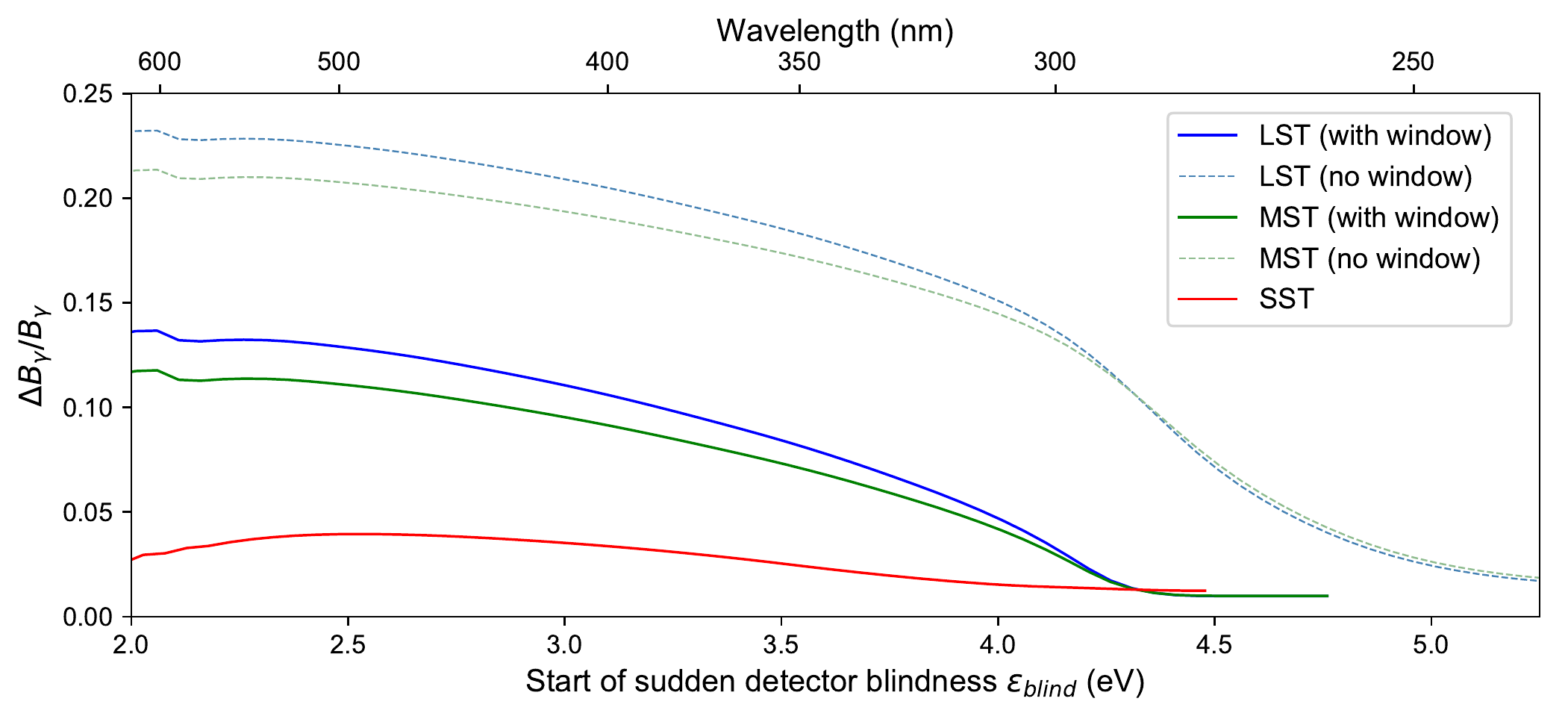}\\
\caption{Calibration error committed in case of a complete loss of sensitivity (``blindness'') above a certain threshold energy $\epsilon_\mathrm{blind}$. Gamma-ray shower heights of $h=10$~km were chosen for the LST, 8~km for the MST, and 6.5~km for the SST, matched to their respective energy ranges, and the telescope pointing to zenith. The wiggles at the left side are due to molecular absorption lines, not taken into account for the muon light. The absorption of gamma-ray shower light across the Chappuis and the strong Huggins bands of ozone were used for gamma-ray shower light but neglected for local muon light.
\label{fig:chromatic}
}
\end{figure}


The results of Eq.~\ref{eq:blind} are shown in Figure~\ref{fig:chromatic},  for cameras both with and without the protection window.


One first notes that even in the high-energy limit, where supposedly no degradation has occurred,
the error does not converge exactly to zero. This has to do with the 
chosen integration limits $\epsilon_{\mathrm{min,orig}}$ and $\epsilon_{\mathrm{max,orig}}$ for the reference atmosphere.  One can 
calibrate the offset with MC simulations.  

An unrecognized chromatic degradation of the detector elements, expressed by the rise of $\Delta B_\gamma/B_\gamma$ when moving to smaller energies (longer wavelengths), produces an intrinsic error of about 10\%  if the detector becomes blind below wavelengths of 400~nm for the PMT-based telescopes, and below 3\% for the SiPM-based ones. If the camera protection window is left out, errors double.

Several possible causes can contribute to the chromatic changes in optical elements of the telescopes. Their effects are summarized in the following:

\begin{itemize}
\item {\it Chromatic degradation of the light detectors:}  photomultiplier tubes (PMTs) may have considerable quantum efficiency below 290~nm, if UV-transparent glass (fused silica \comn{or borosilicate}) is used for the photocathode~\citep{KILDEA2007,Otte:2011}. This is the range where Cherenkov light from air showers is almost completely absorbed, but the one from muons still gets through to the camera.  Experience with the Whipple Hamamatsu R~1398 photomultipliers has shown that the photon detection efficiency degraded from 5\% to~20\% (depending on the individual PMT tube), in the wavelength range from 290 to 450~nm~\citep{danielpriv}. VERITAS observed a 10\% sensitivity \textit{increase} below 300~nm and a 10\%--30\% \textit{drop} above 550~nm with the Hamamatsu R10560-100-20 and claims a serious aging of spectral sensitivity of the previous Photonis XP2970~\citep{Gazda:2016}. If such a behavior occurs with the CTA PMTs, the muon calibration will {\bf overcorrect} the loss by 1\%--2\% for the PMT-equipped cameras. 
\item {\it Chromatic degradation of the \comn{PMMA}:} the Shinkolite \#000 \comn{PMMA}, currently favored for use in both LST and MST cameras, can permanently lose up to 45\% \comm{transmittance} in the range from 290 and 450~nm,  after strong exposure to UV light, whereas the rest of the light spectrum is unaffected (after some recovery time; \citet{foersterpriv}; \citet[see also Figs. 4.5--4.10 of][]{Motohashi:master}). In this case, the muon calibration will {\bf overcorrect} the loss by $\sim 3-4$\% for an affected MST or LST, both  for the worst-case scenario of an extreme exposure to sunlight.  Direct exposure of the material to the Sun during 10 days on the \comm{CTA-N} observatory has not, however, been able to reproduce such a chromatic \comn{transmittance} loss, but instead a constant drop by 2--3\%~\citep{LST-TDR}. 
\item {\it Chromatic degradation of the focused mirror \comn{reflectance}:} experience from the H.E.S.S. mirrors has shown that \comn{reflectance} losses affect the wavelength range around 300~nm 5--10\% more strongly than at wavelengths around 500~nm~\citep{foersterpriv}. Such a scenario would translate into an error in the muon calibration of the order of 1\% for MSTs and LSTs and $<$1\% for the SSTs \comn{(unless new SiPMs are developed with much higher PDE around 300~nm)}.
\end{itemize}

Chromatic effects can be \comm{characterized} from time to time by external devices, as foreseen for CTA~\citep{segreto2016}. If the telescope's optical throughput is characterized with respect to its relative response to two reference (laser) wavelengths, above and below 400~nm (e.g. the \comm{doubled} and \comm{tripled} emission line of an NdYAG laser at 532 and 355~nm, respectively), with a precision of the order of 10\% for the response ratio between the two wavelengths,   the unrecognized chromaticity effects will add maximally 1\% to the systematic uncertainty of the muon calibration method.

\clearpage
 \section{Secondary Effects Broadening the Ring \label{sec:broadening}}


Broadening of the ring influences the detectability of muons (by both the telescope trigger and the offline analysis) and the reconstructed optical PSF of the telescope. 

Normally, muon rings become detectable if their radii become considerably larger than the ring width, i.e. $\Delta\thetac/\thetac \ll 1$. This is the case for energetic muons with large Cherenkov angles.  
Most effects that cause a broadening of the ring width have been described by~\citet{vacanti}, but they are often overestimated. We review them here and estimate their magnitudes for the CTA telescopes. Additionally, we add further, hitherto overlooked, effects. 



\subsection{Instrumental Effects}

\subsubsection{Finite Camera Focuses}

Finite camera focuses were not treated by \citet{vacanti}. We calculate the variance of the ring displacement produced by a camera focused at $u_f$ (Eq.~\ref{eq:varfocus}) and obtain

\begin{equation}
\sqrt{\mathrm{Var}\left[\ddfrac{\Delta\thetac}{\thetac}\right]} \simeq 
\ddfrac{R}{\sqrt{3}\thetac} \cdot \ddfrac{1}{u_f}\cdot
   \sqrt{1 - \ddfrac{3}{4}\cdot E_0^2(\rho_R)} ~.
\end{equation}

For telescopes focused at 10~km, this effect broadens the ring of an LST by $<5$\%, that of an MST by $<3$\%, and $<1$\% for the case of an SST.

\subsubsection{Optical Aberrations}
\label{sec:mirroraberration}

The contribution of the spherical mirror aberration to the broadening of the ring has been described as constant by~\citet{vacanti}. 
Additionally, coma and astigmatism must be taken into account. 
According to \citet{Fegan:2018} (see also \citet{Schliesser:2005} for the pure parabolic case and \citet{Vassiliev2007} for the ideal Davies-Cotton (DC) telescope), aberrations for a single-mirror telescope can be approximated from third-order aberration theory as
\begin{align}
\Delta \xi  \approx 
      \sqrt{ 
            \ddfrac{\upphi^2}{3072} \left(\ddfrac{\mathcal{D}}{f}\right)^4 \cdot f_\xi(\kappa)
           + N_\mathrm{mirrors} \cdot \ddfrac{9\upphi^2}{16}\left(\ddfrac{d^2}{\mathcal{D}\cdot f}\right)^2 
           + \Delta_0^2 
           } \quad,  \span
     \label{eq:xi} \\[0.3cm]
\Delta \eta   \approx \sqrt{\ddfrac{\upphi^2}{3072}\left(\ddfrac{\mathcal{D}}{f}\right)^4 \cdot f_\eta(\kappa)
+ N_\mathrm{mirrors} \cdot \ddfrac{9\upphi^2}{16}\left(\ddfrac{d^2}{\mathcal{D}\cdot f}\right)^2 
+ \Delta_0^2} \quad, \label{eq:eta} \span \\[0.3cm]
       \qquad \mathrm{with} \span \nonumber\\
\begin{array}{ccc}
 f_\xi(\kappa) =
   &   \left\{ \begin{array}{c} 
           6   \\[0.55cm]   
           \vspace{-4mm}\\
          \scriptstyle{4\kappa^2-12\kappa+11} +   
          \ddfrac{\scriptstyle{32\kappa^4-48\kappa^3-96\kappa^2+184\kappa-13}}
          {\scriptstyle{64}}\left(\ddfrac{\mathcal{D}}{f}\right)^2
          \end{array}\right.
   &   \begin{array}{c} 
      \mathrm{for~:~LST}  \\[0.45cm]
      \simeq  \left\{ \begin{array}{c@{\,+~}c@{~\mathrm{~for:~}}cl} 
             3 & 0.92\left(\ddfrac{\mathcal{D}}{f}\right)^2 & \mathrm{DC} & (\kappa = 1) \\[0.25cm]
            3.8 & 0.96\left(\ddfrac{\mathcal{D}}{f}\right)^2 & \mathrm{MST} & (\kappa \approx 0.83) \\
            \end{array} 
            \right. 
        \end{array}  \\[0.4cm]
     \span  \qquad \mathrm{and} \span \nonumber\\[0.3cm]
f_\eta(\kappa) =
    & \left\{ \quad\qquad\quad~\begin{array}{c} 
          3  \\[0.55cm]  
          \vspace{-3mm}\\
          2 + \ddfrac{24\kappa+11}{64}\left(\ddfrac{\mathcal{D}}{f}\right)^2 
        \end{array}\right.\quad\qquad\quad~
   &   \begin{array}{c} 
       \mathrm{for~:~LST}  \\[0.45cm]
        \simeq \left\{ \begin{array}{c@{\,+~}c@{~\mathrm{~for:~}}cl} 
             2 & 0.55\left(\ddfrac{\mathcal{D}}{f}\right)^2 & \mathrm{DC} & (\kappa = 1) \\[0.25cm]
             2 & 0.48\left(\ddfrac{\mathcal{D}}{f}\right)^2 & \mathrm{MST} & (\kappa \approx 0.83) \\
             \end{array} 
             \right.   
     \end{array}  
\end{array}
\end{align}
\noindent 

where $\Delta \xi$ denotes the tangential and $\Delta \eta$ the sagittal \comn{standard deviation of the PSF}. The angle $\upphi$ denotes inclination with respect to on-axis. We have introduced here also a common spread of $\Delta_0$ subsuming the intrinsic optical quality of the individual mirror facets (mainly due to form deviations from ideal mirror shapes and to surface roughness;~\citet{Tayabaly:2015}), the aberrations caused by
the fact that most facets are inclined relative to the
optical axis and the precision with which the individual mirror facets are aligned~\citep{Cornils:2003}. 
%
Spherical aberrations of each (spherical) mirror facet scale with $r^3/4R^2$, where $r$ is the radius of the individual mirrors and $R$ its radius of curvature~\citep{Lewis:1990,Cornils:2003,Vassiliev2007}; $N_\mathrm{mirrors}$ denotes the number of mirror facets (see Table~\ref{tab:tellist}). 

Optical aberration will become the dominant effect for the ring broadening of images from the highest-energy muons for all CTA telescopes, although requirements on their optical PSF are rather strict (see Table~\ref{tab.theta80}). 

\begin{table}
\centering
\begin{tabular}{ccc}
\toprule
Telescope  & Requirement  & Application Range \\
   Type    & on $\theta_\mathrm{80}$  & (from Camera Center) \\
\midrule
LST            &  $<0.11^\circ$   & $<1.2^\circ$  \\
MST            &  $<0.18^\circ$   & $<2.8^\circ$  \\
SST            &  $<0.25^\circ$   & $<3.2^\circ$  \\
\bottomrule
\end{tabular}
\caption{Requirements for the spot size encircling 80\% of the light reflected on to the camera ($\theta_\mathrm{80}$). Note that the rms point spread $\Delta$ (Equations~\ref{eq:xi} and~\ref{eq:eta}) encircles only 39.3\% of the light in case of a 2-dimensional normal distribution of light. In that case $\theta_\mathrm{80} \approx 1.8 \cdot \Delta$. 
\label{tab.theta80}}
\end{table}

\begin{figure}
\centering
\includegraphics[width=0.8\linewidth]{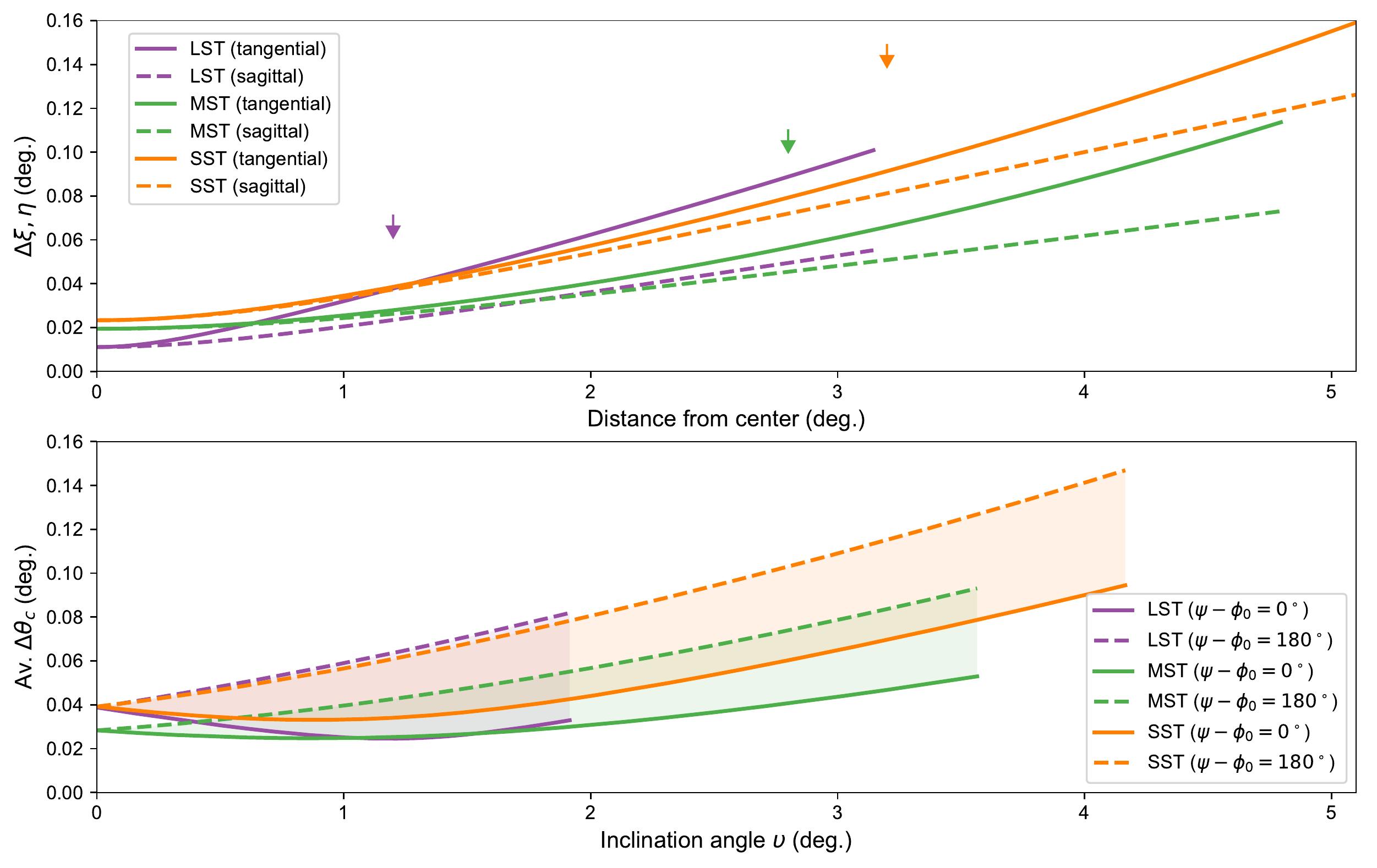}
\caption{Top: expected effect of optical aberrations on the broadening of the muon ring for the CTA telescopes. Values of $\Delta_0$ adopted from~\citet{LST-TDR,MST-TDR,SST-1M-TDR}.
  The small arrows show the requirements introduced in Table~\protect\ref{tab.theta80}, translated to the corresponding values for the standard deviation point spread (assuming a two-dimensional normal distribution of light).
  Bottom: the average ring width for a muon of maximal energy $\thetac=1.23^\circ$ and impact distance $\rho_R = 0.9$, as a function of the muon inclination angle. The lines correspond to the side of the ring with the largest amount of light at the opposite side of the camera center ($\psi-\phi_0 = 0^\circ$, dashed lines)
  and at the side closest to the camera center ($\psi-\phi_0 = 180^\circ$, solid lines). The shaded areas in between show the general case of any azimuthal angle between impact parameter and inclination angle. 
\label{fig:aberration}
}
\end{figure}


Moreover, the PSF of a typical IACT varies with pointing zenith angle $\theta$, because a \comn{support} structure that does not deform under the influence of gravity would be too costly. \citet{Cornils:2003} found that the size of the PSF of the H.E.S.S.-I telescopes can be parameterized as
\begin{equation}
\Delta(\xi,\eta) \approx \sqrt{\Delta(\xi,\eta)_0^2 + d^2 \cdot (\sin\theta - \sin\theta_0)^2} \quad, 
\end{equation}\noindent
where $\Delta(\xi,\eta)_0$ is the PSF value with the telescope pointing to zenith and $d \approx 1$~mrad (in the case of H.E.S.S.-I) is a parameter describing the influence of the deformation of the mirror \comn{support} under inclination. The LST 
will, however, make use of an active mirror control system~\citep{GarczPhD,Biland:AMC,magicperformance1} that effectively removes the zenith angle dependency $d$. 

Additionally, wind may degrade the quality of the PSF, particularly the larger telescopes. Careful simulation studies have shown, however, that the magnitude of related effects is of the same order as or smaller than $\Delta_0$~\citep{LST-TDR,MST-TDR}.

Figure~\ref{fig:aberration} (top) shows an example for the expectation of tangential and sagittal rms point spread for each telescope type, together with the requirements. A muon ring of radius $\thetac$ is affected by a wide range of these parameters. The average radial distribution of the ring depends then on the inclination angle, the impact distance, and particularly the azimuthal component of the inclination parameter with respect to the azimuthal component of the impact parameter. 

We will quantify this effect now by weighting each ring segment with the chord $D$ and the projected component of the tangential $(\sin\delta)$ and sagittal $(\cos\delta)$ aberration onto the ring, evaluated at the distance $d$ of each point on the ring to the camera center (see Appendix~\ref{sec:tangsatweights}): 

\begin{align}
<\!\Delta\thetac(\rho_R,\psi-\phi_0)\!> &\approx \ddfrac{\int_0^\pi\sqrt{\Delta\xi^2(d(\phi))\sin^2\delta(\phi)+\Delta\eta^2(d(\phi))\cos^2\delta(\phi)}\cdot D(\rho_R,\phi,\psi-\phi_0)\ud \phi}{\int_0^\pi\sqrt{\sin^2\delta(\phi)+\cos^2\delta(\phi)}\cdot D(\rho_R,\phi,\psi-\phi_0)\ud\phi}
\end{align}

Figure~\ref{fig:aberration} (bottom) shows the resulting average ring width, for an impact distance of $\rho_R = 0.9$. As expected, the ring width depends not only on the muon inclination angle but also on the relative azimuthal components of the impact parameter and inclination angle. The effect of the latter can even create a peak-to-peak spread larger than the average ring width at a given inclination angle, particularly for the LST. 

These results show that in order to improve precision, both inclination and impact parameter must be taken into account in a ring width analysis, leading to an estimate of the PSF of a CTA telescope.

\subsubsection{Discrete Pixel Width}
\label{sec:discretepixelwidth}

The discrete pixel is mentioned here for completeness but introduces a negligible uncertainty on the reconstruction of the muon ring in our case~\citep{vacanti}:
\begin{eqnarray}
\sigma_\mathrm{pix} &\simeq& \frac{\omega/2}{\sqrt{N_\mathrm{hit}}} \nonumber\\
\frac{\Delta\theta_c}{\theta_c} &=&  \frac{\sigma_\mathrm{pix}}{\theta_c} \simeq \left(  \frac{\omega}{2\pi^{1/3}\theta_c} \right)^{3/2} ~,
\end{eqnarray}
\noindent 
where the number of hit pixels, Eq.~\ref{eq:npix}, has been used. 
Given the fine pixelation of the CTA cameras, the relative ring broadening introduced by this effect is always below 5\%. 

\subsection{Atmospheric Effects}
\subsubsection{Multiple Scattering}

Multiple Coulomb scattering of particles in the air along their trajectory causes the emitted Cherenkov photons to appear
scattered around their mean positions in the focal plane, according to an approximately Gaussian distribution. The standard deviation
 of that distribution can be described as~(\citet{lynch1991}; \citet[sect.~34.3~of][]{pdg2017})

%
%

\begin{align}
\sigma_\mathrm{mul.scat.} &\simeq\ddfrac{13.6~\mathrm{MeV}}{\beta pc} \cdot \sqrt{\frac{x\cdot \rho_\mathrm{air}}{X_0}} \cdot \big(1 + 0.038\ln(\frac{x\cdot \rho_\mathrm{air}}{X_0}) \big)~
\end{align}
\noindent
where $p$ is the momentum of the muon and $X_0$ its radiation length ($\approx$367~kg/m$^2$).
The density of air is approximated by $\rho_0 \cdot \exp(-h/H_0)$, where $\rho_0 \approx$1.225~kg~m$^{-3}$ and $R_0 = X_0/(\rho_0 \cdot \exp(-H_\mathrm{obs}/H_0)) \approx 378$~m for an observatory altitude $H_\mathrm{obs}$ of 2200~m a.s.l. and $H_0 \approx 9.7$~km.
Since the ring probes different path lengths in the atmosphere, according to its azimuth  value, we calculate an expectation value for the \textit{average ring width} (see Appendix~\ref{sec:varmultscat}): 
\begin{align}
E\left[\ddfrac{\sigma_\mathrm{mul.scat.}}{\thetac}\right]  & \simeq  0.129  \cdot \sqrt{\ddfrac{R}{R_0}} \cdot \sqrt{\ddfrac{\theta^2_\infty-\thetac^2}{\thetac^3}} \cdot \left( E_0(\sqrt{\rho_R}) - \sqrt{\ddfrac{R \cdot \cos\vartheta }{\thetac \cdot H_0} } \cdot E_0(\rho_R) \right)~.\label{eq:multscat}
\end{align}
\noindent

\begin{figure}
\centering
\includegraphics[width=0.8\linewidth]{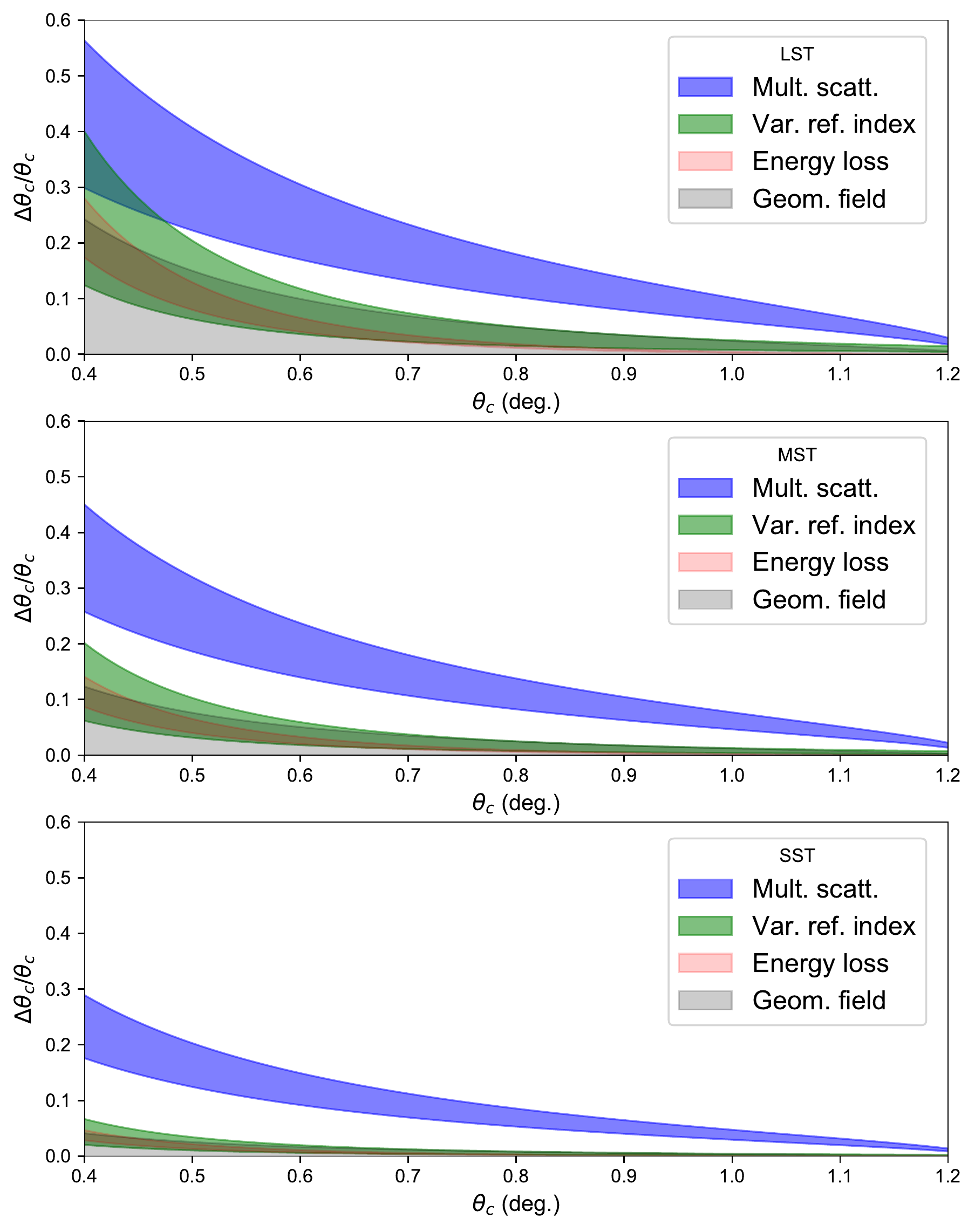}
\caption{Expected atmospheric effects on the broadening of the muon ring for the CTA telescopes. The band shows the possible variations due to the different impact distances and telescope pointing. Only full rings have been considered. 
\label{fig:multscat}}
\end{figure}

The new Equation~\ref{eq:multscat} yields only about half the contribution predicted by~\citet{vacanti}, who assumed that the full muon path is seen by the telescope  through the entire ring, and a constant atmospheric density. 
Figure~\ref{fig:multscat} shows an example for its expectation for each telescope type. One can see that the peak-to-peak variation, due to different impact distances, amounts to a bit less than half the mean ring width produced by this effect. 

\subsubsection{Variation of the Refractive Index\label{sec:varn}}

The refractive index of air varies with altitude, temperature, atmospheric pressure, relative humidity, and wavelength.
Dependencies  of the refractive index of air on these parameters have been discussed in great detail in~\citet{ciddor1996,ciddor2002} and later in~\citet{tomasi}.
Atmospheric conditions, especially temperature and water vapor content, vary with time and altitude, but also, to a lesser extent, with the concentration of CO$_2$. 
In general, a good description is obtained by assuming a variation of $\varepsilon = (n-1)$ with the density of air, which in turn can be approximated by an exponential decrease with a scale height of $H_0=9.7$~km\,\footnote{Note here that~\protect\citep{vacanti,leroyphd} assumed erroneously a scale height of 8.4~km which is valid only for the atmospheric pressure.} at La Palma~\citep{Gaug:2017site}. 
Assuming an exponential scaling law $\epsilon$, and $\Delta \theta_c \simeq \Delta x \cdot \partial \thetac/\partial h$, we obtain for the standard deviation of the angular deviation (see Appendix~\ref{sec:varrefindex}):
\begin{align}
\sqrt{\mathrm{Var}\left[\ddfrac{\Delta\thetac}{\thetac}\right]} 
   &\simeq \ddfrac{\cos\vartheta \cdot R}{2H_0}\cdot \ddfrac{\theta^2_\infty}{\thetac^3}\cdot  \sqrt{  1/3 + \ddfrac{R\cdot\cos\vartheta}{12\cdot \thetac \cdot H_0}\cdot  E_0(\rho_R) - \ddfrac{1}{4} \cdot E_0^2(\rho_R) } ~.
\label{eq:refind}
\end{align}
Eq.~\ref{eq:refind} differs from the one proposed by~\citet{vacanti} by the square root factor, which ranges from $\sim$0.3 for $\rho_R=0$ to $\sim$0.5 for $\rho_R=1$. 

\begin{figure}
\centering
\includegraphics[width=0.8\linewidth]{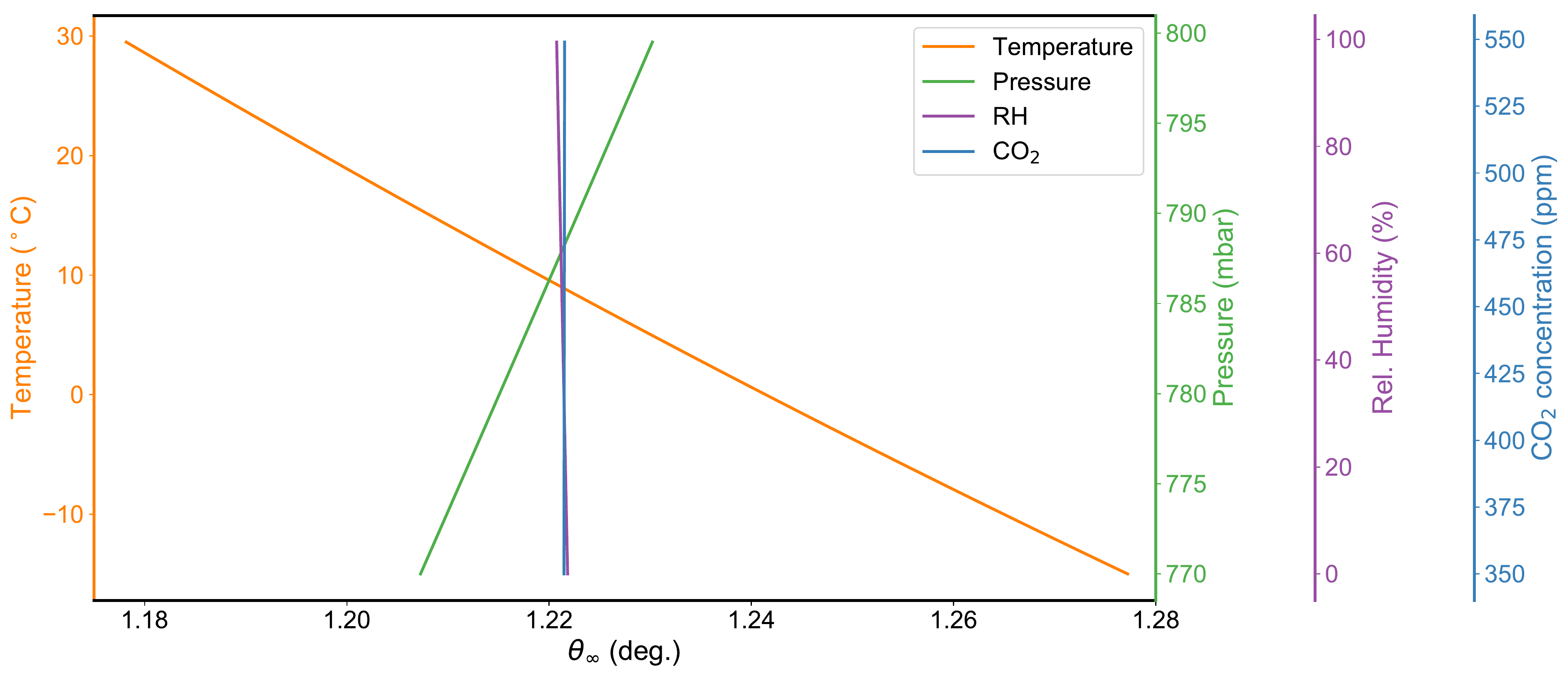}
\caption{\label{fig:rhovsatmos}Expected drifts of $\theta_\infty$ with atmospheric parameters. }
\end{figure}

Additionally, since $\theta_\infty^2 \approx 2\varepsilon$, dependencies of the refractive index and hence the Cherenkov angle on the atmospheric conditions can be expected. 
Drifts of $\theta_\infty$ calculated using formulae in~\citet{tomasi} are shown in Figure~\ref{fig:rhovsatmos}.
The temperature dependency is the largest effect, affecting $\theta_\infty$ by up to 7\%, whilst only 
negligible dependencies of the refractive index on humidity and CO$_2$ concentration are observed.

Since the Cherenkov angle  is reconstructed directly from ring radius and the accumulated signal in the camera, $Q_\mathrm{tot}$, scales linearly with the Cherenkov angle, the residual bias on the reconstructed optical bandwidth is expected to be negligible. 
Nevertheless, \citet{bolzphd} found measurable effects with H.E.S.S. data
of the same size as those suggested in Figure~\ref{fig:rhovsatmos}, implying that temperature- and pressure-dependent effects should be monitored,  at least during the commissioning phase.

\subsubsection{Energy Loss of the Muon}

Muons lose energy through ionization, which reduces the Cherenkov angles with path length. 
This effect has been neglected in Eqs.~\ref{eq:dNdphi} and~\ref{eq:Qtot}, but introduces a small bias into the results.

It is convenient to
write the average rate of muon energy loss as~\citep{barrett1952}
\begin{equation}
-\ud E/\ud x = a(E) + b(E)\cdot E \qquad.
\end{equation}
 Here $a(E)$ is the ionization energy loss and $b(E)$ the sum of $e\!^+e\!^-$-pair production, bremsstrahlung, and photonuclear interactions. The second term  $b(E)\cdot E$ dominates energy loss only above 1.1~TeV in air~\citep{groom}, where the muon flux is already very small, compared to energies in the range from 10 to 100~GeV. The ionization energy is given by the Bethe-Bloch formula~\citep[see, e.g., sect.~34.6 of][]{pdg2017} and amounts to about $-\ud E/\ud x = 2.15$~MeV~g$^{-1}$cm$^{2}$ for a muon producing a Cherenkov angle of $\thetac = 0.4^\circ$ and  $-\ud E/\ud x = 2.43$~MeV~g$^{-1}$cm$^{2}$ 
for $\thetac = 1.2^\circ$ . 

Assuming $\Delta\theta_c \simeq \Delta E_\mu \cdot \partial\thetac/\partial E_\mu$, one obtains the variation of $E_\mu$ while the particle traverses the
path length $L$  losing $\ud E/\ud x$. Calculating expectation values for the changes on the ring width (see Appendix~\ref{sec:vareloss}), one obtains for the average ring width
\begin{align}
\sqrt{\mathrm{Var}\left[\ddfrac{\Delta\thetac}{\thetac}\right]} 
&\simeq 1.68\times 10^{-3} \cdot \left(\ddfrac{R}{\mathrm{(m)}}\right) \cdot \left( \ddfrac{\theta_\infty^2}{\thetac^2} -1 \right)^{3/2} \cdot  \sqrt{  1/3 + \ddfrac{R\cdot\cos\vartheta}{12\cdot \thetac \cdot H_0}\cdot  E_0(\rho_R) - \ddfrac{E_0^2(\rho_R)}{4} } ~,
\label{eq:eloss}
\end{align}
where the muon rest mass $m_\mu \simeq 105.7$~MeV has been used and the relationship $\thetac^2 \simeq \theta^2_\infty - (m_\mu/E_\mu)^2$. As before, the average ring width results are smaller than that  estimated by~\citet{vacanti} by a factor of 0.3 to~0.5.

Figure~\ref{fig:multscat} shows examples for its expectation for each telescope type. One can see that this contribution to the ring broadening is smallest at high muon energies.

\newpage
\subsubsection{Bending of the Muon Trajectory in the Geomagnetic Field}

We calculate the variance of the angular shift (Eq.~\ref{eq:deltatheta_magn}), introduced by the bending of the muon path in the Earth's magnetic field \comn{(see Appendix~\ref{sec:varmagnetic})}:  
\begin{align}
\sqrt{\mathrm{Var}\left[\ddfrac{\Delta\thetac}{\thetac}\right]} 
&\simeq 4.7 \times 10^{-5}     
\cdot \left(\ddfrac{R}{\mathrm{m}}\right) 
\cdot \sqrt{\ddfrac{\theta^2_\infty-\thetac^2}{1-(\theta^2_\infty-\thetac^2)}} \cdot \frac{1}{\thetac^2}
\cdot\left(\ddfrac{B_\perp(\vartheta,\varphi)}{40~\mu\mathrm{T}}\right) \nonumber\\
     & \qquad \cdot  \sqrt{ 1 + \ddfrac{\rho_R^2}{\comn{8}} \cdot \comn{\left(1-5\sin^2(\phi_0-\varphi)\right)}}~ .
\label{eq:muondeltac}
\end{align}

For a muon of $p = 10$~GeV in a magnetic field perpendicular to its velocity, 
the average fractional ring broadening amounts to up to 0.3  for an LST, for the worst case of a telescope pointing perpendicular to the Earth's magnetic field and detecting a small ring (see Figure~\ref{fig:multscat}).  


\subsubsection{Combined Effects}

Natural ring broadening is dominated by multiple scattering of the muon, with broader rings at small impact distances and large observation zenith angles. On the other hand, variation of the refractive index, energy loss, and the magnetic field create broader rings for larger impact distances. The net effect yields slightly broader rings for smaller impact distances and large observation zenith angles. 

\begin{figure}\centering
\includegraphics[width=0.8\linewidth]{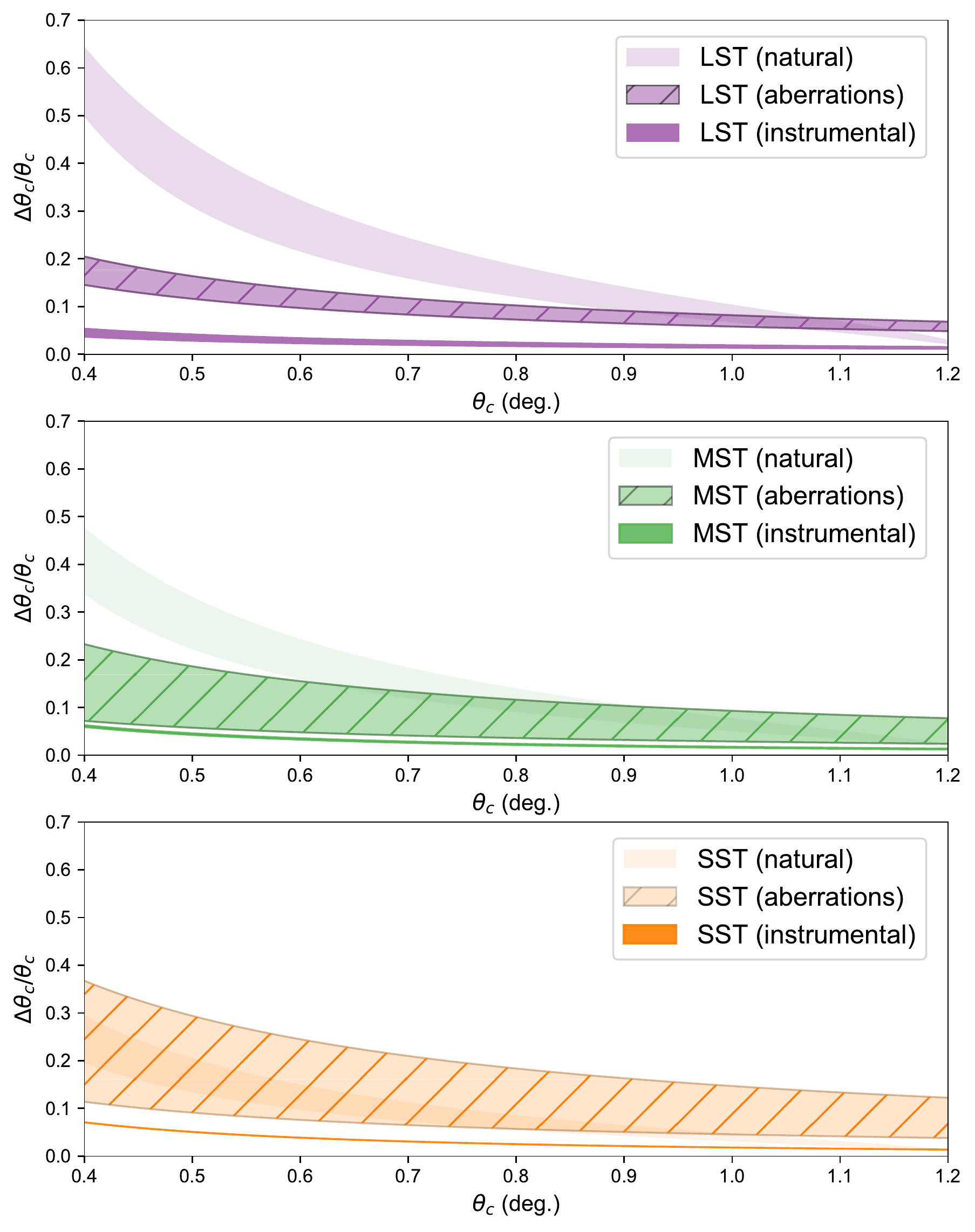}
\caption{Expected combined ring-broadening effects for the different telescope types. The band shows possible variations due to different impact distances and telescope pointing. Instrumental effects include the finite pixel size and a finite camera focus of 10~km. Only full rings have been considered. 
\label{fig:broadening}}
\end{figure}

Figure~\ref{fig:broadening} shows the combined effects of the ring broadening for all telescope types, together with the range of values expected from aberrations. As expected, natural broadening dominates small rings, whereas telescope aberrations dominate the largest rings. There is, however, a broad transition region where either one or the other may dominate, according to the muon impact distance and inclination angle. 

\clearpage
\section{Muon Spectra and Expected Event Rates\label{sec:rates}}

The muon spectrum at ground level in the range 10~GeV --
1~TeV  was published by~\citet{Hebbeker2002107} and modeled with a parameterization 
based on the theoretical formula proposed by~\citet{Bugaev1998}:

\begin{align}
\ddfrac{\ud \Phi_\mu}{\ud p_\mu}  &= C \times 10^{H(y)} \qquad \mathrm{m^{-2}\,s^{-1}\,sr^{-1}\,GeV^{-1}} \label{eq:muonspectrum}\\[0.1cm]
       \textrm{with} \span \nonumber\\[0.2cm]
H(y) &= H_1 \cdot (y^3 /2 - 5 y^2 /2 + 3 y) \nonumber \\
      & +H_2 \cdot (-2 y^3 /3 + 3y^2 - 10 y/3 + 1) \nonumber \\
      & +H_3 \cdot (y^3 /6 - y^2/2  + y/3 ) \nonumber \\
      & +S_2  \cdot  (y^3 /3 - 2 y^2 + 11 y/3 - 2)  \nonumber\\[0.2cm]
           \textrm{and} \span \nonumber\\[0.2cm]
      y    &= \log_{10}(p_\mu/\textrm{GeV})  \quad,\nonumber
\end{align}
\noindent
where \comn{$\Phi$ is the muon flux, $p_\mu$ is the muon momentum, and} $H_1=0.133\pm 0.002$, $H_2=-2.521\pm 0.002$, $H_3=-5.78\pm 0.004$, $S_2=-2.11\pm 0.03$, and $C=0.86\pm 0.06$.
The parameters $H_1$, $H_2$ and $H_3$ denote the logarithm of the differential flux at 10, 100, and 1000 GeV, $S_2$ represents the exponent of the differential flux at 100 GeV, while $C$ is the absolute normalization of the muon spectrum.

More recent data are given by~\citet{Schmelling20131} for the range 100~GeV -- 2.5~TeV and 
by the IceCube Collaboration ~\citep{Aartsen:AP2016a} at higher energies ($E_\mu>15$~TeV ).  The latter is 
reproduced by the following formula: 


\begin{equation}
\ddfrac{\ud \Phi_\mu}{\ud E_\mu} 
= 1.06^{+0.42}_{-0.32} \times 10^{-3}  \left( \ddfrac{E_\mu}{10~\mathrm{TeV}} \right)^{-3.78 \pm 0.02} 
 \quad \mathrm{m^{-2}\,s^{-1}\,sr^{-1}\,GeV^{-1}} ~.
\end{equation}

The muon flux dependence with altitude has been measured by balloon experiments \citep[e.g.][]{Pascale1993,Belotti:muons,Haino2004}. 
The flux for muon momenta above 10~GeV  can be parameterized by the following formula at altitudes less than 1000~m \citep{Hebbeker2002107}:
\begin{align}
\ddfrac{\ud \Phi_\mu}{\ud p_\mu}\left(h\right)  &= \ddfrac{\ud \Phi_\mu}{\ud p_\mu}(0) \cdot \exp(h / L) \quad  \label{eq:muonheight}\\[0.2cm]
     & \textrm{with} \nonumber\\[0.2cm]
  L &\approx 4900 + 750 \cdot \left(p_\mu / \mathrm{GeV}\right) ~\mathrm{m}\nonumber \quad,
\end{align}
\noindent
where $h$ is the altitude in meters.

The dependency of the muon flux on zenith angles computed using air shower simulations can be described as
\begin{align}
\ddfrac{\ud \Phi_\mu}{\ud \cos\vartheta}  &\sim 1+a(p_\mu)(1-\cos\vartheta)\label{eq:muontheta1} 
\end{align}
where the parameter $a(p_\mu)$ is $-1.5$ at $p_\mu=10$~GeV and $-1.28$ at $p_\mu=30$~GeV
\citep{Hebbeker2002107}.  The behavior for energies $E_\mu \cdot\cos\vartheta > 100$~GeV and zenith angles $<70^\circ$ is better described by the following formula~\citep[Sect. 29 of][]{pdg2017}:
\begin{align}
\ddfrac{\ud \Phi_\mu}{\ud E_\mu}\left(\cos\vartheta\right) &= 1.4 \times 10^{3} \cdot E_\mu^{-2.7}  
  \cdot \left(
   \ddfrac{1}    {1+\ddfrac{1.1 \cdot E_\mu\cdot\cos\vartheta}{115\,\mathrm{GeV}}}
 + \ddfrac{0.054}{1+\ddfrac{1.1\cdot E_\mu\cdot\cos\vartheta}{850\,\mathrm{GeV}}}
\right)  \quad \mathrm{m^{-2}\,s^{-1}\,sr^{-1}\,GeV^{-1}\label{eq:muontheta}} 
\end{align}

\begin{figure}[h!]
\centering
\includegraphics[width=0.6\linewidth]{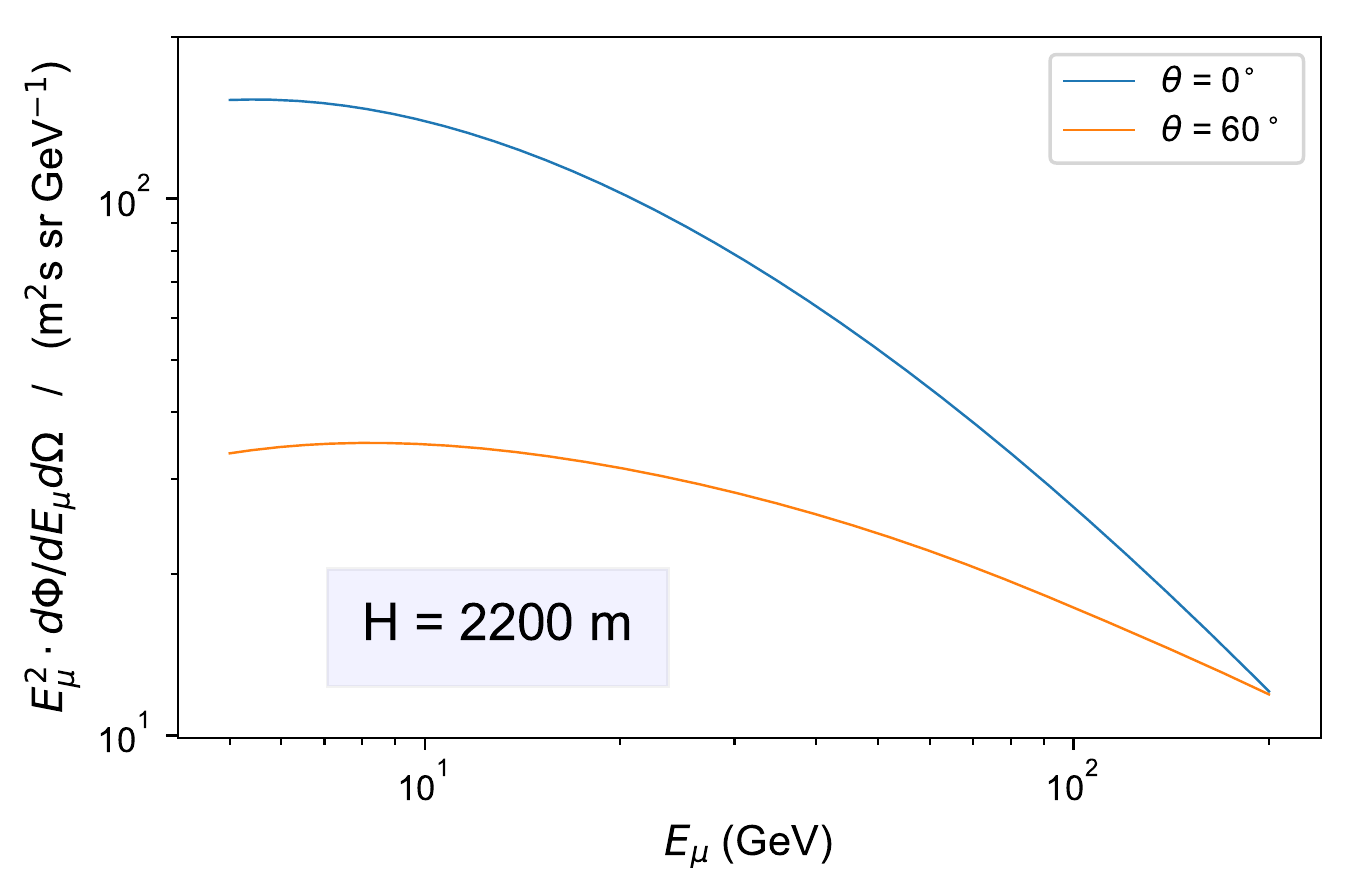} 
\caption{Assumed  differential energy spectrum of atmospheric muons,  multiplied by $E_\mu^{2}$ for better visibility, for an observatory altitude of 2200~m, as derived from Eqs.~\ref{eq:muonspectrum},~\ref{eq:muonheight}, and \ref{eq:muontheta}.   
\label{fig:muspectra}}
\end{figure}

We estimate, in the following, per-telescope \textit{mono trigger} muon event rates, required by CTA to get triggered in an unbiased way and flagged, unless another neighboring telescope triggers at the same time and activates a \textit{stereo trigger}. Such a requirement ensures that the \textit{stereo trigger bias} is eliminated and muon rates are kept at a useful level for calibration. Only the LSTs will be allowed to detect muon images only under a stereo trigger. 

To compute an estimate for the maximally achievable muon rates, we assumed a reference observatory altitude of 2200~m a.s.l., a minimum muon energy of $E_\mathrm{min}=8$~GeV (i.e. $\theta_c \gtrsim 1.0^\circ$), an impact distance value contained within $0.5 < \rho_R < 1$, and two reference observation zenith angles: $\vartheta = 0^\circ$ and $\vartheta=60^\circ$. 
The energy spectra relative to the two considered angles are plotted in Fig.~\ref{fig:muspectra} as a function of the muon energy. Using Eqs.~\ref{eq:muonspectrum} and~\ref{eq:muontheta1}  for energies lower than 100~GeV and Eq.~\ref{eq:muontheta} above this energy, we calculate a maximally achievable rate $R_\mathrm{max}$ of useful muon images:

\begin{align}
 R_\mathrm{max}\left(\cos\vartheta\right)  &= \int_{E_\mathrm{min}}^\infty \ddfrac{\ud \Phi_\mu}{\ud E_\mu}\left(\cos\vartheta\right) \cdot \left(\rho_\mathrm{max}^{\,2}-\rho_\mathrm{min}^{\,2}\right)  \cdot 2 \pi^2 \cdot \left(1 - \cos\upsilon_\mathrm{max}\right) ~\ud E_\mu \quad, \label{eq:muonintegral}
\end{align}
\noindent
where $\rho_\mathrm{max}$ is the maximum impact parameter and $\upsilon_\mathrm{max}$ the maximum muon angle to achieve a fully contained ring in the camera. The latter has been defined as \comn{$\textit{FOV}/2 - \theta_\infty-0.2^\circ$} to account for the broadening of the ring.

Under these assumptions, the estimated maximally available rate of useful muons ranges from $\sim$0.4~Hz (for the SSTs) to $\lesssim$10~Hz (for the LSTs) (see Table~\ref{tab:ratelistfinal}). 
These values are sufficiently high  to yield acceptable statistics for nightly calibration.

Requiring a statistical precision of  $\leq$1\% for the derived optical bandwidth with muons $B_\mu$, and using 5\% for the events-wise resolution  of $B_\mu$ for LST and MST (see, e.g., Table~\ref{tab:ipbwresults}) and 20\% for the SSTs, 
we estimate a required minimum number of $\sim$25 and $\sim$400 usable muon images for the larger telescopes and for the 
SST, respectively. With these estimates, we derive minimum time scale $T_\mathrm{monitor}$ within which optical bandwidth estimates can be updated for monitoring purposes. These time scales lie always below the typical run time of an observation, albeit one order of magnitude larger for the SSTs, as expected. 
All results are shown in Table~\ref{tab:ratelistfinal}.

We calculate now the rates of muon images that are required to contain a given pixel in the camera, such that we can use it for an alternative camera flat-fielding approach~\citep{bolzphd}. To do so, we use Eqs.~\ref{eq:npix} and~\ref{eq:Qtot}, and obtain
\begin{align}
R^\mathrm{pix}_\mathrm{max}\left(\cos\vartheta\right)  &= \ddfrac{1}{\omega \cdot N_\mathrm{pixels}} \cdot \int_{E_\mathrm{min}}^\infty \ddfrac{\ud \Phi_\mu}{\ud E_\mu}\left(\cos\vartheta\right) \cdot \sqrt{\theta_\infty^2 -\left(m_\mu/E_\mu\right)^2} \cdot \left(\rho_\mathrm{max}^{\,2}-\rho_\mathrm{min}^{\,2}\right)  \cdot 4 \pi^3 \cdot \left(1 - \cos\upsilon_\mathrm{max}\right) ~\ud E_\mu \quad, \label{eq:muonintegralpix}
\end{align}
\noindent
 where $\omega$ and $N_\mathrm{pixels}$ can be read off from Table~\ref{tab:tellist}. In order to estimate the monitoring time scales, we realize that the resolution is now dominated by the intrinsic Poissonian fluctuations of $Q_\mathrm{pix}$, which will be of the order of $\pm \sqrt{60}~\mathrm{p.e.}$ for the LST and MST and $\pm\sqrt{10}~\mathrm{p.e.}$ for the SST, assuming a ring width smaller than one pixel.  This entails again $E_\mathrm{min} \approx 8$~GeV and an event-wise resolution of $\sim$0.13 (LST and MST) and $\sim$0.32 for the SST. Here we ask for 3\% resolution of $Q_\mathrm{pix}$ (a typical value for the accuracy of the pixel-wise gain adjustments~\citep{aharonian2004,magicperformance1}) and end up again with at least 25 events for the larger telescopes and 150 events for the SST.
As shown in Table~\ref{tab:ratelistfinal}, considerably longer monitoring time scales are necessary for a pixel-wise calibration update, albeit well within one typical observation night. 

In order to assess rates of muon images that can be used to determine 
the average time delay of a given pixel in the camera, with respect to
the rest of the camera, we use Eq.~\ref{eq:muonintegralpix}, albeit with a
slightly relaxed minimum ring radius of $\thetac > 0.7^\circ$, hence
 $E_\mathrm{min}\approx 6$~GeV, and no lower impact distance ($\rho_\mathrm{min}=0$). To determine the number of events needed to be accumulated, we realize that one event is already sufficient to obtain a time resolution of $<$1~ns for an LST or MST pixel, given the isochronous parabolic reflector (or the modified Davies-Cotton design of the MST, optimized to minimize timing dispersion; \citet{MST-TDR,Fegan:2018}), and the transit time spread of the  PMTs used~\citep{TOYAMA2015280}. For the SST, we assume a time spread $<$1.5~ns across the camera~\citep{SST-1M-TDR,Fegan:2018}, and the excellent timing characteristics of the SiPM~\citep{Piemonte:2012} and arrive at the same conclusions. This is sufficient to detect a single pixel's time behavior change with respect to its camera average (however, not for gradual drifts across the camera, for which more statistics are required).



\begin{table}[h!]
\centering
\begin{tabular}{lcccccc} \toprule
 & \multicolumn{2}{c}{LST}  &  \multicolumn{2}{c}{MST} & \multicolumn{2}{c}{SST} \\
 & $\vartheta=0^\circ$ & $\vartheta=60^\circ$  & $\vartheta=0^\circ$ & $\vartheta=60^\circ$  & $\vartheta=0^\circ$ & $\vartheta=60^\circ$ \\ \midrule 
Maximum impact distance $\rho_\mathrm{max}$ (m) &  \multicolumn{2}{c}{11.8} & \multicolumn{2}{c}{4.9}   &  \multicolumn{2}{c}{2.0} \\
Maximum muon angle $\upsilon_\mathrm{max}$ (deg) & \multicolumn{2}{c}{0.8}  &  \multicolumn{2}{c}{2.5}  &   \multicolumn{2}{c}{3.1}    \\
 \midrule
\multicolumn{2}{l}{Determination of optical bandwidth $B_\mu$}   &&&&& \\[0.1cm]   
 \hspace{0.5cm} $R_\mathrm{max}$ (Hz)  &  1.8  & 0.5  & 5 & 1.4  & 0.8    & 0.2  \\
 \hspace{0.5cm} $T_\mathrm{monitor}$ for 1\% resolution & 14~s  & 48~s &  5~s   & 18~s  & 28~min  & 18~min  \\[0.1cm]
 \midrule
 \multicolumn{2}{l}{Determination of optical bandwidth (per pixel) $B_\mathrm{pix}$}   &&&&& \\[0.1cm]   
 \hspace{0.5cm} $R_\mathrm{max}$ (Hz)  & 0.07   & 0.02   &  0.10   & 0.03 
        & 0.02   & 0.006\\
 \hspace{0.5cm} $T_\mathrm{monitor}$ for 3\% resolution & 6~min & 21~min  &  4~min & 13~min  & 2.3~h & 7.6~h\\[0.1cm] 
\midrule
 \multicolumn{2}{l}{Determination of time offset (per pixel) $\Delta t_\mathrm{pix}$}   &&&&& \\[0.1cm]   
 \hspace{0.5cm} $R_\mathrm{max}$ (Hz)  & 0.1  & 0.03 & 0.2 & 0.04 & 0.03 & 0.008 \\ 
 \hspace{0.5cm} $T_\mathrm{monitor}$ &  10~s & 36~s & 6~s & 23~s & 36~s & 2.2~min \\[0.1cm]
\midrule
 \multicolumn{2}{l}{Determination of optical PSF}  &&&&& \\[0.1cm]   
 \hspace{0.5cm} $R_\mathrm{max}$ (Hz)  & 5 & 1.6     & 8    & 2.5    & 2.4   & 0.7 \\
 \hspace{0.5cm} $T_\mathrm{monitor}$ for 5\% resolution &  9~s & 25~s & 5~s  & 16~s & 17~s&  1~min \\[0.1cm]  
\bottomrule
\end{tabular}  
\caption{\label{tab:ratelistfinal}List of maximally achievable rates of useful muon images for calibration for the proposed telescopes of CTA and corresponding monitoring time scales (see text for details). 
}
\end{table}

Finally, Table~\ref{tab:ratelistfinal} also lists maximum rates with which the PSF of a telescope can be monitored. For this, we use the results of the previous section and require muon rings with $\thetac > 1.15^\circ$ ($E_\mu > 14.3$~GeV for the LST), $\thetac > 1.1^\circ$ ($E_\mu > 11.2$~GeV for the MST), and $\thetac > 0.9^\circ$ ($E_\mu > 7.2$~GeV for the SST). On the other hand, impact distance cuts do not need to be so strict for this purpose, allowing $\rho_R > 0.2$. 
Here we assume 30\% event-wise resolution for all telescopes (see, e.g., Fig.~\ref{fig:quantile})\footnote{%
 We multiply the \comn{standard deviation} by $\sqrt{\pi/2}$ to account for the variance of the 0.2 quantile instead of the mean, as outlined in Section~\ref{sec:optical_psf}.
}, but only 5\% desired overall resolution, given the systematics outlined in the previous section. Hence, about 40 events needed to be accumulated within $T_\mathrm{monitor}$ for all telescopes.

The estimates of Table~\ref{tab:ratelistfinal} have to be understood as upper limits since neither trigger losses nor analysis cut efficiencies have been included. However, realistic estimates for these effects cannot be too far from unity, if an unbiased trigger and analysis scheme needs to be achieved. For this reason, we believe that the rates and monitor times of Table~\ref{tab:ratelistfinal} predict at least a correct order of magnitude.

\clearpage

\section{Discussion and available precision \label{sec:discussion}}


\begin{table}[htp]
\centering
\begin{tabular}{p{6.6cm}cccp{7.2cm}} \toprule
Item & \multicolumn{1}{c}{LSTs} & \multicolumn{1}{c}{MSTs} & \multicolumn{1}{c}{SSTs} & Comments \\  
\midrule
{\small Instrumental part $U_0$}                      &   \%   &   \%   &   \%   &  \\
{\scriptsize \hspace{0.4cm} 1 Determination of average $R$} & $<$0.5 & $<$0.5 & $<$0.5 & {\scriptsize Account for hexagonal mirrors (Fig.~\ref{fig:modulation}),     can be calibrated with MC simulations.} \\ 
\addlinespace[-0.2cm] 
{\small Reconstr. Cherenkov angle $\theta_c$ }       &        &        &        &   \\ 
{\scriptsize \hspace{0.4cm} 2 All ring broadening effects}  & $<$0.5  & $<$0.3 & $<$0.1 & {\scriptsize Residual biases.} \\ 
{\scriptsize \hspace{0.4cm} 3 Optical aberration }  &   \textcolor{blue}{$<$2}  &  \textcolor{blue}{$<$1}    &   \textcolor{blue}{$<$1}   & {\scriptsize Assuming plate-scale calibration, otherwise: 4\% (LST),  2\% (MST and SST).} \\
{\scriptsize \hspace{0.4cm} 4 Finite camera focus }  &   \textcolor{blue}{$<$0.5}  &  \textcolor{blue}{$<$0.5}    &   \textcolor{blue}{$<$0.5}   & {\scriptsize Assuming offline correction, otherwise: 8\% (LST),  4\% (MST), 2\%  (SST).} \\
{\scriptsize \hspace{0.4cm} \comn{5 Bending in geomagnetic field }}  &   \textcolor{blue}{$<$1}  &  \textcolor{blue}{$<$0.5}    &   \textcolor{blue}{$<$0.01}   & {\scriptsize Assuming  averageing pointings parallel and perpendicular to geomagn. field, otherwise: 2\% (LST),  1\% (MST).} \\
{\scriptsize \hspace{0.3cm} 6 Cherenkov light emission between first and 
\rule[0pt]{9mm}{0pt}secondary mirror } &   \textcolor{lightgray}{0}  &  \textcolor{lightgray}{0}   &  \textcolor{blue}{$\boldsymbol{<0.5}$}   & {\scriptsize Assume a suitable correction or cut, outlined in Sect.~\ref{sec:secondary}. } \\ 
{\small Reconstr. impact distance and $E_0(\rho_R)$ }   &    &  &   &  \\ 
{\scriptsize \hspace{0.4cm} 7 Neglect muon inclination }                      & $<$0.1 & $\boldsymbol{<0.1}$ &   $\boldsymbol{<0.2}$   & {\scriptsize } \\ 
{\scriptsize \hspace{0.4cm} 8 Overall shadow}  &  \textcolor{blue}{$<$1}  &  \textcolor{blue}{$<$1}  &  \textcolor{blue}{$\boldsymbol{<3}$}  & {\scriptsize Requires MC calibration using simulations. }\\ 
{\scriptsize \hspace{0.4cm} 9 Changing shadows with  inclination angle}                      & \textcolor{blue}{\comn{$\boldsymbol{
<1}$}} & \textcolor{blue}{\comn{$\boldsymbol{<2}$}}   &   $\boldsymbol{<1}$  & {\scriptsize Assume careful \comn{treatment}  of effect of \comn{displaced} quadratic camera shape (Sect.~\ref{sec:biasinclination})} \\
\addlinespace[-0.2cm]
{\small Atmospheric transmission $T$  }           &     &    &     &  \\ 
{\scriptsize \hspace{0.3cm} 10 Molecular part $T_\mathrm{mol}$} & \textcolor{blue}{$<$0.1} & \textcolor{blue}{$<$0.1} & \textcolor{blue}{$<$0.04} & {\scriptsize Assuming correction (Sect.~\ref{sec:moltransmission}), otherwise $<$3\% (LST), $<$2\% (MST), $<$0.3\% (SST).} \\
{\scriptsize \hspace{0.3cm} 11 Aerosol part $T_\mathrm{aer}$}   & \textcolor{blue}{$<$2} & \textcolor{blue}{$<$1}   & $<$0.1  & {\scriptsize Assume exclusion of bad nights (Sect.~\ref{sec:aertransmission}), or correction from atmospheric monitoring data. } \\ 
\addlinespace[-0.2cm]
{\small Reconstructed image size $N_\mathrm{tot}$} &     &     &      &  \\ 
{\scriptsize \hspace{0.3cm} 12 Mono trigger biases}                  & \textcolor{lightgray}{0} &  $<$1  & $<$2 & {\scriptsize  Stereo trigger assumed for the LST, but can be corrected using mono runs.} 
\\
{\scriptsize \hspace{0.3cm} 13 Stereo trigger bias}                  & \textcolor{blue}{$<$2} &  \textcolor{lightgray}{$<$2}  & \textcolor{lightgray}{$<$5} & {\scriptsize  Assuming that a bias calibration using mono runs is carried out regularly, otherwise $3-10$\%.} \\
{\scriptsize \hspace{0.3cm} 14 Pulse leakage}  &  \textcolor{blue}{$<$1}  & \textcolor{blue}{$<$1}  &  \textcolor{blue}{$<$2}  & {\scriptsize Requires signal estimators of sufficient length}. 
\\ 
{\scriptsize \hspace{0.3cm} 15 Signal estimator biases}  &  \textcolor{blue}{$<$0.1}  &  \textcolor{blue}{$<$0.1}  &  \textcolor{blue}{$<$0.1}  & {\scriptsize Requires un-biased signal estimators, otherwise (8--15)\% (Sect.~\ref{sec:fadc} and~\ref{sec:systbias})} \\ 
{\scriptsize \hspace{0.3cm} 16 Image selection biases}  &  \textcolor{blue}{$<$0.1}  &  \textcolor{blue}{$<$0.1}  &  \textcolor{blue}{$<$3}  & {\scriptsize Requires un-biased analysis cuts and  image cleaning algorithms optimized for muons.}\\ 
{\scriptsize \hspace{0.3cm} 17 Pixel baselines}         & \textcolor{blue}{$<$0.1} & \textcolor{blue}{$<$0.2} & \textcolor{blue}{$<$1} & {\scriptsize Assuming the baselines are controlled to  $<$0.3 p.e., as required for CTA} \\
{\scriptsize \hspace{0.3cm} 18 Non-active pixels}       & $<$0.5 &$<$0.5 &$<$0.5 &  {\scriptsize Requires advanced impact distance fits and less than 10\% broken pixels.} \\
\addlinespace[-0.2cm]
{\small Translation muon to gamma efficiency $\varepsilon_\mu \rightarrow \varepsilon_\gamma$} &   &   &   &  \\ 
{\scriptsize \hspace{0.3cm} 19 Chromaticity of degradation}   &   \textcolor{blue}{$<$2}  &  \textcolor{blue}{$<$1.5} &  $<$1 & {\scriptsize Assume spectral cutoff by camera protection window, additional monitoring of chromaticity of degradation otherwise: (10--15)\% for PMT-based cameras.} \\
{\scriptsize \hspace{0.3cm} 20 Dependencies on camera pixel incidence angles }   &    \textcolor{blue}{$<$0.5}  &  \textcolor{blue}{$<$0.5} &  \textcolor{blue}{$<$1} & {\scriptsize Assume corrections (Section~\ref{sec:incidence}), otherwise 3--5\%. }    \\ 
{\scriptsize \hspace{0.3cm} 21 Mis-focused mirrors}          &   \textcolor{blue}{$<$0.5}  &  \textcolor{blue}{$<$1}  &  \textcolor{lightgray}{0} & {\scriptsize Assume only one mirror mis-aligned.} \\ 
\midrule
Total                                    &   $<$4--5  & \comn{$<$3--4}  & $<$5--6  &    {\scriptsize Assuming un-correlated uncertainties} \\ 
{\scriptsize (only items 9--20:)}   &   $<$4  & \comn{$<$3.5} &  $<$5    &  {\scriptsize (Assuming previous comparison with simulations)} \\
\bottomrule
\end{tabular}  
\caption{Summary of systematic uncertainties in percent for CTA. The bold numbers correspond to those which are sizeably larger for CTA than for traditional IACTs, due to the large field-of-view, or new technology employed.  
The  uncertainties marked in blue require additional treatment beyond the analysis of \citet{vacanti} to be applicable. The numbers in light gray do not apply for standard operation of a given telescope. 
\label{tab:summaryuncertainties}}
\end{table}

We summarize the expected systematic uncertainties in Table~\ref{tab:summaryuncertainties} 
and obtain an overall accuracy of the optical throughput calibration using muon rings, after adding each contribution quadratically. The obtained numbers require careful analysis and corrections, going well beyond the algorithms presented in the past. Nevertheless, the resulting accuracy is considerably worse than that estimated in the past. 

The situation is, however, not as dramatic if we assume that telescopes arrive very well calibrated and undegraded on site, and a careful first correction can be made by comparing MC simulated muon images with those obtained during telescope commissioning.  The first eight items in Table~\ref{tab:summaryuncertainties} can then be assessed and corrected. For the subsequent monitoring of the degradation of the telescope, only the latter parts are important (provided that the hardware is not modified or replaced).

The last row of Table~\ref{tab:summaryuncertainties} 
shows these uncertainties applicable to the monitoring of the optical bandwidth. 
One can see that the numbers are already quite acceptable for the LST and the MST. 

In order to achieve this accuracy, several hardware requirements have to be met by the telescope designers: 

\begin{itemize}
\item An unbiased event flagging and forced readout has been required for muon images triggering only one single  MST or SST telescope. This measure has been taken to ensure that sufficient high-quality ring images are obtained in order for the muon calibration to provide at least a monitoring value per observation night
and to avoid trigger biases (particularly the stereo trigger bias).   
\item PMT-based cameras have selected a camera protection window that is  sufficiently opaque to light below 290~nm wavelength. This choice ensures that the overcorrection produced by muon calibration due to chromatic degradation of the optical elements of a telescope remains below the values provided in item~19 of Table~\ref{tab:summaryuncertainties}. 
\item Additionally, CTA will be equipped with a dedicated device to monitor the wavelength dependency of the telescopes' optical throughput~\citep{segreto2016}. 
\item Generally, requirements on the quality of the optical system, required to avoid significant biases in the reconstructed ring parameters, are much stricter than that achieved with the current generation of IACTs.
\end{itemize}

Moreover, the muon analysis must apply several additional calibration steps, which were not among the standard procedures in the past: 
\begin{itemize}
\item A plate-scale calibration must be carried out to a precision better than 1\%, in order to eliminate biases due to coma and astigmatism of the primary mirrors.
\item The effect of the finite camera focus must be corrected, according to Eq.~\ref{eq:tanvf}. 
\item A strong impact distance cut of $\rho_R > 0.6$ must be applied to reduce the effect of the shadows from the large-FOV cameras.
\item The atmospheric conditions, particularly aerosol extinction, must be regularly monitored, and data with large boundary layer extinction excluded. Atmospheric transmission must be included (at least on average) in the derivation of the optical bandwidth and its transformation to the bandwidth seen by Cherenkov light emitted from gamma-ray showers.
\item The readout hardware and pulse integration software must ensure that the full muon light pulse gets integrated in an unbiased way. This requires modifications of the typical low-level analysis carried out for gamma-ray analysis.
\end{itemize}

A selection of the analyzed muon images can  also be  used for monitoring purposes: the overall quality of the telescope PSF can be monitored for muon rings with $\thetac > 1.15^\circ$ (for the LST), $\thetac > 1.1^\circ$ (MST) and $\thetac > 0.9^\circ$ (SST). We predict event rates that allow such monitoring to be carried out with sufficient precision of $<$1\% on time scales of minutes to the LST and the MST and less than 5 minutes for each SST. Preferably, events are selected that were recorded at low observation zenith angle and have large 
impact distances and small inclination angles. 

With very large statistics of selected muon events, a pixel-wise monitoring of relative photon detection efficiencies should be envisaged to obtain an independent and complementary cross-check for the standard flat-fielding procedures.  We predict that such a data set will be available after less than 10~minutes (LST and MST), and within one average observation night (SST) with a resolution of 3\%.

\clearpage
\section{Conclusions \label{sec:conclusions}}

Muon calibration has been used as both an absolute calibrator and a relative monitor of the optical throughput of all IACTs in the past. Using the \citet{vacanti}'s algorithms and a series of suitable estimators for the ring parameters and the impact distance of the muon, several authors have claimed a theoretical accuracy of as good as $\approx 2$\% in the past.  We have carefully reviewed all documented algorithms and systematic effects  and found several that have not been taken into account so far, or which appear non-negligible for the large mirrors or large square-shaped cameras of CTA, or the dual-mirror telescope designs. 
Most of these effects can be corrected by modifications of the standard analysis at several critical steps, like the pulse integration, pixel selection, carefully selecting the remaining ring images, and using new models for the effect of camera shadows and central holes in the reflector surface. Nevertheless, a few of the systematic effects
will require dedicated studies or even adaptation in the design of the hardware employed for the CTA telescopes to achieve the desired accuracy. 
These effects are mostly related to the differences in the observed spectrum of Cherenkov light received from local muons and distant gamma-ray showers. The conversion of measured optical bandwidth with the muon method to its equivalent for the telescopes' response to Cherenkov light from gamma-ray showers, to account for wavelength-dependent degradation, requires hardware that efficiently cuts the camera response to photon 
wavelengths below about 290~nm.
Finally, close comparison with MC simulated telescope response to Cherenkov light emitted by local muons is necessary at the beginning of operation, to correct for residual shadows and spaces between individual mirrors. 

We find that after such a carefully prepared calibration scheme, the optical bandwidth of the full telescope can be determined with a resolution of $<$1\% (LST and MST) and $<$5\% (SST) on time scales of better than a minute and an accuracy of better than  4\% (LST and MST) and 5\% (SST).
This is considerably better than other proposed methods so far~\citep{gaug2014,Brown:2018,Stefanik:2018} and only slightly worse than the cross-calibration scheme proposed by~\citet{mitchell2015}, which allows, however, only for a relative calibration between telescopes or telescope types. Other direct methods~\citep{segreto2016} perform even better  and show more flexibility but require a dedicated setup and cannot be carried out \textit{at the same time as the telescopes perform science observations}. The muon calibration method is therefore the first option for  \textit{online} and \textit{offline} monitoring of the optical bandwidth of all CTA telescopes~\citep{gaugSPIE2014}.

Muon calibration allows one to \textit{monitor additionally, and without much effort} the  flat-fielding of the camera 
with a resolution of 3\% on time scales of at least one typical 20-minute data run (for the LST and MST) and within one typical observation night for the SSTs.  
For those telescopes that record the event arrival times, the muon images will provide the time resolution of each pixel and can serve to determine the relative time offsets with a precision of better than 1~ns within time scales of less than 2.5~minutes.
Similarly, differences in \comn{reflectance} among the mirrors can be detected by plotting the retrieved optical bandwidth as a function of reconstructed impact point.

Following the approach of the MAGIC analysis, 
the 0.2 quantile of the ring width distribution shall be used to monitor changes in the telescopes' optical PSF. 
Contrary to stars, the light from muon rings gets registered within sub-nanosecond time windows, which allows one to greatly reduce residual backgrounds from the night sky and unresolved stars. The stronger concentration of Cherenkov light from local muons close to their impact points may even enable  point-spread monitoring resolved to individual mirror facets or groups of mirrors. 
We have shown that the larger telescope mirrors, together with very strict requirements on the optical quality of the reflector, make such a monitoring scheme more challenging for CTA than in previous IACTs. Particularly, the LSTs need to select only the largest rings with a radius $\thetac > 1.0^\circ$ to be able to disentangle degradations of the optical point spread from the natural ring-broadening effects. Further selection of impact distances and the orientation of the telescope with respect to the Earth's magnetic field may be necessary. 
Even under these conditions, monitoring points can be obtained at least per minute with a precision of 5\%.

\acknowledgments

The authors thank the internal CTA referees for attentive reading and comments on the paper. 
This work has been funded by the grant FPA2015-69210-C6-6-R of the Spanish MINECO/FEDER,~EU. 
The CTA Consortium gratefully acknowledges financial support from the agencies and organisations listed at \url{https://www.cta-observatory.org/consortium_acknowledgments}.
This paper has gone through internal review by the CTA Consortium.

\clearpage
\begin{appendix}

\section{Useful Relations  \label{sec:usefulrelations}}

Several useful relations for calculations of muon Cherenkov light have been used in the following sections of this appendix. \comn{All of them assume that in air $(n-1) = \epsilon \ll 1$},
\begin{align}
\theta_\infty &\approx  \sqrt{2\epsilon}  \\
\thetac^2 &\simeq \theta^2_\infty - \left(\ddfrac{m_\mu c^2}{E_\mu}\right)^2 \\ 
  p_\mu  
          &\simeq m_\mu c\cdot\sqrt{1/(\theta^2_\infty-\thetac^2)-1} \\[0.2cm]
 \ddfrac{1}{\beta_\mu^2} &\simeq 1 - \theta_\infty^2 + \thetac^2 
\end{align}

\section{Derivation of the Expected Photon Yield for Inclined Muons  \label{sec:inclinedmuons}}

\begin{figure}[h!]
\centering
\includegraphics[width=0.55\linewidth,trim={0cm 1.5cm 0cm 0cm},clip]{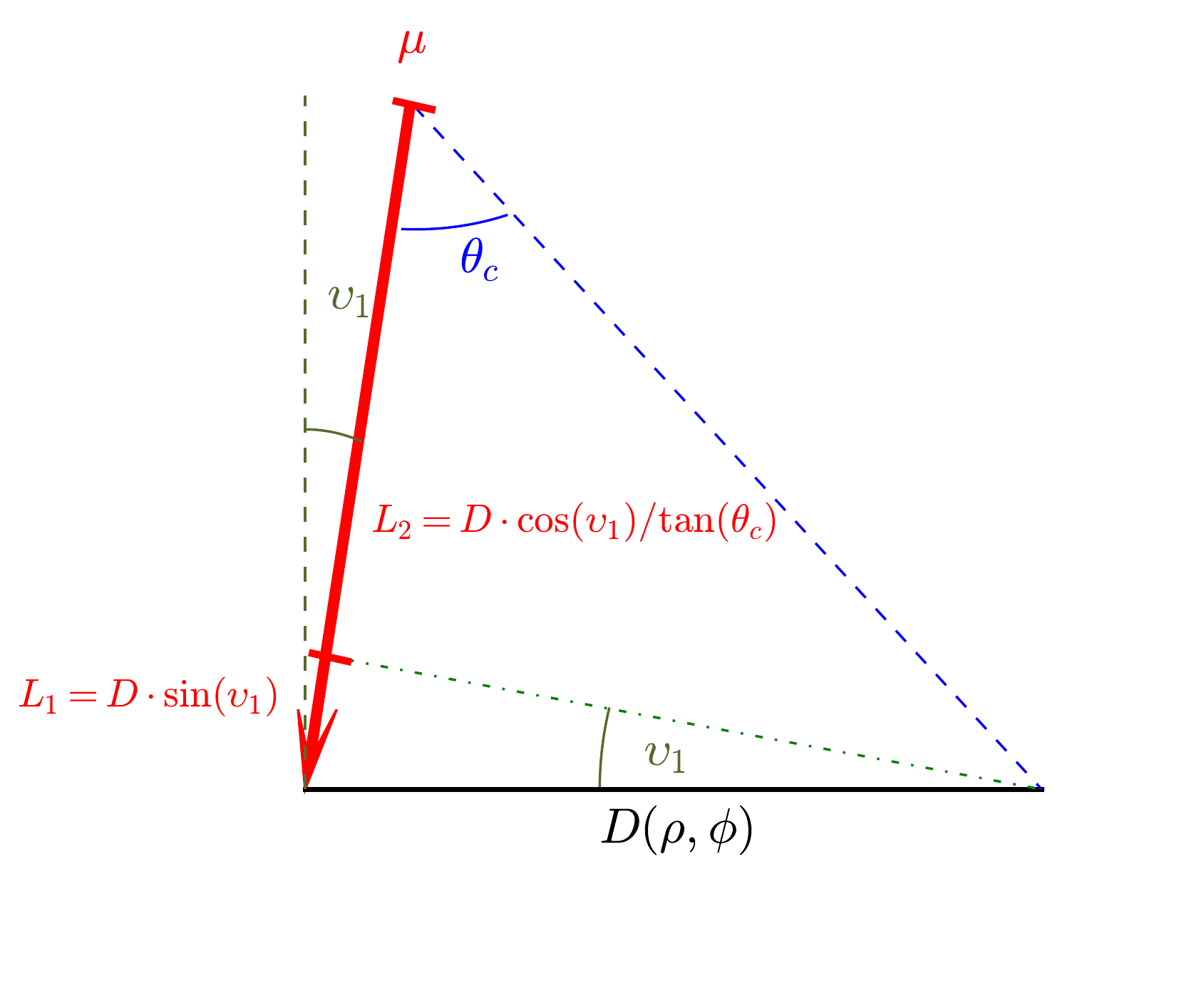}
\caption{\comn{Sketch of the parameters introduced to describe the geometry of a local \textit{inclined} muon $\mu$. 
The muon emits
Cherenkov light under the Cherenkov angle $\thetac$ along its trajectory, which is inclined by the angle $\vec{\upsilon}$ with respect to the optical axis of the telescope. The projection of $\vec{\upsilon}$ onto a vertical plane rotated by its azimuthal component $\psi$ with respect to $\phi$ yields $\upsilon_1$.  
The full length $L$ within which muon light emitted under the Cherenkov angle $\thetac$ is captured by the telescope mirror can be understood as composed of two segments, $L = L_1 + L_2$. 
For reasons of visibility, all angles and lengths are not to scale.}
\label{fig:geometry3} }
\end{figure}

Inclined muons are observed over a path length $L$ different from $D(\vec{\rho},\vec{\upsilon})/\tan\thetac$ (see Fig.~\ref{fig:geometry3}), namely,

\begin{align}
L &= D \cdot \ddfrac{\cos(\thetac-\upsilon_1)}{\sin\thetac} = \ddfrac{D}{\tan\thetac} \cdot \left( \cos(\upsilon_1) + \tan\thetac \cdot \sin(\upsilon_1)\right)~.\nonumber\\[0.25cm]
  & \mathrm{Using:\quad} \tan(\upsilon_1) = \tan\upsilon \cdot \cos(\psi-\phi)~, \nonumber\\[0.25cm]
 L &= \ddfrac{D}{\tan\thetac} \cdot  \ddfrac{1+\tan\thetac\tan\upsilon\cos(\psi-\phi)}{\sqrt{1+\tan^2\upsilon\cos^2(\psi-\phi)}} ~, \\
  &\simeq \ddfrac{D}{\tan\thetac} \cdot \left( 1 + \cos(\psi-\phi) \tan\thetac \tan\upsilon  - \cos^2(\psi-\phi) \ddfrac{\tan^2\upsilon}{2}  + O(\tan^3\upsilon) \right)~,  \label{eq:Lupsilon}
\end{align}
\noindent
where $\upsilon_1$ is the projection of the angle $\vec{\upsilon}$ onto a vertical plane rotated by $\psi$ with respect to the $\phi$. 

Since the cosine of the azimuthal projection of the incidence angle minus the azimuth part of the chord varies in the range $[-1,+1]$, the received signal gets modified by maximally
\begin{align}
\left|\ddfrac{\Delta(\ud N/\ud \phi)}{\ud N/\ud \phi}\right|_\mathrm{max}  &=  \tan\thetac \tan\upsilon  + \ddfrac{\tan^2\upsilon}{2} \\
  &\approx 0.02\,\upsilon + \upsilon^2/2 \quad,
\end{align}
\noindent
where a Cherenkov angle of $1.23^\circ$ has been inserted in the last line. 

\section{\comn{Derivation of the Expected Photon Shadow for a Quadratic Camera } \label{sec:quadraticcamera}}

\subsection{\comn{Impact Point within the Square}}

\begin{figure}[h!]
\centering
\includegraphics[width=0.48\linewidth,trim={0.5cm 1.5cm 0cm 0cm},clip]{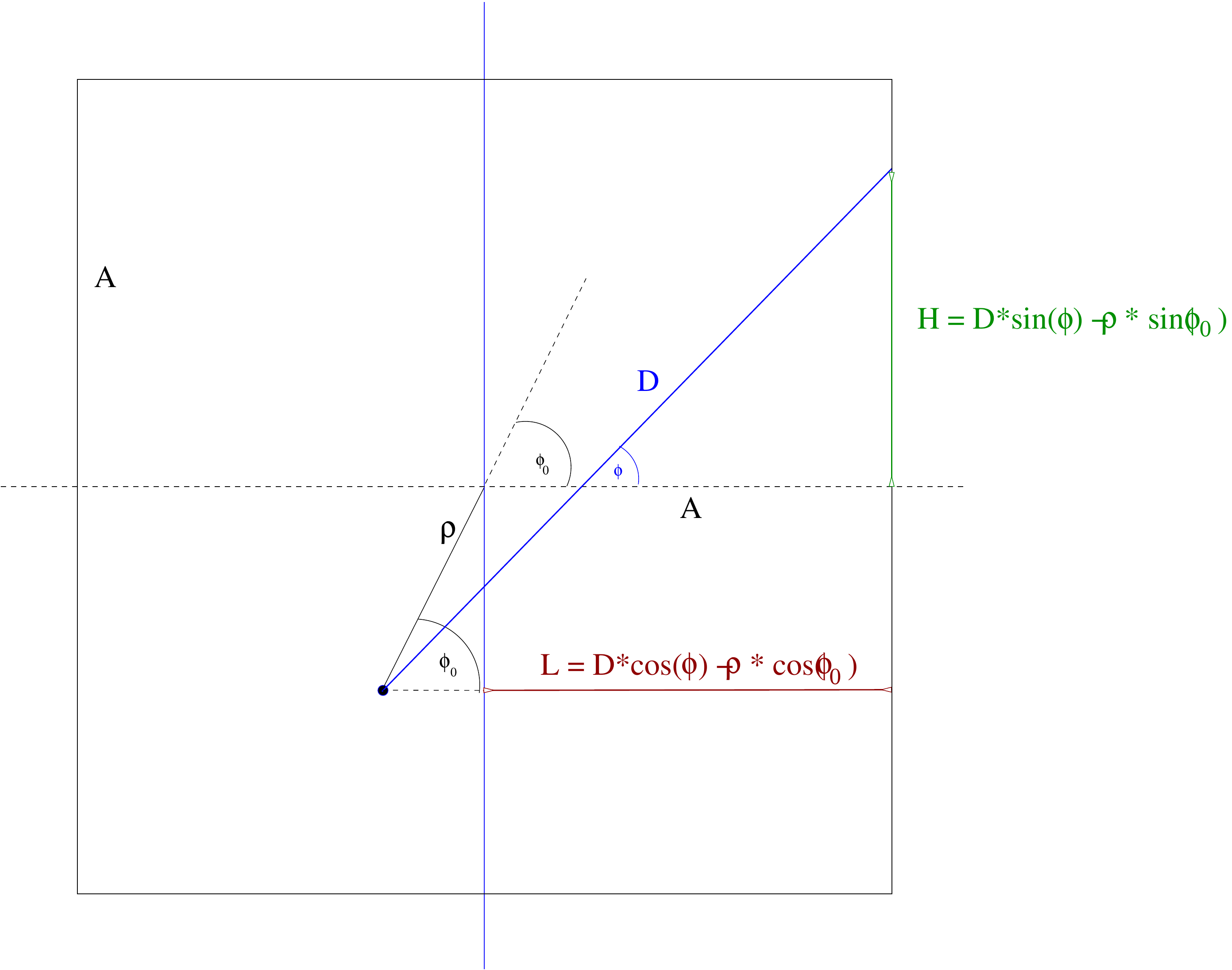}
\includegraphics[width=0.48\linewidth,trim={0cm 1.5cm 0.5cm 0cm},clip]{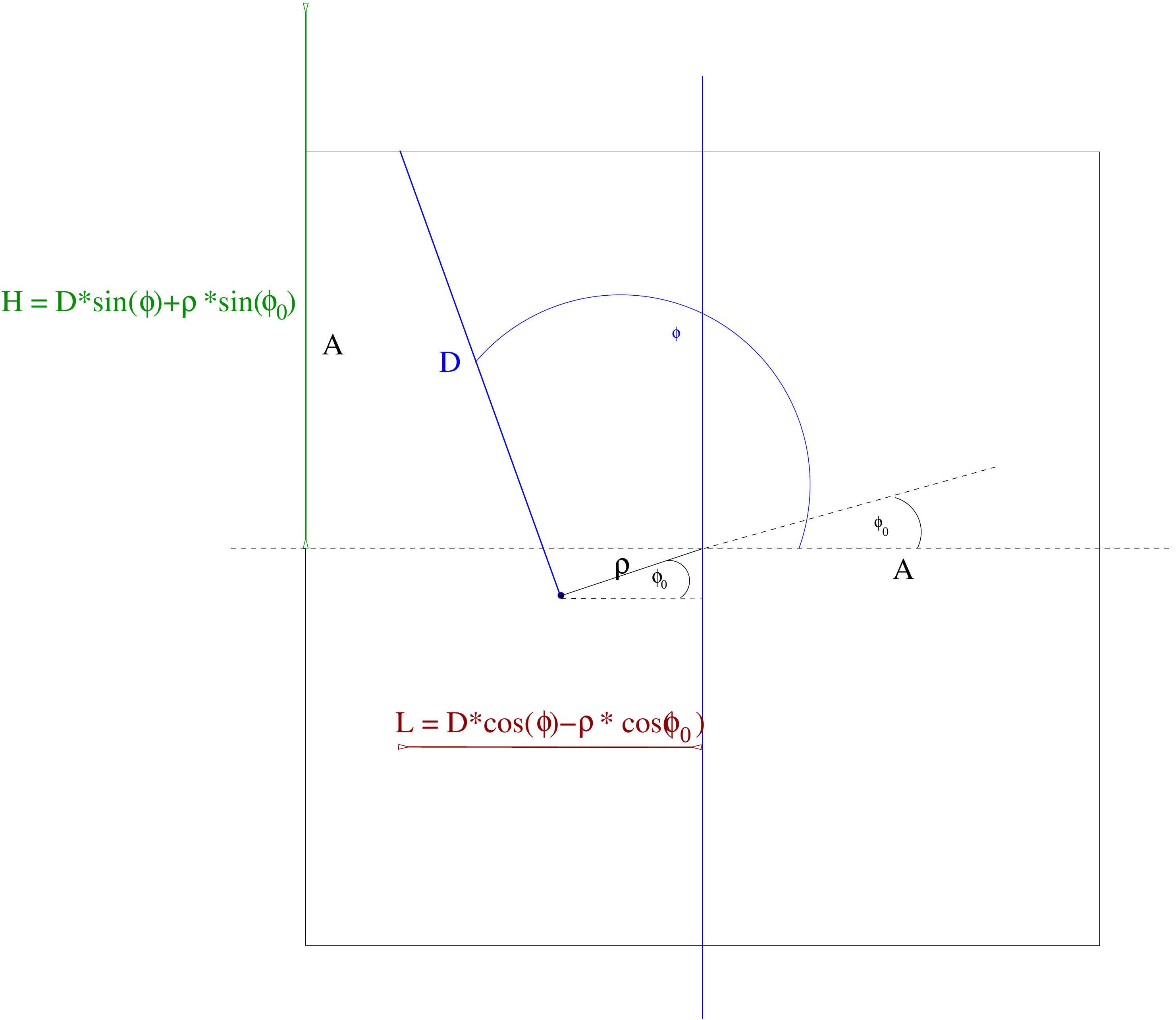}
\caption{Sketch of the parameters introduced to describe the geometry of a quadratic shadow. 
\label{fig:quadratic} }
\end{figure}

\comn{The chord $D(\rho,\phi,\phi_0)$ of a square-shaped camera with half-diameter $A$ can be expressed as the one of a roundish camera of equivalent radius $A$ (see e.q. Eq.~\ref{eq:Thetaapprox}), plus the contribution $b$ between circle and square (see Fig.~\ref{fig:quadratic}). }
\comn{Whenever the length $L$ achieves $A$, the triangles are swapped and $b$ is calculated until the upper/lower edge of the square or the right/left edge, respectively. The swap condition $S$ can be obtained by requiring that the height $H$ is found in the range $[-1,1]$:} 

\begin{align}
\comn{S := \left|\tan\phi \cdot \left(1 + \sgn(\cos\phi)\cdot\rho_A \cdot \cos\phi_0 \right) - \sgn(\cos\phi)\cdot \rho_A \cdot \sin\phi_0 \right| > 1 } \label{eq:swapcondition}
\end{align}

\comn{from which the }
\noindent
\comn{new full chord $D(\rho,\phi,\phi_0)$ can be obtained, }
\begin{align}
\comn{D(\rho,\phi,\phi_0)} &=  A \cdot 
 \begin{cases}
    \left| \ddfrac{1 + \sgn(\cos\phi)\cdot\rho_A \cdot \cos\phi_0}{\cos\phi}\right| & (S < 1) \\[0.35cm]
    \left| \ddfrac{1 + \sgn(\sin\phi)\cdot\rho_A \cdot \sin\phi_0}{\sin\phi} \right| & (S \ge 1) 
\end{cases} \quad,  \label{eq:Dquadratic}
\end{align}

\begin{figure}[h!]
\centering
\includegraphics[width=0.75\linewidth]{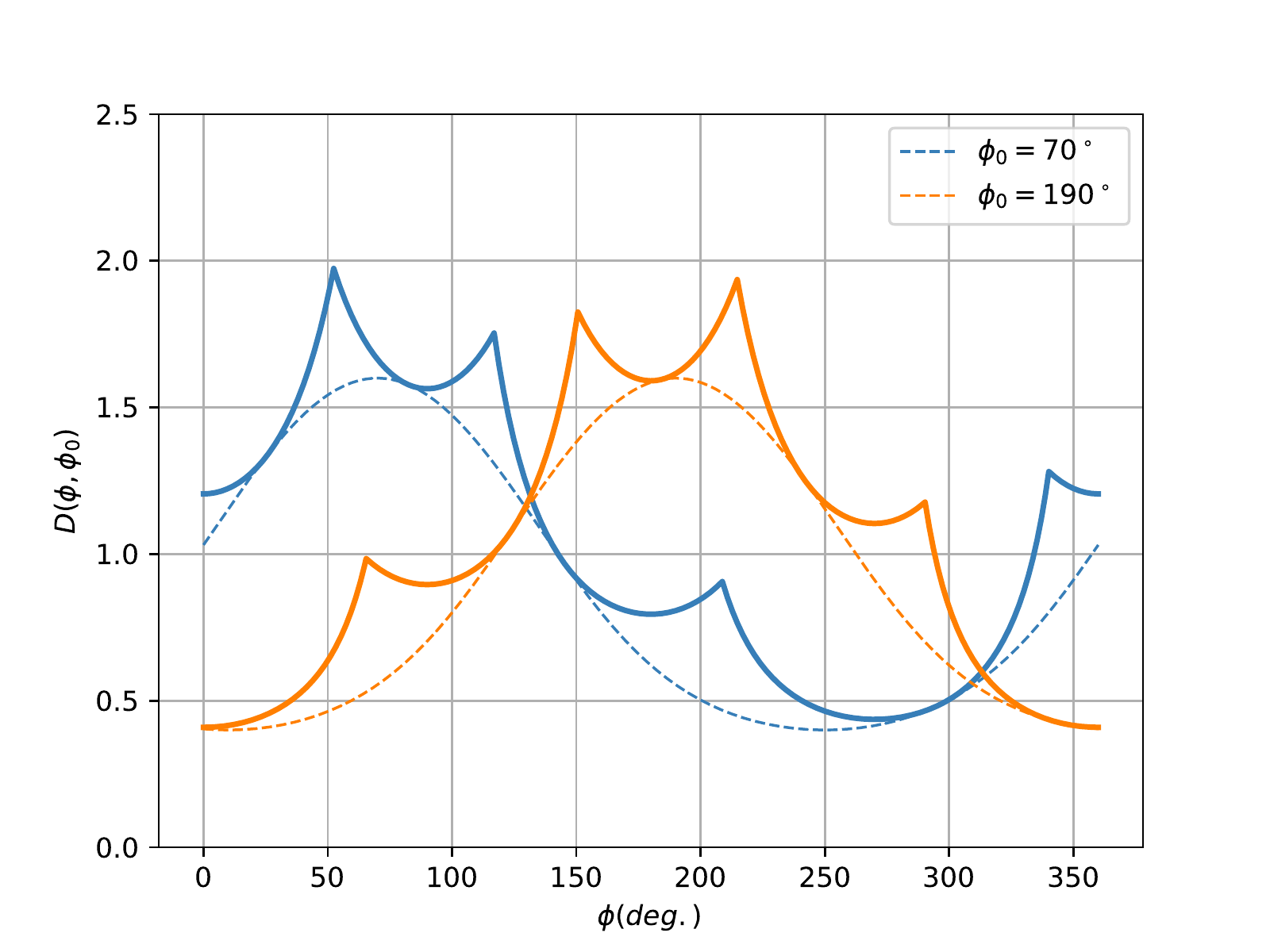}
\caption{Comparison of the shadow of a quadratic camera of half-side length $A=1$ (solid lines, using Eq.~\ref{eq:Dquadratic}) with a roundish one of radius $R=1$ (dashed lines, using Eq.~\ref{eq:D}) for two exemplary azimuth angles of the impact distance. Both cases have been calculated using a relative impact distance of $\rho_{R/A} = 0.6$. 
\label{fig:quadratic_comp} }
\end{figure}

\subsection{\comn{Impact Point outside the Square}}

\comn{For an impact point outside the square ($\rho_A > 1$), we determine first the four angles $\phi_\textit{XY}$ under which the corners of the square are seen from the impact point: }

\begin{align}
\mathrm{Upper~right~corner:} \quad
\phi_\textit{UR} &= \arctan\left(\ddfrac{\rho_A \sin\phi_0 + 1}{\rho_A \cos\phi_0 + 1}\right)  + \pi \cdot \Theta(\rho_A\cos\phi_0 < -1)   \\
\mathrm{Upper~left~corner:} \quad
\phi_\textit{UL} &= \arctan\left(\ddfrac{\rho_A \sin\phi_0 + 1}{\rho_A \cos\phi_0 - 1}\right)  + \pi \cdot \Theta(\rho_A\cos\phi_0 < +1)   \\
\mathrm{Lower~right~corner:} \quad
\phi_\textit{LR} &= \arctan\left(\ddfrac{\rho_A \sin\phi_0 - 1}{\rho_A \cos\phi_0 + 1}\right)  + \pi \cdot \Theta(\rho_A\cos\phi_0 < -1)   \\
\mathrm{Lower~left~corner:} \quad
\phi_\textit{LL} &= \arctan\left(\ddfrac{\rho_A \sin\phi_0 - 1}{\rho_A \cos\phi_0 - 1}\right)  + \pi \cdot \Theta(\rho_A\cos\phi_0 < +1)   \quad, 
\end{align}
\noindent
\comn{where $\Theta(x)$ is the Heaviside function. In order to ensure that all four angles are correctly sorted, we require, additionally, that if any of the four angles are negative \textit{and} any of the four angles are larger than $\pi$, then }
\begin{align}
\comn{\phi_\textit{XY}} &\rightarrow \phi_\textit{XY} + 2\pi  \qquad \mathrm{if}~\phi_\textit{XY} < 0
\end{align}

\comn{The visible angular range is then found between $[\min(\phi_\textit{XY}),\max(\phi_\textit{XY})]$. The sorted four angles enclose three angular ranges: within the central range, the square is fully crossed by the chord, whereas the other two ranges lead to a chord that only cuts through an edge of the square. We need to treat the three cases separately. As above,  a swap condition will be established to distinguish whether the chord crosses the square from the bottom to the top, or from the left to the right: }

\begin{align}
\comn{S :=  \left|\tan\phi_0 \right| > 1  }\label{eq:swapcondition2}
\end{align}

\comn{Then, the length of the chord fully crossing the square is }
\begin{align}
D(\rho,\phi,\phi_0) &=   A \cdot
 \begin{cases}
    \left| \ddfrac{2}{\cos\phi} \right| & (S < 1) \\[0.35cm]
    \left| \ddfrac{2}{\sin\phi} \right| & (S \ge 1) 
\end{cases} \quad,  \label{eq:Dquadratic_ext_full}
\end{align}

\comn{and the length of the chord partly crossing the square is (i.e. between the first and second sorted corner angles and between the third and fourth) }
\begin{align}
D(\rho,\phi,\phi_0) &=  A \cdot 
 \begin{cases}
     \left| \ddfrac{1+\sgn(\sin\phi)\cdot \rho_A\sin\phi_0}{\sin\phi}\right| -  \ddfrac{\rho_A|\cos\phi_0|-1}{|\cos\phi|} & \mathrm{if~chord~exits~crossing~upper~or~lower~line~of~square} \\[0.35cm]
     \left| \ddfrac{1+\sgn(\cos\phi)\cdot \rho_A\cos\phi_0}{\cos\phi}\right| - \ddfrac{\rho_A|\sin\phi_0|-1}{|\sin\phi|} & \mathrm{otherwise}
\end{cases} \quad,  \label{eq:Dquadratic_ext_part}
\end{align}

\comn{The condition if the chord exits crossing an upper/lower or a left/right side of the square was determined with the following condition for $\phi_0$ and the corner angles $\min(\phi_{XY})$ (for the chord crossing between first and second corner angles) and $\max(\phi_{XY})$ (for the chord crossing between third and fourth corner angles), respectively,}
\begin{align}
S :=  & 
\begin{cases}
(\rho_A \cdot |\sin\phi_0| < 1. ~ \wedge ~ \rho_A \cdot |\cos\phi_0| > 1)  & \mathrm{true}  \\
(\rho_A \cdot |\cos\phi_0| < 1. ~ \wedge ~ \rho_A \cdot |\sin\phi_0| > 1)  & \mathrm{false} \\[0.25cm] 
\mathrm{otherwise:}   & \\[0.25cm] 
 |\tan(\min/\max(\phi_\textit{XY})| > 1   & \mathrm{true}  \\
 |\tan(\min/\max(\phi_\textit{XY})| <= 1  & \mathrm{false}  \\
\end{cases} \quad,  \label{eq:condition_ext_part}
\end{align}
\comn{This condition determines whether the impact point is found beside the square or above or below. If not (i.e. in one of the four quarters in diagonal direction from the square), the orientation of the first and last corner point angles determines the crossing mode.  }

\begin{figure}[h!]
\centering
\includegraphics[width=0.75\linewidth]{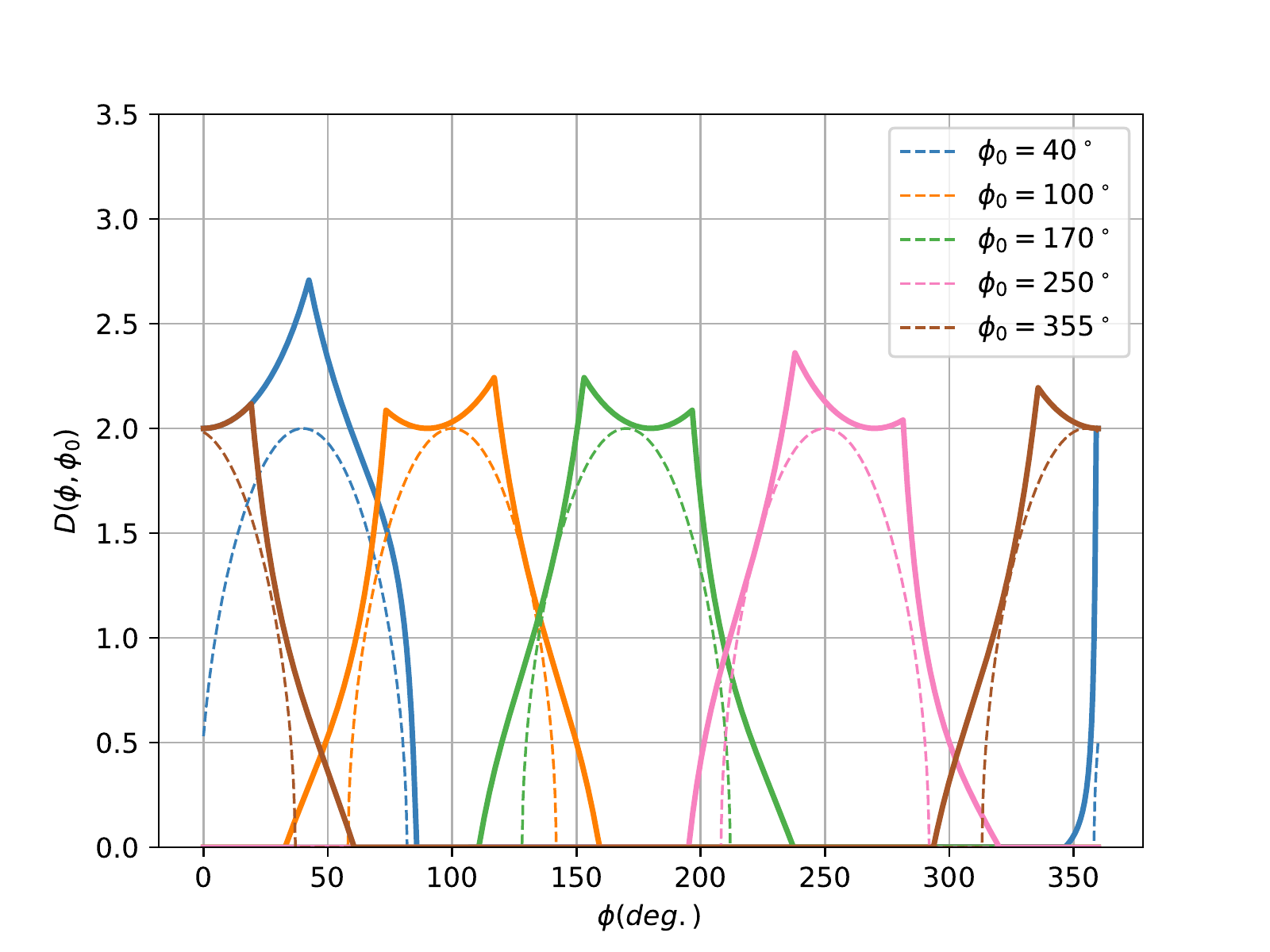}
\caption{Comparison of the shadow of a quadratic camera of half-side length $A=1$ (solid lines, using Eq.~\ref{eq:Dquadratic_ext_part}) with a roundish one of radius $R=1$ (dashed lines, using Eq.~\ref{eq:D}) for several exemplary azimuth angles of the impact distance. Both cases have been calculated using a relative impact distance of $\rho_{R/A} = 1.5$. 
\label{fig:quadratic_comp_ext} }
\end{figure}

\comn{Figure~\ref{fig:quadratic_comp_ext} compares the square shadow with a roundish one for different azimuthal angles. }

\section{Derivation of the Tangential and Sagittal Weight along the Muon Ring  \label{sec:tangsatweights}}

We consider that a muon analysis projects the ring's radial distribution along the azimuthal angle $\phi$. The distance $d$ of a point on the ring to the camera center and the angle $\delta$ of the radial projection with the line connecting the ring with the camera center can be expressed as
\begin{align}
     d       &=  \sqrt{\upsilon^2+\thetac^2-2\upsilon\thetac\cos\phi}  \\[0.3cm]
\delta(\phi) &= \arctan\left(\ddfrac{\upsilon-\thetac\cos\phi}{\thetac\sin\phi}\right) - \phi \\[0.3cm]
    & \quad \mathrm{and}  \nonumber\\[0.3cm]
    \sin(\delta) &= \left|\ddfrac{\upsilon\cos\phi-\thetac}{d}        \right|  \\[0.3cm]
   \cos(\delta)  &=   \ddfrac{\upsilon\sin\phi}{d}
\end{align}

The weight given to the tangential component, evaluated at the corresponding distance of the ring from the camera center, is then $\sin(\delta)$, whereas the sagittal component gets weighted by $\cos(\delta)$. 

\section{Model of a Nonuniform Pixel Response to Light under Different Incidence Angles}
\label{app:beta}

The following model has been adopted from \citet{fegan2007} and slightly modified. 
We can model the amount of light in a muon image
under the reasonable assumption that the response of the instrument, $\Omega(r)$, approximates to a quadratic function of the distance from the center of the reflector: 
\begin{equation}
\Omega(r_R) = \ddfrac{1 + 2\beta r^2_R}{1+\beta}\quad ,
\label{eq:Omega}
\end{equation}
where the distance from the mirror center, in units of mirror radius $r_R = r/R$ and $\beta$ describes the enhancement $(\beta > 0)$ or suppression $(\beta < 0)$ of the response for
photons reflected from the edge of the reflector. The function is normalized such that the response to a distant source that illuminates the full reflector uniformly is independent of $\beta$.
\begin{equation}
\ddfrac{1}{\pi}\int_0^1 \Omega(r_R) \, 2\pi r_R \, \ud r_R = 1 \quad.  
\end{equation}

The response function is shown in Figure~\ref{fig:Qpixbeta} (left). 
With $\beta < 0$ the response to photons reflected from the
center of the telescope is enhanced with respect to those from the edge. 
If $\beta > 0$, the efficiency from the edge of the reflector is enhanced. 

The length of the chord $D$ (Eq.~\ref{eq:D}), convolved with the response of the instrument, is then
\begin{equation}
  \mathfrak{D}(\rho_R,\phi-\phi_0,\beta) = R \cdot \int_{\comn{l_\mathrm{min}}}^{\comn{l}_\mathrm{max}}
  \ddfrac{1 + 2\beta r^2_R\comn{(l,\rho_R,\phi-\phi_0)}}{1+\beta} \, \ud \comn{l} \quad, \label{eq:DOmega}
\end{equation}
\noindent 
where
\begin{align}
\comn{r^2_R(l,\rho_R,\phi-\phi_0)} &= \comn{\rho_R^2 + l^2 - 2 \rho_R l \cos(\phi-\phi_0)}  \\
\comn{l}_\mathrm{max} &= \rho_R \cos(\phi-\phi_0) + \sqrt{1-\rho^2_R\sin^2(\phi-\phi_0)} \\
\comn{l}_\mathrm{min} &= \left\{ 
\begin{array}{lll}
0 & \mathrm{for} & \rho_R \leq  1 \nonumber\\
\rho_R \cos(\phi-\phi_0) - \sqrt{1-\rho^2_R\sin^2(\phi-\phi_0)} ~~~ 
& \mathrm{for} & \rho_R > 1 \\
\end{array}\right. \\ 
\end{align}
\noindent
Solving the integral Eq.~\ref{eq:DOmega} yields
\begin{align}
\mathfrak{D}(\rho,\phi-\phi_0,\beta) &= \ddfrac{R}{1+\beta} \cdot \left\{ 
\begin{array}{lll}
\sqrt{1-\rho^2_R\sin^2(\phi-\phi_0)} \cdot \left(1 + \ddfrac{2}{3}\beta 
\comn{\left(1+2\rho^2_R\sin^2(\phi-\phi_0)\right) }\right) + && \nonumber\\[0.3cm]
 +  \rho_R \cos(\phi-\phi_0) \cdot \left(1   
 \comn{ + \ddfrac{2}{3}\beta \rho^2_R \left( 1 + 2\sin^2(\phi-\phi_0) \right)} \right)   &
 \mathrm{for} & \rho_R \leq  1 \nonumber\\[0.4cm]
2 \sqrt{1-\rho^2_R\sin^2(\phi-\phi_0)} \cdot \left(1 + \ddfrac{2}{3}\beta 
\comn{\left(1+2\rho^2_R\sin^2(\phi-\phi_0)\right) }\right)~~~ 
& \mathrm{for} & \rho_R > 1
\end{array}\right. \\\label{eq:mathfrakD}
\end{align}
Then, integration over $\phi$ yields
\begin{align}
  Q_\mathrm{tot}(\theta_c,\rho,\beta) &=  2\frac{\alpha}{hc} \cdot \sin(2\theta_c) \cdot B_\mu  \cdot \int_0^\Phi \mathfrak{D}(\rho,\phi-\phi_0,\beta)~ \mathrm{d}\phi \nonumber\\[0.2cm]
            {}  & \qquad \mathrm{with} ~ \Phi =  \left\{ \begin{array}{lcl}
    \mathrm{arcsin}(1/\rho_R) & ~~\mathrm{for:~} & \rho_R > 1 \\[0.2cm]
    \pi / 2       & ~~\mathrm{for:}~ & \rho_R \leq 1
\end{array}
\right. \\[0.2cm]
        &\approx U_0 \cdot \theta_c \cdot \mathcal{E}(\rho_R,\beta)\label{eq:throughputcalibrationbeta}\\[0.2cm]
       {} & \qquad \mathrm{with}  \nonumber\\[0.2cm]
        \mathcal{E}(\rho_R,\beta) &= 
     \ddfrac{1}{1+\beta} \cdot \left[  (1+\ddfrac{2}{3}\beta)E_0(\rho_R) 
       \comn{+ \ddfrac{2}{3} \beta \rho_R^2E_2(\rho_R)}\right]  \\
        {} & \qquad \mathrm{and:} \left\{ \begin{array}{lcl} 
        E_0(\rho_R) &=& \frac{2}{\pi} \int\limits_0^\Phi \sqrt{1 -\rho_R^2\sin^2\phi}~ \ud\phi  \\[0.25cm]
        E_2(\rho_R) &=& \frac{4}{\pi} \int\limits_0^\Phi \sin^2\phi \cdot \sqrt{1 -\rho_R^2\sin^2\phi}~ \ud\phi \quad, 
        \end{array}
\right. \\
        \label{eq:e2}
\end{align}\noindent
and $U_0$ and $\Phi$ as defined in Eq.~\ref{eq:throughputcalibration}. \comn{For $\rho_R < 1$, }the integral $E_2$ can be translated into a combination of Legendre elliptic integrals of first $F(\rho_R)$ and second $E(\rho_R)$ kind, namely,
\begin{align}
E_2(\rho_R) &= \left\{ \begin{array}{ll}
1 & \mathrm{for:}~~ \rho_R = 0 \\[0.25cm]
\ddfrac{4}{3\pi} & \mathrm{for:~~} \rho_R = 1 \\[0.25cm]
\ddfrac{2}{3\rho^2_R}   \left( \left(2\rho^2_R-1\right)E_0(\rho_R) - \left(\rho^2_R-1\right)F_0(\rho_R)  \right) & \mathrm{~~otherwise}  \\[0.25cm]
\end{array}
\right. \nonumber\\
& \mathrm{where}  \nonumber\\
F_0(\rho_R) &= \frac{2}{\pi} \int\limits_0^{\pi/2} \ddfrac{1}{\sqrt{1 -\rho_R^2\sin^2\phi}}~ \ud\phi  
\end{align}

\section{Derivation of Bias and Variance of Ring Width from Finite Camera Focuses \label{sec:varfocus}}

From Eq.~\ref{eq:tanvf}, we derive the net change in observed Cherenkov angle, if the telescope is focused to a distance $x_f$, instead of infinity. Then, 
\begin{align}
  \ddfrac{\Delta\thetac}{\thetac} &\simeq -\ddfrac{x}{x_f} ~,
\end{align}
\noindent
where $x$ denotes the distance of the photon emission point to the telescope.

We calculate the expectation values for the net shift: 
\begin{align}
E\left[\ddfrac{\Delta\thetac}{\thetac}\right]   &= \ddfrac{1}{2\pi} \int_0^{2\pi} \!\!\ddfrac{1}{L(\rho,\phi)} \int_0^{L(\rho,\phi)} \ddfrac{\Delta \thetac(x,\phi)}{\thetac}\, \ud x\,\ud\phi \\
           &= - \ddfrac{1}{4\pi\thetac} \cdot\ddfrac{1}{x_f}  \int_0^{2\pi} \!\! D(\rho,\phi-\phi_0)  \,\ud\phi\\
          &\approx  - \ddfrac{R}{2\thetac} \cdot \ddfrac{1}{\comn{x}_f} \cdot E_0( \rho_R) ~.\label{eq:finitefocus}
\end{align}

To derive the variance, we calculate the expectation value of $(\Delta\thetac/\thetac)^2$: 
\begin{align}
E\left[\left(\ddfrac{\Delta\thetac}{\thetac}\right)^2\right]   
   &= \ddfrac{1}{2\pi} \int_0^{2\pi} \!\!\ddfrac{1}{L(\rho,\phi)} \int_0^{L(\rho,\phi)} \left(\ddfrac{\Delta \thetac(x,\phi)}{\thetac}\right)^2\, \ud x \,\ud\phi \\
           &= \ddfrac{1}{6\pi\thetac^2} \cdot\ddfrac{1}{\comn{x}_f^2}  \int_0^{2\pi} \!\! D(\rho,\phi-\phi_0)^2 \,\ud\phi \\
          &\approx  \ddfrac{R^2}{3\thetac^2}\cdot\ddfrac{1}{\comn{x}_f^2}   ~.
\end{align}

\begin{align}
\mathrm{Var}\left[\ddfrac{\Delta\thetac}{\thetac}\right] &= E\left[\left(\ddfrac{\Delta\thetac}{\thetac}\right)^2\right] - E\left[\ddfrac{\Delta\thetac}{\thetac}\right]^2  \\
   &\simeq \ddfrac{R^2}{3\thetac^2} \cdot \ddfrac{1}{\comn{x}_f^2}
   \left(1 - \ddfrac{3}{4}\cdot E_0^2(\rho_R)   \right) ~. \label{eq:varfocus}
   \end{align}

\section{Derivation of Bias and Variance of Ring Width from Muon Bending in the Earth's Magnetic Field \label{sec:varmagnetic}}

The angle under which the muon Cherenkov light is imaged in the camera is shifted by (see, e.g., Eq.~\ref{eq:Lupsilon}, \comn{using $\tan\thetac \simeq D/L$})
\begin{align}
\Delta \thetac(x,\phi)  &= \cos(\varphi-\phi) \cdot \left(\tan(\ddfrac{\ud \theta}{\ud x} \cdot x) -\ddfrac{ \cos(\varphi-\phi)}{\comn{2\tan\thetac}} \cdot \tan^2(\ddfrac{\ud\theta}{\ud x} \cdot x) \right) \\
                &\simeq     \cos(\varphi-\phi) \cdot  \left( \ddfrac{\ud\theta}{\ud x}\right) \cdot x ~, \label{eq:deltatheta_magn:a}        
\end{align}
\noindent
\comn{where a small bending effect (i.e. $\ud\theta/\ud x  \cdot L \ll \thetac$) has been assumed in the last line. Using realistic values for the Earth's magnetic field at the CTA sites, the approximation is then better than 5\% for all telescopes and useful muon Cherenkov angles.}. 
\comn{Moreover, assuming constant bending (i.e. $\ud \theta / \ud x = \mathrm{const}$),} the net shift, averaged over the ring, has an expectation value  of
\begin{align}
E\left[\ddfrac{\Delta\thetac}{\thetac}\right]   &= \ddfrac{1}{2\pi} \int_0^{2\pi} \!\!\ddfrac{1}{L(\rho,\phi)} \int_0^{L(\rho,\phi)} \ddfrac{\Delta \thetac(x,\phi)}{\thetac}\, \ud x\,\ud\phi \\
           &\approx \ddfrac{1}{4\pi\thetac^2} \cdot\left(\ddfrac{\ud\theta}{\ud x}\right)  \int_0^{2\pi} \!\! D(\rho,\phi-\phi_0) \cdot \cos(\varphi-\phi) \,\ud\phi\\
          &\approx  \ddfrac{R}{\comn{4}\thetac^2} \cdot \left(\ddfrac{\ud\theta}{\ud x}\right) \cdot \rho_R \cdot \cos(\phi_0-\varphi) ~.\label{eq:muonbending2:a}
\end{align}

To derive the variance, we calculate the expectation value of $(\Delta\thetac/\thetac)^2$: 
\begin{align}
E\left[\left(\ddfrac{\Delta\thetac}{\thetac}\right)^2\right]   
   &= \ddfrac{1}{2\pi} \int_0^{2\pi} \!\!\ddfrac{1}{L(\rho,\phi)} \int_0^{L(\rho,\phi)} \left(\ddfrac{\Delta \thetac(x,\phi)}{\thetac}\right)^2\, \ud x\,\ud\phi \\
           &\approx \ddfrac{1}{6\pi\thetac^4} \cdot \left(\ddfrac{\ud\theta}{\ud x}\right)^2  \int_0^{2\pi} \!\! D(\rho,\phi-\phi_0)^2 \cdot \cos^2(\varphi-\phi)\,\ud\phi \\
          &\approx  \ddfrac{R^2}{6\thetac^4} \cdot \left(\ddfrac{\ud\theta}{\ud x}\right)^2 \cdot \left(1 +  \ddfrac{\rho_R^2}{\comn{2}}\cos(2(\phi_0-\varphi)) \right) ~.
\end{align}

\begin{align}
\mathrm{Var}\left[\ddfrac{\Delta\thetac}{\thetac}\right] &= E\left[\left(\ddfrac{\Delta\thetac}{\thetac}\right)^2\right] - E\left[\ddfrac{\Delta\thetac}{\thetac}\right]^2  \\
   &\approx \ddfrac{R^2}{6\thetac^4} \cdot \left(\ddfrac{d\theta}{dx}\right)^2 \cdot
  \left(1 + \comn{ \ddfrac{\rho_R^2}{8}\cdot \left(1 - 5\sin^2(\phi_0-\varphi)\right) }  \right) ~.
   \end{align}

\section{Derivation of the Variance of Ring Width from Multiple Scattering of the Muons \label{sec:varmultscat}}

We denote with $h$ the altitude of the emitted Cherenkov light above the telescope and with $\vartheta$ the telescope's pointing zenith angle. The distance of the photon emission point to the telescope is $x$. Multiple scattering of the muon produces an azimuth-dependent average widening of the ring according to
\begin{align}
\ddfrac{\Delta\thetac(x)}{\thetac} &= \ddfrac{1}{\thetac}  \cdot \ddfrac{13.6~\mathrm{MeV}}{\beta p_\mu c} \cdot \sqrt{\ddfrac{H_0 \cdot(1 - e^{-x \cdot \cos\vartheta /H_0})}{R_0}}\cdot \left( 1 + 0.038 \ln \left(\ddfrac{H_0 \cdot (1 - e^{-x \cdot \cos\vartheta /H_0})}{R_0}\right) \right) \\
 \ddfrac{\Delta\thetac(\phi)}{\thetac} &\simeq \ddfrac{13.6~\mathrm{MeV}}{m_\mu} \cdot \sqrt{\ddfrac{\theta^2_\infty-\thetac^2}{\thetac^3}} \cdot \sqrt{\ddfrac{x(\rho_R,\phi)\cdot (1 - x(\rho_R,\phi)\cdot \cos\vartheta/H_0)}{R_0}} 
 \cdot \left(1 + 0.038 \ln \left(\ddfrac{x(\rho_R,\phi)}{R_0}\right) \right) 
\end{align}
where the density of air is approximated by $\rho_0 \cdot \exp(-h/H_0)$, and $\rho_0 \approx$1.225~kg~m$^{-3}$ and $R_0 = X_0/(\rho_0 \cdot \exp(-H_\mathrm{obs}/H_0)) \approx 378$~m for an observatory altitude $H_\mathrm{obs}$ of 2200~m a.s.l. 
We calculate expectation values for $\Delta\thetac/\thetac$: 
\begin{align}
E\left[\ddfrac{\Delta\thetac}{\thetac}\right]  &=
\ddfrac{1}{2\pi} \int_0^{2\pi} \ddfrac{\Delta\thetac(\phi)}{\thetac} \,\ud\phi \\ 
 & \simeq 0.129 \cdot \sqrt{\ddfrac{R}{R_0}} \cdot \sqrt{\ddfrac{\theta^2_\infty-\thetac^2}{\thetac^3}}  \ddfrac{1}{2\pi} \int_0^{2\pi} \!\! \sqrt{D(\rho,\phi)\cdot (1 - D(\rho_R,\phi)\cdot \cos\vartheta/H_0)} 
 \,\ud\phi \\
& \simeq  0.129  \cdot \sqrt{\ddfrac{R}{R_0}} \cdot \sqrt{\ddfrac{\theta^2_\infty-\thetac^2}{\thetac^3}} \cdot \left( E_0(\sqrt{\rho_R}) - \sqrt{\ddfrac{R \cdot \cos\vartheta }{\thetac \cdot H_0} } \cdot E_0(\rho_R) \right)~. 
\end{align}

\section{Derivation of the Variance of Ring Width from Variations of the Refractive Index with Altitude  \label{sec:varrefindex}}

A change in the emission altitude produces a relative change of the Cherenkov angle, according to
\begin{align}
\ddfrac{\Delta\thetac(x)}{\thetac} &= \ddfrac{1}{\thetac}  \cdot \ddfrac{\partial \thetac}{\partial \varepsilon} \cdot \ddfrac{\partial\varepsilon}{\partial h}\cdot \Delta h \\
&= \ddfrac{\theta^2_\infty}{\thetac^2} \cdot \ddfrac{\cos\vartheta \cdot x}{2H_0} \cdot e^{-x \cdot \cos\vartheta /H_0}  
\end{align}
where the density of air is approximated by $\rho_0 \cdot \exp(-h/H_0)$, and $\rho_0 \approx$1.225~kg~m$^{-3}$. 
We calculate expectation values for $\Delta\thetac/\thetac$ and for $(\Delta\thetac/\thetac)^2$: 
\begin{align}
E\left[\ddfrac{\Delta\thetac}{\thetac}\right]  &=
\ddfrac{1}{2\pi} \int_0^{2\pi} \ddfrac{1}{L(\rho,\phi)} \int_0^{L(\rho,\phi)} \ddfrac{\Delta\thetac(x)}{\thetac} \,\ud x \,\ud\phi \\ 
& \simeq \ddfrac{\theta^2_\infty}{\thetac} \cdot \ddfrac{\cos\vartheta}{2H_0} \ddfrac{1}{2\pi} \int_0^{2\pi} \!\!\!\! \ddfrac{1}{D(\rho,\phi)} \int_0^{D(\rho,\phi)/\thetac} x \cdot e^{-x \cdot \cos\vartheta /H_0}  \,\ud x \,\ud\phi \\
& \simeq \ddfrac{\theta^2_\infty}{\thetac} \cdot \ddfrac{H_0}{2\cos\vartheta} \ddfrac{1}{2\pi} \int_0^{2\pi} \!\!\!\! \ddfrac{1}{D(\rho,\phi)} \left(1 - (\ddfrac{D(\rho,\phi)\cdot \cos\vartheta}{\thetac \cdot H_0}+1)\cdot \exp(-\ddfrac{D(\rho,\phi)\cdot \cos\vartheta}{\thetac H_0} \right) \,\ud\phi \\
& \simeq \ddfrac{\theta^2_\infty}{\thetac^3} \cdot \ddfrac{\cos\vartheta}{4H_0} \ddfrac{1}{2\pi} \int_0^{2\pi} D(\rho,\phi) \cdot \left( 1 - \ddfrac{2\cdot D(\rho,\phi) \cdot \cos\vartheta}{3\cdot \thetac H_0} \right) \,\ud\phi \\
& \simeq \ddfrac{\cos\vartheta \cdot R}{4H_0}\cdot \ddfrac{\theta^2_\infty}{\thetac^3} \cdot  \left( E_0(\rho_R) - \ddfrac{2 \cdot R\cdot\cos\vartheta}{3\cdot \thetac \cdot H_0} \right) ~. 
\end{align}
\begin{align}
E\left[\left(\ddfrac{\Delta\thetac}{\thetac}\right)^2\right]  &=
\ddfrac{1}{2\pi} \int_0^{2\pi} \ddfrac{1}{L(\rho,\phi)} \int_0^{L(\rho,\phi)} \left(\ddfrac{\Delta\thetac(x)}{\thetac} \right)^2 \,\ud x \,\ud\phi \\ 
& \simeq \ddfrac{\theta^4_\infty}{\thetac^3} \cdot \ddfrac{\cos^2\vartheta}{4H_0^2} \ddfrac{1}{2\pi} \int_0^{2\pi}\!\!\!\!  \ddfrac{1}{D(\rho,\phi)} \int_0^{D(\rho,\phi)/\thetac} x^2 \cdot e^{-2x \cdot \cos\vartheta /H_0}  \,\ud x \,\ud\phi \\
& \simeq \ddfrac{\theta^4_\infty}{\thetac^3} \cdot \ddfrac{H_0}{16\cos\vartheta} \ddfrac{1}{2\pi} \int_0^{2\pi}\!\!\!\!  \ddfrac{1}{D(\rho,\phi)} \left(1 - (\ddfrac{2D(\rho,\phi)\cdot \cos\vartheta}{\thetac \cdot H_0}\cdot (\ddfrac{2D(\rho,\phi)\cdot \cos\vartheta}{\thetac \cdot H_0}+1)+1)\cdot \exp(-\ddfrac{2D(\rho,\phi)\cdot \cos\vartheta}{\thetac H_0} \right) \,\ud\phi \\
& \simeq \ddfrac{\theta^4_\infty}{\thetac^6} \cdot \ddfrac{\cos^2\vartheta}{12H_0^2} \ddfrac{1}{2\pi} \int_0^{2\pi} D^2(\rho,\phi) \cdot \left( 1 - \ddfrac{3\cdot D(\rho,\phi) \cdot \cos\vartheta}{4\cdot \thetac H_0} \right) \,\ud\phi \\
& \simeq \ddfrac{\cos^2\vartheta \cdot R^2}{12H_0^2}\cdot \ddfrac{\theta^4_\infty}{\thetac^6} \cdot  \left( 1 - \ddfrac{3 \cdot R\cdot\cos\vartheta}{4\cdot \thetac \cdot H_0}\cdot  E_0  \right) \quad. 
\end{align}

\begin{align}
\mathrm{Var}\left[\ddfrac{\Delta\thetac}{\thetac}\right] &= E\left[\left(\ddfrac{\Delta\thetac}{\thetac}\right)^2\right] - E\left[\ddfrac{\Delta\thetac}{\thetac}\right]^2  \\
   &\simeq \ddfrac{\cos^2\vartheta \cdot R^2}{12H_0^2}\cdot \ddfrac{\theta^4_\infty}{\thetac^6}\cdot  \left(  1 + \ddfrac{R\cdot\cos\vartheta}{4\cdot \thetac \cdot H_0}\cdot  E_0(\rho_R)  - \ddfrac{3E_0^2(\rho_R)}{4}  \right) ~.
 \end{align}
 
 \section{Derivation of the Variance of Ring Width from Muon Energy Loss  \label{sec:vareloss}}

\begin{align}
\ddfrac{\Delta\thetac(x)}{\thetac} &= \ddfrac{1}{\thetac} \cdot \ddfrac{\partial \thetac}{\partial E_\mu} \cdot \Delta E_\mu(x) \\
&= \ddfrac{\thetac}{m_\mu} \cdot \left(\theta^2_\infty/\thetac^2-1\right)^{3/2} \cdot \ddfrac{\ud E}{\ud x} \cdot \rho_\mathrm{air}(H_\mathrm{obs})   \cdot x \cdot \exp(-x \cdot \cos \vartheta /H_0) \\
\end{align}
Then, analog to the previous section, we can calculate the variance of $\ddfrac{\Delta\thetac(x)}{\thetac}$.
\begin{align}
\mathrm{Var}\left[\ddfrac{\Delta\thetac}{\thetac}\right] &= E\left[\left(\ddfrac{\Delta\thetac}{\thetac}\right)^2\right] - E\left[\ddfrac{\Delta\thetac}{\thetac}\right]^2  \\
   &\simeq \left(\ddfrac{\ud E}{\ud x}\right)^2 \cdot \rho_\mathrm{air}^2(H_\mathrm{obs}) \cdot \ddfrac{R^2}{3m_\mu^2}\cdot \left(\theta^2_\infty/\thetac^2-1\right)^{3} \cdot  \left(  1 + \ddfrac{R\cdot\cos\vartheta}{4\cdot \thetac \cdot H_0}\cdot  E_0(\rho_R) - \ddfrac{3E_0^2(\rho_R)}{4}  \right) ~.
\end{align}

\end{appendix}

\clearpage

\bibliographystyle{elsarticle-harv_srt}
\bibliography{ccf}

\end{document}